%% file: gA_fv_mdwf_hisq.tex
\documentclass[prd,twocolumn,tightenlines,preprintnumbers,showpacs,superscriptaddress,notitlepage,nofootinbib,eqsecnum,floatfix,longbibliography]{revtex4-1}
\input{preamble}
\input{command_list}

\input{def}

\input{affiliations}
\newcount\hour \newcount\hourminute \newcount\minute
\hour=\time \divide \hour by 60
\hourminute=\hour \multiply \hourminute by 60
\minute=\time \advance \minute by -\hourminute

\begin{document}

\title{
Signs of Non-Monotonic Finite-Volume Corrections to \texorpdfstring{$\boldsymbol{g_A}$}{}
}

\setlength{\parskip}{1.8pt}

\author{Zack~Hall}
\affiliation{\unc}
\affiliation{\lblnsd}

\author{Dimitra A.~Pefkou}
\affiliation{\ucb}
\affiliation{\lblnsd}

\author{Aaron S. Meyer}
\affiliation{\llnl}

\author{Thomas R. Richardson}
\affiliation{\ucb}
\affiliation{\lblnsd}

\author{Ra\'ul~A.~Brice\~{n}o}
\affiliation{\ucb}
\affiliation{\lblnsd}

\author{M.A.~Clark}
\affiliation{\nvidia}

\author{Martin~Hoferichter}
\affiliation{\bern}

\author{Emanuele~Mereghetti}
\affiliation{\lanl}

\author{Henry~Monge-Camacho}
\affiliation{\ornl}

\author{Colin~Morningstar}
\affiliation{\cmu}

\author{Amy~Nicholson}
\affiliation{\unc}

\author{Pavlos~Vranas}
\affiliation{\llnl}
\affiliation{\lblnsd}

\author{Andr\'{e}~Walker-Loud}
\affiliation{\lblnsd}
\affiliation{\ucb}
\affiliation{\llnl}


\begin{abstract}
We study finite-volume (FV) corrections to determinations of $g_A$ via lattice quantum chromodynamics (QCD) using analytic results and numerical analysis. We observe that $SU(2)$ Heavy Baryon Chiral Perturbation Theory does not provide an unambiguous prediction for the sign of the FV correction, which is not surprising when one also considers large-$N_c$ constraints on the axial couplings.  We further show that non-monotonic FV corrections are naturally allowed when one considers either including explicit $\Delta$-resonance degrees of freedom or one works to higher orders in the chiral expansion.
We investigate the potential impact of these FV corrections with a precision study of $g_A$ using models of FV corrections that are monotonic and non-monotonic. Using lattice QCD data that is approximately at the 1\% level of precision, we do not see significant evidence of non-monotonic corrections. Looking forward to the next phase of lattice QCD calculations, we estimate that calculations that are between the $0.1\%$--$1\%$-level of precision may be sensitive to these FV artifacts.   Finally, we present an update of the CalLat prediction of $g_A$ in the isospin limit with sub-percent precision, $g_A^{\rm QCD} = 1.2674(96)$.
\end{abstract}


\maketitle

\section{Introduction \label{sec:intro}}
The nucleon axial coupling, $g_A$, is a critical quantity for determining many nuclear physics processes, from the primordial abundances of hydrogen and helium formed during Big Bang Nucleosynthesis~\cite{Fields:2019pfx}, to our theoretical understanding of Solar Fusion~\cite{Acharya:2024lke}, and to relating precision measurements of nuclear $\beta$ decay rates to Standard Model (SM) parameters~\cite{Hayen:2024xjf}.
Measurements of $g_A$ from the $\beta$-asymmetry of ultracold neutron decays have yielded an experimental value at the 0.1\% precision level, but with some tension between the results~\cite{Bopp:1986rt,Erozolimsky:1990ui,Liaud:1997vu,Mostovoi:2001ye,Schumann:2007hz,Mund:2012fq,UCNA:2017obv,Markisch:2018ndu,Beck:2019xye,Hassan:2020hrj} that go into the PDG average~\cite{ParticleDataGroup:2024cfk}.  The most precise measurement has a total uncertainty of 0.04\%~\cite{Markisch:2018ndu}.

There is currently a $\sim\!4\s$ discrepancy between between beam~\cite{Kossakowski:1989dc,Byrne:1996zz,Nico:2004ie,Yue:2013qrc} and bottle~\cite{Serebrov:2004zf,Pichlmaier:2010zz,Steyerl:2012zz,Arzumanov:2015tea,Serebrov:2017bzo,Pattie:2017vsj,Ezhov:2014tna,UCNt:2021pcg,Musedinovic:2024gms} measurements of the neutron lifetime ($\t_n$), which depends upon $g_A$, inspiring suggestions of a light dark-matter particle~\cite{Fornal:2018eol}. 
However, a recent beam measurement at J-PARC~\cite{Fuwa:2024cdf} has been found to be consistent with the bottle measurements.
Ref.~\cite{Czarnecki:2018okw} points out that the bottle measurements are consistent with recent measurements of $g_A$.  A theoretical prediction of $g_A$ with a 0.2\% precision could discriminate between the two lifetime measurements with the same precision as the beam measurements, using the relation~\cite{Cirigliano:2023fnz,VanderGriend:2025mdc}, 
\begin{equation}\label{eq:tau_n_gA}
\t_n\big(1+3\lambda^2\big)|V_{ud}|^2=4902.3(1.3)\,\text{s}\, ,
\end{equation}
where $\lambda=g_A/g_V$ and $V_{ud}$ is the Cabibbo-Kobayashi-Maskawa (CKM) matrix element.

However, the value of $g_A$ in this formula, and as reported by experiments and in the PDG, contains radiative QED corrections.  While many of these corrections are accounted for~\cite{Sirlin:1967zza,Sirlin:1977sv,Seng:2018yzq, Seng:2018qru,Czarnecki:2019mwq,Hayen:2020cxh, Hayen:2021iga,Gorchtein:2023srs,Hayen:2024xjf}, it has recently been pointed out that there are previously unrealized structure-dependent QED corrections to $g_A$ that may be as large as 2\%~\cite{Cirigliano:2022hob}. This amount is estimated with $SU(2)$ Heavy Baryon Chiral Perturbation Theory (HB$\chi$PT)~\cite{Jenkins:1990jv,Jenkins:1991es}, and thus dependent upon unknown low-energy constants (LECs).  This means that presently lattice QCD (LQCD) calculations of $g_A$~\cite{Bhattacharya:2016zcn,Chang:2018uxx,Gupta:2018qil,Alexandrou:2019brg,Bali:2023sdi,Djukanovic:2024krw,Alexandrou:2024ozj}, which are performed in the isospin limit, cannot be directly compared to the experimental determination within the current percent-level of LQCD precision.  Refs.~\cite{Seng:2024ker,Cirigliano:2024nfi} propose strategies, such as LQCD+QED calculations that can be used to determine these QED corrections, and thus determine the unknown LECs.

The importance of obtaining sub-percent theoretical predictions for $g_A$ is further exemplified in the context of precision tests of the first row of the CKM matrix.
In the SM, $\Delta_{\rm CKM }=|V_{ud}|^2 + |V_{us}|^2 + |V_{ub}|^2 - 1 = 0$, yet there currently exists a $3\sigma$ deficit from unitarity~\cite{Cirigliano:2022yyo}  which could be due to beyond-the-SM (BSM) physics. Precision measurements of $\tau_n$ and $g_A$ are critical to render the determination of $|V_{ud}|$ from neutron decay competitive with superallowed $\beta$ decays~\cite{Hardy:2020qwl,Cirigliano:2024rfk,Cirigliano:2024msg}.
While the aforementioned QED corrections to $g_A$ do not impact the determination of $|V_{ud}|$, since they cancel in \eqnref{eq:tau_n_gA}, comparison of LQCD predictions of $g_A$ to the experimental measurements places competitive constraints on BSM right-handed charged currents (RHCCs)~\cite{Alioli:2017ces}.
A recent study considering high-energy collider observables, low-energy charged-current processes, and electroweak precision observables has shown that, within an SM effective field theory (EFT) framework, scenarios including RHCCs are most preferred for explaining this unitarity tension~\cite{Cirigliano:2023nol}. At present, LQCD calculations of $g_A$ allow one to test RHCCs at the level of $\epsilon_R=-0.2(1.2)\times10^{-2}$~\cite{Chang:2018uxx,Gupta:2018qil,Cirigliano:2022hob} (using the normalization of Ref.~\cite{Cirigliano:2021yto}), while the unitarity deficit points to $\epsilon_R=-0.69(27)\times 10^{-3}$~\cite{Cirigliano:2022yyo}, strongly motivating further improvements. 

Originally, $g_A$ served as an important benchmark quantity to demonstrate that LQCD could precisely reproduce important, known experimental results of relevance to nuclear physics~\cite{Edwards:2005ym}. In recent years, it has become a key quantity for precision tests of the SM and for testing our low-energy understanding of QCD through stringent constraints on LECs of HB$\chi$PT~\cite{Drischler:2019xuo}.

Motivated by a desire for sub-percent precision results for $g_A$, we explore the effects of the FV on LQCD calculations of this quantity and examine the behavior of the results as a function of volume ($L^3$) and pion mass ($m_{\pi}$).  
For sufficiently large volumes, $m_\pi L \gg 1$, artifacts due to the volume are exponentially suppressed, $\Order(e^{-m_\pi L})$~\cite{Gasser:1987zq}.
As a result, one might na\"ively expect these systematic effects to be simple to parameterize and monotonically vanish as $L\rightarrow\infty$.
However, prior to more recent results, there was a systematic deficit in LQCD results for $g_A$ as compared with the PDG value which became more discrepant as $m_\pi$ was reduced towards the physical pion mass.  This led to speculation that $g_A$ had particularly large FV corrections~\cite{Jaffe:2001eb} (which was refuted in Ref.~\cite{Cohen:2001bg}) which may be the source of this discrepancy~\cite{Yamazaki:2008py} and that $m_\pi L>6$ may be required to keep FV corrections $\lesssim1\%$~\cite{Yamazaki:2009zq}.  It is now understood that this discrepancy arose from uncontrolled excited state contamination~\cite{Capitani:2012gj,Bhattacharya:2016zcn,Chang:2018uxx}.

Because the precision of LQCD calculations of $g_A$ has increased, we scrutinize FV effects as they become more relevant as statistical uncertainties decrease.
In particular, we note that the sign of the approach to infinite volume (IV) at heavy pion mass differs with the expectation from HB$\chi$PT suggesting that these effects might be \emph{non-monotonic} with respect to variations $ m_\pi$ and/or $m_\pi L$.  
With this context in mind, our main goal is to understand the potential shortcomings of previous treatments of FV that have ignored potential non-monotonic FV effects and to present a new strategy to arrive at a final uncertainty that accounts for this FV systematic.  We investigate the pion mass dependence of such FV trends and explore the sensitivity of a final quoted result for $g_A$ to the use of various IV extrapolation formulae, including those predicted by HB$\chi$PT.
We take the opportunity to update the CalLat prediction from Refs.~\cite{Chang:2018uxx,Walker-Loud:2019cif}.

\section{Theoretical expectations and numerical observations \label{sec:FV_gA}}

\subsection{Numerical observations of FV corrections}
 
\begin{figure}
\includegraphics[width=\columnwidth]{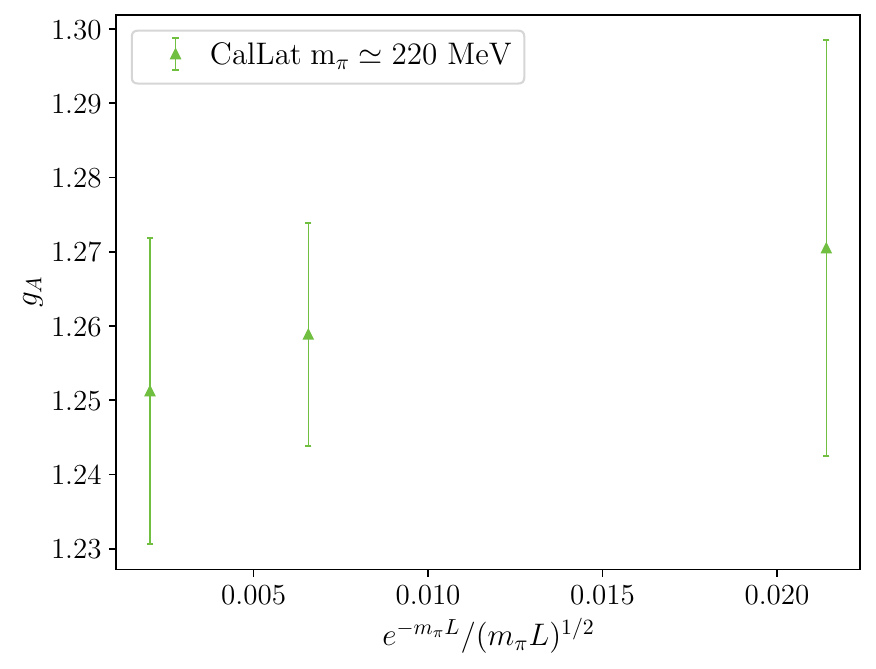}
\includegraphics[width=\columnwidth]{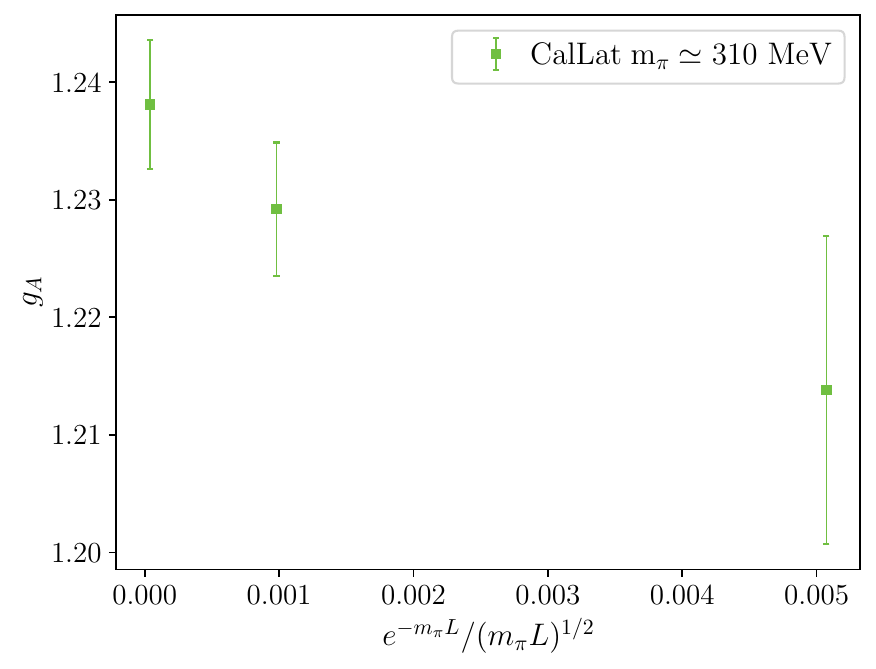}
\includegraphics[width=\columnwidth]{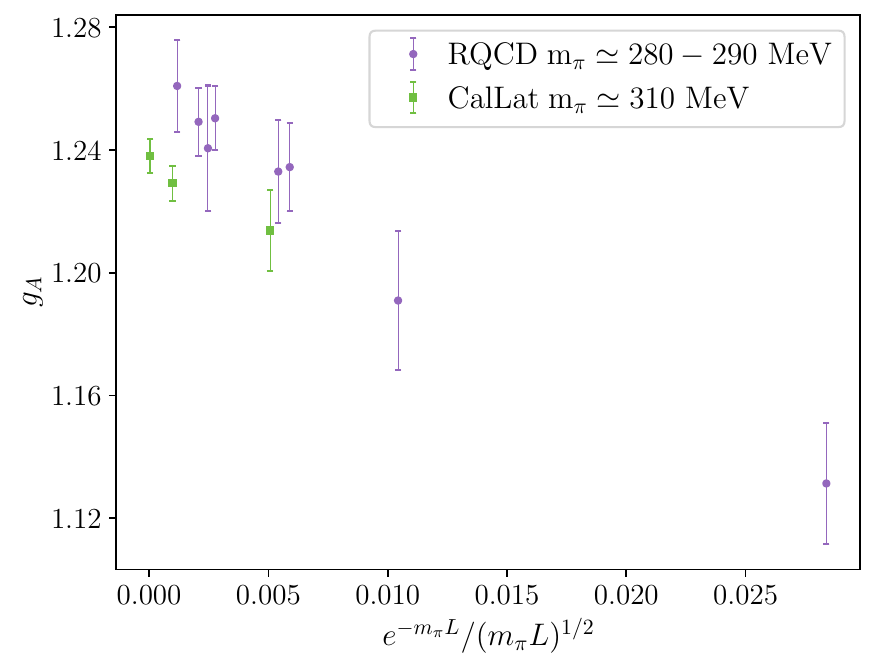}
\caption{\label{fig:a12m220_RQCD}
Top: We plot the results from Ref.~\cite{Chang:2018uxx} for $m_\pi \simeq 220$ MeV and $a \simeq 0.12$ fm.
Middle: The data point at the smallest volume with $m_\pi \simeq 310$ MeV and $a \simeq 0.12$ fm from Ref.~\cite{Chang:2018uxx}, with two new ensembles at larger volumes analyzed for this work.
Bottom: The volume dependence of data published by RQCD ~\cite{Bali:2023sdi} at $m_\pi\simeq (280\text{--}290)$ MeV.
}
\end{figure}

In Ref.~\cite{Chang:2018uxx}, CalLat observed the central values of $g_A$ trending upwards with decreasing volume at $m_\pi\approx220$~MeV,  consistent with the leading FV prediction from $SU(2)$ HB$\chi$PT without explicit $\D$ degrees of freedom (HB$\chi$PT$(\D{\hskip-0.5em/})$)~\cite{Beane:2004rf}, though the trend was not statistically significant (see the upper panel of \figref{fig:a12m220_RQCD}).
The previously unpublished data at $m_\pi\simeq 310$~MeV plotted in the middle panel of \figref{fig:a12m220_RQCD} shows the opposite trend with mild statistical significance. 
In both cases, the precision of these results is also compatible with no FV dependence at the $\sim\! 1\s$ level. All CalLat data used in this study is tabulated in Table~\ref{tab:results}.
RQCD has computed $g_A$ at several volumes (and lattice spacings in the range $a\simeq 0.064-0.086$~fm) for $m_\pi\simeq 285$~MeV, showing a statistically significant trend that the FV corrections to $g_A$ are negative at this pion mass~\cite{Bali:2023sdi}. This is seen on the bottom panel of \figref{fig:a12m220_RQCD}, while the plotted RQCD results are listed in Table~\ref{tab:RQCDresults}.


\subsection{Expectations from \texorpdfstring{$\boldsymbol{\chi}$}{}PT and large \texorpdfstring{$\boldsymbol{N_c}$}{}\label{sec:XPT_largeN}}

From an EFT perspective, it is not surprising that the FV corrections to $g_A$ are more complex than for other quantities, particularly in comparison with meson quantities.
First, the chiral expansion for nucleon quantities is a series in $\e_\pi$, defined as
\begin{align}
&\e_\pi \equiv \frac{m_\pi}{\L_\chi}\, ,&
&\L_\chi \equiv 4\pi F_\pi\, ,&
\end{align}
where $F_\pi$ is the pion decay constant.
For pions, the expansion is in powers of $\e_\pi^2$ and so higher-order chiral corrections are relatively more important for nucleons.

Second, several resonances are strongly coupled to the nucleon and of relatively low mass excitation.  In particular, the mass gap to the $\D$-resonance at the physical point is small, $\D \equiv m_\D - m_N \simeq  293$~MeV, and decreases as the pion mass increases~\cite{Walker-Loud:2008rui}.  The $\D$-resonances are also strongly coupled to the nucleon, both through QCD dynamics as well as through large electroweak couplings that govern $N\rightarrow\D$ matrix elements.  Therefore, the $\D$s make significant contributions to many quantities, adding another channel of important radiative virtual corrections~\cite{Jenkins:1991es,Jenkins:1995gc,Hemmert:1996rw,Flores-Mendieta:2000ljq,Jenkins:2002rj,Siemens:2016jwj}.

Third, for $g_A$, as for other quantities, the large-$N_c$ expansion places constraints on radiative corrections.  These constraints do not need to be imposed, but emerge naturally from the Feynman rules in $SU(2)$ HB$\chi$PT with $\Delta$ degrees of freedom (HB$\chi$PT($\D$)).
This theory has an additional small parameter to control the perturbative expansion and is sometimes referred to as the small-scale expansion~\cite{Hemmert:1997ye},
\begin{equation}
\e_\D = \frac{\D}{\L_\chi}\, .
\end{equation}
For $g_A$, there are significant cancellations between contributions with virtual $\D$-resonance states and those with virtual nucleon states \cite{Jenkins:1991es}.  These cancellations are evident with the use of phenomenological values of the LECs and become manifest when the large-$N_c$ expansion is used to relate them~\cite{Dashen:1993as, Dashen:1993ac, CalleCordon:2012xz, Jenkins:1995gc,Flores-Mendieta:2000ljq}.  These cancellations also hold for the FV corrections, as we will now discuss.

In $SU(2)$ HB$\chi$PT($\D$), the expression for $g_A$ at next-to-leading order (NLO) in the chiral expansion is~\cite{Hemmert:2003cb,Beane:2004rf}%
\footnote{This expression is derived in  App.~\ref{app:gA_NLO_Delta} from that given in Ref.~\cite{Beane:2004rf}. $d_{16}^{r,\D}$ is defined in such a way that for $\D\to\infty$ it coincides with $d_{16}^r$ in the conventions of Ref.~\cite{Kambor:1998pi,Becher:2001hv,Gasser:2002am}. The $\D$ parameters are related to those of Ref.~\cite{Siemens:2016jwj} by $g_{N\D}=\sqrt{2}h_A$ and $g_\D=-g_1$.}
\begin{align}\label{eq:gA_SU2Delta}
g_A^{(\D)} &= g_0 \left[1
    -\e_\pi^2 \ln \e_\pi^2
    \right]
    +\e_\pi^2 (4\tilde{d}_{16}^{r,\D} -g_0^3)
\nonumber\\&\phantom{=}
    -2\e_\pi^2 \ln \e_\pi^2 \left[
        g_0^3+\frac{1}{9}g_0 g_{N\D}^2 +\frac{25}{81}g_{\D}g_{N\D}^2
    \right]
\nonumber\\&\phantom{=}
    -R\left(\frac{m_\pi^2}{\D^2}\right) g_{N\D}^2\left[
        \e_\pi^2\frac{32}{27}g_0
        +\e_\D^2 \left(\frac{76}{27}g_0 + \frac{100}{81}g_{\D}\right)
    \right]
\nonumber\\&\phantom{=}
    +\frac{32}{27}g_0 g_{N\D}^2 \e_\pi^2 \left[
    1+
    \ln \frac{4\e_\D^2}{\e_\pi^2}
    +\frac{\pi\e_\pi}{\e_\D}
    \right]
\nonumber\\&\phantom{=}
    +\frac{2}{81} g_{N\D}^2 \e_\pi^2 \left(9g_0+25g_\D\right)\left[1+\ln(4\e_\D^2)\right]\, ,
\end{align}
where the non-analytic function is
\begin{equation}\label{eq:R}
R(z) = \left\{
    \begin{matrix}
    \sqrt{1-z}\, \ln\left(\frac{1-\sqrt{1-z}}{1+\sqrt{1-z}}\right) +\ln(4/z),& z \leq 1\\
    2\sqrt{z-1}\arctan(z) +\ln(4/z), & z>1
    \end{matrix}
\right.\, ,
\end{equation}
and we have defined $\tilde{d}_{16}^{r,\D} = (4\pi F)^2 d_{16}^{r,\D}$.
Each of the axial couplings, $g_0$, $g_{N\D}$, and $g_{\D}$, represent the leading-order (LO) contribution to the axial charges of the nucleon, the nucleon-to-$\D$ transition, and $\D$, respectively.  As such, they all scale like $N_c$ \cite{Dashen:1993as, Dashen:1993jt, Dashen:1994qi}.
Because $g_A$ scales like $N_c$, and the radiative corrections scale like $g^3/F^2\sim N_c^2$ ($F^2\sim N_c$) with $g\in\{g_0, g_\D, g_{N\D}\}$, all corrections with three powers of these axial couplings must exactly cancel in the large-$N_c$ limit.

One can consider the large-$N_c$ expansion about $N_c\rightarrow\infty$ or about large but finite $N_c$, which leads to two different relations at LO in the large-$N_c$ expansion between the axial couplings~\cite{Karl:1984cz}:
\begin{align}
\label{eq:largeN_g}
&N_c=\infty:&
&\!\D = 0\,,&
&\!g_{N\D} = \frac{3}{2}g_0\,,&
&\!g_{\D} = -\frac{9}{5}g_0\,,&\\
\label{eq:largeN_g_finite}
&N_c\neq\infty:&
&\!\D = 0\,,&
&\!g_{N\D} = \frac{6}{5}g_0\,,&
&\!g_{\D} = -\frac{9}{5}g_0\, \asmadd{.}\old{,}&
\end{align}
Both the former~\cite{Dashen:1993as, Dashen:1993jt, Dashen:1993ac, Jenkins:1998wy, Siemens:2016jwj}
and the latter~\cite{Dashen:1994qi,Flores-Mendieta:2000ljq, CalleCordon:2012xz, Flores-Mendieta:2006ojy, Flores-Mendieta:2012fxp, Flores-Mendieta:2021wzh, Flores-Mendieta:2024kvj}
commonly appear in the literature.
The difference between these two is formally $\Order(N_c^{-2})$.
Using the expansion about $N_c\rightarrow\infty$, Eq.~\eqref{eq:largeN_g}, the cancellation between all terms that scale as $g_0^3$
is explicit. This
can be verified by inserting these relations into \eqnref{eq:gA_SU2Delta}, yielding the result
\begin{equation}
\lim_{N_c\rightarrow\infty}g_A^{(\D)} = g_0
 +\e_\pi^2\left[- g_0\ln\e_\pi^2 
    +4\tilde{d}_{16}^{r,\D}
    \right]\, .
\end{equation}
This expression is contrasted with that in $SU(2)$ HB$\chi$PT$(\D{\hskip-0.5em/})$ at same order in $\e_\pi$,
\begin{equation}
g_A^{(\D{\hskip-0.5em /})}
=
    g_0
    +\e_\pi^2 \left[
    -g_0(1+2g_0^2)\ln \e_\pi^2
    +4\tilde{d}_{16}^r -g_0^3
    \right]\, .
\end{equation}
As $g_0\simeq 1.2$, the coefficient of the $\e_\pi^2\ln\e_\pi^2$ term in HB$\chi$PT$(\D{\hskip-0.5em/})$ is much larger than one would expect starting from HB$\chi$PT$(\D)$.  That LQCD results for $g_A$ show such mild pion mass dependence is evidence that the large-$N_c$ expansion provides at least a good qualitative guide for understanding chiral corrections to $g_A$ as it naturally reduces the size of corrections at a given order through the large-$N_c$-predicted opposite-sign contributions from virtual $\D$ corrections~\cite{Jenkins:1991es,Hemmert:2003cb}.

Similar constraints hold when considering FV corrections, which at NLO  are given by $\d_{\rm FV}^{(2)}+\d_{\rm FV}^{(2,\D)}$~\cite{Beane:2004rf}, where
\begin{align}
\d_{\rm FV}^{(2)}&=g_0\frac{8}{3}\e_\pi^2 \bigg[
    g_0^2 F_1^{(2)}(m_\pi L) +F_3^{(2)}(m_\pi L)\bigg]\,,
        \label{eq:NLO_FV}
\\
    \d_{\rm FV}^{(2,\D)}&=g_0\frac{8}{3}\e_\pi^2 \bigg[g_{N\D}^2\left(1 +\frac{25}{81}\frac{g_\D}{g_0}\right)
    F_2^{(2)}(m_\pi L,\D L)\notag
\\
    &+g_{N\D}^2\, F_4^{(2)}(m_\pi L, \D L)
    \bigg]\, ,
    \label{deltaNLO}
\end{align}
and the FV functions $F_i^{(2)}$ are defined in Ref.~\cite{Beane:2004rf},
\begin{widetext}
\begin{align}
F_1^{(2)}(x) &= \sum_{\vec{n}\neq0} \left[ 
    K_0(x|\vec{n}|)
    -\frac{K_1(x|\vec{n}|)}{x|\vec{n}|}
    \right]\, ,\qquad
F_3^{(2)}(x) = -\frac{3}{2} \sum_{\vec{n}\neq0} 
    \frac{K_1(x|\vec{n}|)}{x|\vec{n}|}\, ,
\nonumber\\
F_2^{(2)}(x,y) &= -\sum_{\vec{n}\neq0}\bigg[
    \frac{K_1(x|\vec{n}|)}{x|\vec{n}|}
    +\frac{y^2-x^2}{x^2}K_0(x|\vec{n}|)
    -\frac{y}{x^2}\int_x^\infty dz
    \frac{2zK_0(z|\vec{n}|)+(y^2-x^2)|\vec{n}|K_1(z|\vec{n}|)}{\sqrt{z^2+y^2-x^2}}
    \bigg]\,,
\nonumber\\
F_4^{(2)}(x,y) &= \frac{8}{9}\sum_{\vec{n}\neq0}\bigg[
    \frac{K_1(x|\vec{n}|)}{x|\vec{n}|}
    -\frac{\pi e^{-x|\vec{n}|}}{2y|\vec{n}|}-\frac{y^2-x^2}{x^2y}
    \int_x^\infty dz
    \frac{zK_0(z|\vec{n}|)}{\sqrt{z^2+y^2-x^2}}
\bigg]\, .
\label{eq:Fn2} 
\end{align}
\end{widetext}
$K_n(x)$ are modified Bessel functions of the second kind.
As with the IV corrections, a careful implementation of the large-$N_c$ limit at fixed quark mass, \eqnref{eq:largeN_g}, leads to an exact cancellation of all corrections that scale like $g_0^3$.  In this limit, one would predict that the leading FV correction to $g_A$ is given by
\begin{equation}
\d_{\rm FV}^{(2)}+\d_{\rm FV}^{(2,\D)} \to
    -4\e_\pi^2 g_0 \sum_{\vec{n}\neq0} \frac{K_1(m_\pi L|\vec{n}|)}{m_\pi L|\vec{n}|}\, .
\end{equation}
In contrast, the leading FV correction in HB$\chi$PT$(\D{\hskip-0.5em/})$ is
\begin{align}
\label{eq:FV_SU2_DeltaLess}
\d_{\rm FV}^{(2)} &= 
    4\e_\pi^2 g_0\sum_{\vec{n}\neq0} \bigg[
    \frac{2}{3}g_0^2 K_0(m_\pi L|\vec{n}|)
\notag\\
    &-\left( 1 + \frac{2}{3}g_0^2 \right) \frac{K_1(m_\pi L|\vec{n}|)}{m_\pi L|\vec{n}|}
    \bigg]\, .
\end{align}
For values of $m_\pi L\gtrsim3$ and $g_0\simeq 1.2$,
the first term is larger than the second term and so the predicted FV correction is positive,
while the leading prediction from HB$\chi$PT$(\D)$ after application of the large-$N_c$ expansion is negative for any value of $m_\pi L$.

\begin{figure}
\includegraphics[width=\columnwidth]{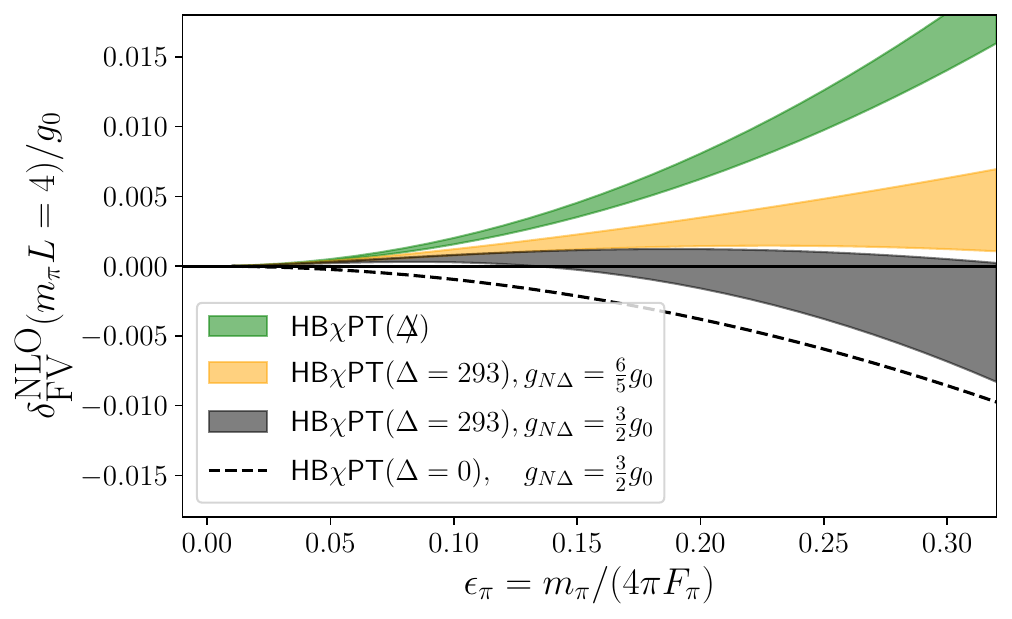}
\caption{\label{fig:NLO_FV_Delta}
We plot the predicted NLO FV corrections to $g_A$ using $SU(2)$ HB$\chi$PT($\D$) and the large-$N_c$ expansion, with two different large-$N_c$ relations for $g_{N\D}$.  We also plot the predicted FV correction from HB$\chi$PT$(\D{\hskip-0.5em/})$ and the large-$N_c$ expansion also with $\D=0$.  All corrections are plotted with $m_\pi L=4$.
}
\end{figure}

The full NLO prediction for the HB$\chi$PT$(\D)$ FV correction will lie somewhere between these limits.  We estimate them as follows.
First, we plot the predicted FV correction from HB$\chi$PT$(\D{\hskip-0.5em/})$ as a function of $\e_\pi$ while setting $m_\pi L=4$ and $g_0=1.20(05)$, see \figref{fig:NLO_FV_Delta}.
Next, we set $\D=293$~MeV and then approximate $g_{N\D}$ and $g_\D$ according to \eqnref{eq:largeN_g} while adding an extra uncertainty from the large $N_c$ expansion, e.g. $g_{N\D}=\frac{3}{2}(g_0+(0,1/N_c^2))$.
We plot the same correction except using the finite $N_c$ relation $g_{N\D} = \frac{6}{5}(g_0+(0,1/N_c^2))$, \eqnref{eq:largeN_g_finite} and, finally, the prediction with \eqnref{eq:largeN_g} and $\D=0$.
The spread of these predictions demonstrates that our present knowledge is insufficient to predict the sign of the NLO FV correction without further input.  It would be ideal to have LQCD calculations of the $N\rightarrow\D$ axial transition matrix elements and the $\D$ axial charge~\cite{Bulava:2022vpq,Barca:2022uhi}, which could be used to make more precise predictions for these FV corrections, to be compared with more precise LQCD results of $g_A$ at different volumes and pion masses. To this end, we note that also the full next-to-next-to-leading-order (\nxlo{2}) results including $\D$ degrees of freedom and FV corrections are available, see App.~\ref{app:FV_n2lo_dleta} for the complete  expressions. 

\section{Quantifying the FV Uncertainty\label{sec:reduced_unc}}
When accounting for FV corrections to $g_A$, the strategy used by most groups is to use the asymptotic form of the leading prediction from $SU(2)$ HB$\chi$PT($\D{\hskip-0.5em /}$)
\begin{equation}
\d_{\rm FV} = c_V m_\pi^2 \frac{e^{-m_\pi L}}{\sqrt{m_\pi L}}\, ,
\end{equation}
which can be obtained from \eqnref{eq:FV_SU2_DeltaLess} by expanding the Bessel functions for large argument and replacing the coefficient predicted by HB$\chi$PT$(\D{\hskip-0.5em/})$ with $c_V$.  This parameter is then determined by fitting results at heavier than physical pion mass, typically in a global analysis simultaneously with discretization corrections as well as quark mass dependence~\cite{Bhattacharya:2016zcn,Gupta:2018qil,Bali:2023sdi,Djukanovic:2024krw}.
Under the assumption that the FV corrections are non-monotonic versus pion mass, an important question to resolve is at what level of precision will the above model introduce an error?

In order to estimate the potential impact of the FV model, we perform the following exercise:
first, we pick a specific model to describe the pion mass dependence and two models of the FV corrections, one that is monotonic and one that supports non-monotonicity.
In order to create this model, we use $SU(2)$ HB$\chi$PT($\D{\hskip-0.5em /}$) at \nxlo{2}, for which $g_A$ is given by 
\begin{align}\label{eq:gA_inf}
g_A &= g_0 
    + \d_\chi^{(2)} 
    +\d_\chi^{(3)}
    +\d_{\rm FV}^{(2)}
    +\d_{\rm FV}^{(3)}\, ,
\end{align}
where the superscript, $n$, denotes the order in $\e_\pi^n$ at which a term contributes (see also App.~\ref{app:FV_n2lo_dleta} for a summary of all corrections).  The IV chiral corrections are given by~\cite{Kambor:1998pi}
\begin{align}
\label{eq:deltachis}
\d_\chi^{(2)} &= \e_\pi^2 \left[
    -g_0(1+2g_0^2)\ln \e_\pi^2
    +4\tilde{d}_{16}^r
    -g_0^3
    \right]\,,
\nonumber\\
\d_\chi^{(3)} &= \e_\pi^3 g_0 \frac{2\pi}{3}\left[
    3(1+g_0^2) \frac{4\pi F}{M_0}
    +4(2\tilde{c}_4 -\tilde{c}_3)
    \right]\, ,
\end{align}
where the LECs are defined as
\begin{equation}\label{eq:dimensionless_c_d}
F = \lim_{m_\pi \rightarrow 0} F_\pi \,,
\quad
\tilde{c}_i = (4\pi F) c_i \,,
\quad
\tilde{d}_i^r =(4\pi F)^2 d_i^r\,,
\end{equation}
such that $\tilde{c}_i$ and $\tilde{d}_i$ are dimensionless.
Note that the use of $\e_\pi$ in these expressions, rather than $m_\pi / (4\pi F)$, induces higher-order corrections beginning at $\Order(\e_\pi^4)$, which should be accounted for if the next-to-next-to-next-to-leading (\nxlo{3}) expression is used to fit $g_A$ results (see Refs.~\cite{Bernard:2006te,Chang:2018uxx}).

The NLO FV corrections are given in Eq.~\eqref{eq:NLO_FV}, while the \nxlo{2} ones are
\begin{align}\label{eq:NNLO_FV}
\d_{\rm FV}^{(3)} &= 
    \e_\pi^3 g_0\frac{2\pi}{3}
    \bigg\{
    g_0^2 \frac{4\pi F}{M_0} F_1^{(3)}(m_\pi L)\\
    &-\left[ 
        \frac{4\pi F}{M_0}(3+2g_0^2) 
        +4( 2 \tilde{c}_4 -\tilde{c}_3 ) 
    \right] F_3^{(3)}(m_\pi L)
    \bigg\}\, ,\notag
\end{align}
with
\begin{align} 
F_1^{(3)}(x) &= \sum_{\vec{n}\neq\vec{0}}
    \frac{K_{\frac{1}{2}}(x |\vec{n}|)}{\sqrt{\frac{\pi}{2} x |\vec{n}|}} x|\vec{n}|
    =
    \sum_{\vec{n}\neq\vec{0}} e^{-x |\vec{n}|}\,,\notag\\
F_3^{(3)}(x) &= \sum_{\vec{n}\neq\vec{0}}
    \frac{K_{\frac{1}{2}}(x |\vec{n}|)}{\sqrt{\frac{\pi}{2} x |\vec{n}|}}
    =
    \sum_{\vec{n}\neq\vec{0}} \frac{e^{-x |\vec{n}|}}{x|\vec{n}|}
\,.\label{eq:F13+F33}
\end{align}
With these functional pieces defined, we proceed to study the potential impact of different choices in modeling the FV corrections.

\subsection{FV sensitivity study: a cautionary tale \label{sec:cautionarytale}}

\begin{figure}
\includegraphics[width=\columnwidth,valign=t]{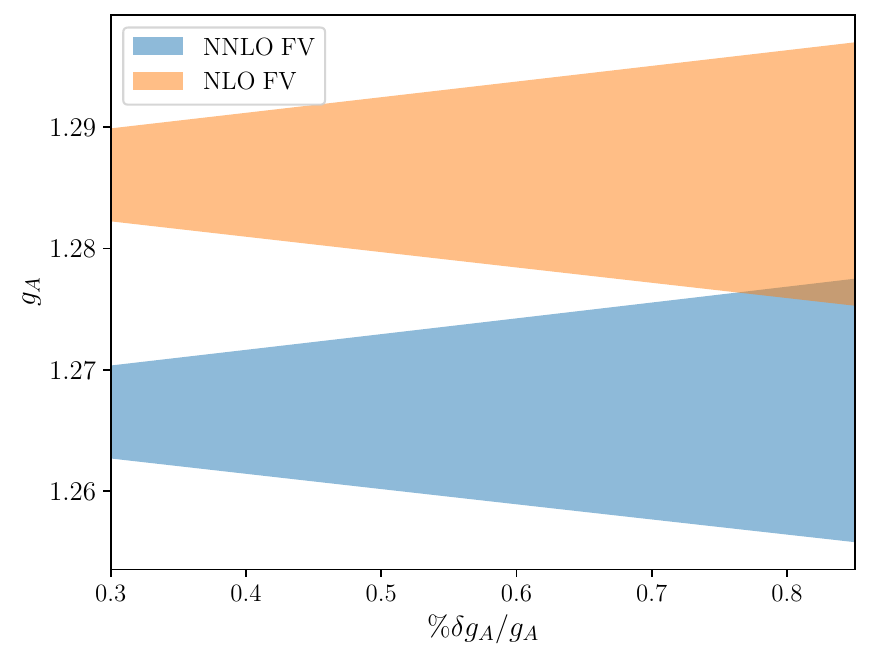}
\caption{\label{fig:precision_results}
Using the strategy outlined in the FV study, we plot the value of $g_A$, as function of the relative percent uncertainty, $\delta g_A/g_A$. The top band is determined by utilizing Eq.~\eqref{eq:fv_model_nlo} with NLO FV corrections, while the bottom band includes corrections up to \nxlo{2}, Eq.~\eqref{eq:fv_model_nnlo}. 
}
\end{figure}

To study the impact of the FV model used, we undertake the following exercise:
\begin{enumerate}[leftmargin=*]

\item We begin by fitting the CalLat results at lattice spacing $a\simeq 0.12~$fm listed in Table~\ref{tab:results} to the IV expression, 
\begin{equation}
g_A = g_0 +\d_\chi^{(2)} + \d_\chi^{(3)}\,,
\end{equation}
utilizing only the largest volume at a given pion mass.  We use this fit to provide an estimate for the LECs
\begin{equation}
g_0^{\rm FV}\,,\quad
\tilde{c}_{34}^{\rm FV} = 2\tilde{c}_4^{\rm FV} -\tilde{c}_3^{\rm FV}\,,
\end{equation}
that we will use to estimate the FV corrections.  In $\d_\chi^{(3)}$, we set $M_0=939$~MeV and $F=92.3$~MeV as the difference using these values and their respective chiral-limit values is higher order in the chiral expansion.  We focus on a single lattice spacing to isolate the FV effects.

\item We use the RQCD results~\cite{Bali:2023sdi} at $m_\pi\simeq(280\text{--}290)$~MeV, listed in Table~\ref{tab:RQCDresults} and shown in the bottom panel of \figref{fig:a12m220_RQCD}, to construct two models of FV corrections:
\begin{enumerate}
\item Monotonic FV model: Using the value of $g_0^{\rm FV}$ from step one, we fit the RQCD data to the model
\begin{equation}
    g_A^{285} = g_0^{285} + f_2\, \d_{\rm FV}^{(2)}(g_0^{\rm FV})\, ,
\end{equation}
where $f_2$ is priored as a Gaussian distribution of zero mean and width of $\s=1$.

\item Non-monotonic FV model: Using the values of $g_0^{\rm FV}$ and $\tilde{c}_{34}^{\rm FV}$ from step 1, as well as the same values of $M_0$ and $F$, we fit the RQCD data to the model
\begin{equation}
g_A^{285} = g_0^{285} 
    + \d_{\rm FV}^{(2)}(g_0^{\rm FV})
    +f_3\, \d_{\rm FV}^{(3)}(g_0^{\rm FV}, \tilde{c}_{34}^{\rm FV})\, ,
\end{equation}
where $f_3$ is priored as a Gaussian distribution of zero mean and width of $\s=1$.

\end{enumerate}

\item We then fit the full $a\simeq 0.12~$fm data set listed in Table~\ref{tab:results} with two models
\begin{subequations}
\begin{align}\label{eq:fv_model_nlo}
g_A^{\rm mono} &= g_0 
    +\d_\chi^{(2)}(g_0^{\rm IV}) 
    +\d_\chi^{(3)}(g_0^{\rm IV},\tilde{c}_{34}^{\rm IV})
\nonumber\\&\phantom{=}\phantom{g_0}
    +f_2 \d_{\rm FV}^{(2)}(g_0^{\rm FV})\, ,
\\\label{eq:fv_model_nnlo}
g_A^{\rm non-mono} &=g_0 
    +\d_\chi^{(2)}(g_0^{\rm IV}) 
    +\d_\chi^{(3)}(g_0^{\rm IV},\tilde{c}_{34}^{\rm IV})
\nonumber\\&\phantom{=}\phantom{g_0}
    +\d_{\rm FV}^{(2)}(g_0^{\rm FV}) 
    +f_3 \d_{\rm FV}^{(3)}(g_0^{\rm FV},\tilde{c}_{34}^{\rm FV})\, ,
\end{align}
\end{subequations}
where the LECs in the IV and FV functions are kept separate, with the FV values determined in step 1 used in the FV functions and the values of $f_2$ or $f_3$ taken from step 2.  The resulting analysis is used to predict $g_A$ at the physical pion mass for which we find
\begin{align}
g_A^{\eqref{eq:fv_model_nlo}} &= 1.284(12)\, ,
\notag\\
g_A^{\eqref{eq:fv_model_nnlo}} &= 1.266(12)\, ,
\end{align}
showing tension between these models for the final prediction even at the $\Order(1\%)$ uncertainty level.

\item We repeat step 3, while artificially decreasing the statistical uncertainty on the $a\simeq 0.12~$fm data by a common rescaling factor and compare the prediction from these two models as a function of the final total uncertainty on the predicted value of $g_A$ at the physical pion mass and IV.  The results from this study are plotted in \figref{fig:precision_results}.

\end{enumerate}
This exercise demonstrates that the choice of FV model can have a statistically significant systematic impact on the predicted final value of $g_A$.  For results at the percent level and more precise, it is important to incorporate a set of models with sufficient flexibility to support non-monotonic FV corrections and compare the extrapolation with monotonic FV models.
Otherwise, one may invariably introduce an unquantified systematic uncertainty that is comparable to, or larger than, the statistical uncertainty.

\begin{table}
\caption{\label{tab:RQCDresults}
Data from Ref.~\cite{Bali:2023sdi} shown in Fig.~\ref{fig:a12m220_RQCD}c, and used in the FV study of Sec.~\ref{sec:cautionarytale}.  We also estimate the value of $\e_\pi$ for each CLS ensemble by scaling $\e_\pi$ on our a12m310XL ensemble with the ratio of pion masses where we convert our pion mass to MeV units using the scale setting in Ref.~\cite{Miller:2020evg}:
$\e_\pi = \e_\pi^{\rm a12m310XL}\times(m_\pi/m^{\rm a12m310XL}_\pi)$.}
\begin{ruledtabular}
\begin{tabular}{lccccc}
Ensemble& $m_\pi~[\text{MeV}]$& $\e_\pi$& $m_\pi L$& $a~[\text{fm}]$&  $g_A$\\
\hline
N101 & $281$ &$0.2216$& $5.86$ & $0.086$ & $1.261(15)$ \\
N451 & $289$ &$0.2279$& $5.34$ & $0.076$ & $1.249(11)$ \\
N401 & $287$ &$0.2263$& $5.30$  & $0.076$ & $1.241(20)$ \\
N450 & $287$ &$0.2263$& $5.30$  & $0.076$ & $1.250(10)$ \\
N201 & $287$ &$0.2263$& $4.49$ & $0.064$ & $1.234(14)$ \\
N200 & $286$ &$0.2255$& $4.47$ & $0.064$ & $1.233(17)$ \\
H105 & $281$ &$0.2216$& $3.91$ & $0.086$ & $1.191(23)$ \\
S201 & $290$ &$0.2287$& $3.02$ & $0.064$ & $1.131(20)$ \\
\end{tabular}
\end{ruledtabular}
\end{table}

This particular study was performed under the assumption that the RQCD results could be used to determine the coefficients of the FV model.  
A more realistic study would add these models in a global analysis, including discretization effects.
One challenge in isolating the FV corrections to $g_A$ is that they are comparable in size to the observed discretization effects.  In principle, these corrections should be independent through the separation of short- and long-distance scales of the problem.
In practice, because the FV corrections are assessed by fitting coefficients multiplying FV correction terms, their contributions become entangled with discretization corrections through the analysis procedure and the two effects can influence each other, including changing the sign of the leading FV or discretization corrections depending upon the model for these systematic effects.

\section{Updated \texorpdfstring{$\boldsymbol{g_A}$}{} result with MDWF/HISQ \label{sec:callat_update}}
We update the results in Ref.~\cite{Chang:2018uxx}, which utilized a mixed lattice action~\cite{Renner:2004ck,Bar:2002nr} with $N_f=2+1+1$ Highly Improved Staggered Quarks (HISQ)~\cite{Follana:2006rc} in the sea and M\"obius Domain Wall Fermion (MDWF)~\cite{Brower:2004xi,Brower:2005qw} quarks in the valence sector.
The gauge links were smoothed with gradient flow~\cite{Luscher:2010iy,Lohmayer:2011si} prior to using them in the Dirac operator, resulting in our MDWF/HISQ action~\cite{Berkowitz:2017opd}.
Some of the HISQ ensembles were generated by the MILC Collaboration~\cite{MILC:2010pul,Bazavov:2012xda}, while the rest were generated by the CalLat Collaboration~\cite{Chang:2018uxx,Miller:2020xhy,Miller:2020evg} including a new ensemble first used in this work.

\begin{table*}[t]
\caption{\label{tab:results}
Results of $\epsilon_\pi$, $m_\pi L$, $a/w_0$, and $g_A$ on each ensemble used in this work.  The results in the top part of the table are from Ref.~\cite{Chang:2018uxx} and the four ensembles in the bottom are updated (a12m130) or new.
The ensembles generated by CalLat used the MILC code with the a12m310L, a12m310XL, and a15m135XL ones also using QUDA~\cite{Clark:2009wm,Babich:2011np}.
We also list the source (first publication) where each ensemble was used.
}
\begin{ruledtabular}
\begin{tabular}{llcccc}
Ensemble& Source& $\epsilon_\pi$& $m_\pi L$& $a/w_0$& $g_A$\\
\hline
a15m400& Ref.~\cite{Chang:2018uxx}
    & 0.30374(54)& 4.8& 0.9201(12)&  1.2154(60)\\
a12m400& Ref.~\cite{Chang:2018uxx}
    & 0.29840(52)& 5.8& 0.73335(43)&  1.2162(97)\\
a09m400& Ref.~\cite{Chang:2018uxx}
    & 0.29819(54)& 5.8& 0.53516(77)& 1.2097(80)\\
a15m350& Ref.~\cite{Chang:2018uxx}
    & 0.27412(52)& 4.2& 0.9101(11)&  1.197(14)\phantom{0}\\
a12m350& Ref.~\cite{Chang:2018uxx}
    & 0.27063(69)& 5.1& 0.72701(58)&  1.235(14)\phantom{0}\\
a09m350& Ref.~\cite{Chang:2018uxx}
    & 0.26949(58)& 5.1& 0.52910(73)&  1.227(15)\phantom{0}\\
a15m310& Ref.~\cite{MILC:2010pul}
    & 0.24957(36)& 3.8& 0.90261(73)&  1.215(12)\phantom{0}\\
a12m310& Ref.~\cite{MILC:2010pul}
    & 0.24486(51)& 4.5& 0.72134(62)&  1.214(13)\phantom{0}\\
a09m310& Ref.~\cite{MILC:2010pul}
    & 0.24619(44)& 4.5& 0.52449(61)&  1.235(11)\phantom{0}\\
a15m220& Ref.~\cite{Bazavov:2012xda}
    & 0.18084(31)& 4.0& 0.88849(63)&  1.274(14)\phantom{0}\\
a12m220& Ref.~\cite{Bazavov:2012xda}
    & 0.18220(44)& 4.3& 0.70942(55)&  1.259(15)\phantom{0}\\
a12m220L& Ref.~\cite{Bazavov:2012xda}
    & 0.18155(42)& 5.4& 0.70972(35)&  1.251(21)\phantom{0}\\
a12m220S& Ref.~\cite{Bazavov:2012xda}
    & 0.18418(57)& 3.3& 0.71078(81)&  1.270(28)\phantom{0}\\
a09m220& Ref.~\cite{Bazavov:2012xda}
    & 0.18197(37)& 4.7& 0.51578(32)&  1.2519(85)\\
\hline
a12m310L& This work
    & 0.24295(40)& 6.0& 0.72084(46)&  1.2292(57)\\
a12m310XL& Ref.~\cite{Miller:2020evg}
    & 0.24318(39)& 9.0& 0.72124(31)&  1.2381(55)\\
a12m130& Ref.~\cite{Bazavov:2012xda}
    & 0.11343(32)& 3.9& 0.70368(30)&  1.266(13)\phantom{0}\\
a15m135XL& Ref.~\cite{Miller:2020xhy}
    & 0.11486(20)& 4.9& 0.88347(31)&  1.273(13)\phantom{0}\\
\end{tabular}
\end{ruledtabular}
\end{table*}

Table~\ref{tab:results} lists the resulting values of $g_A$, $\e_\pi$, $m_\pi L$, and $a/w_0$ on each ensemble needed for the extrapolation analysis, where $w_0$ is the gradient flow scale defined in Ref.~\cite{BMW:2012hcm}.
The short-hand naming key~\cite{Bhattacharya:2015wna} indicates the approximate lattice spacing and pion mass on the given ensemble, with a12m130 indicating a lattice spacing of $a\simeq0.12$~fm and a pion mass of $m_\pi\simeq130$~MeV, for example.
For details of the input parameters of the action, see Refs.~\cite{Berkowitz:2017opd,Miller:2020xhy,Miller:2020evg}.
The analysis of the correlation functions is described in Ref.~\cite{Chang:2018uxx}:
a constant is fit to the $m_{\rm res}$ correlation function and 
multi-state fits are performed to the pion correlation function to determine $m_\pi$ and $F_\pi$ on each ensemble;
the values of $\mathring{g}_A$ and $\mathring{g}_V$, which give $g_A = (Z_A / Z_V) (\mathring{g}_A / \mathring{g}_V)$, with $Z_A/Z_V -1 \lesssim 1\times10^{-4}$~\cite{Chang:2018uxx}, are determined with a multi-state fit to the nucleon two-point correlation function and the ``Feynman-Hellmann'' correlation functions described in Ref.~\cite{Bouchard:2016heu}.
Further details of the correlator analysis will be provided in a forthcoming publication.

The results in the upper part of the table are from Ref.~\cite{Chang:2018uxx}.
The results on the a12m310L and a12m310XL ensemble are new to this work and enable a volume study at $m_\pi\simeq310$~MeV in addition to that possible with the a12m220 set of ensembles.
The results on the a12m130 and a15m135XL ensembles were first reported in Ref.~\cite{Walker-Loud:2019cif}.  For a12m130, we changed the quark smearing from that in Ref.~\cite{Chang:2018uxx} to that in Ref.~\cite{Miller:2020xhy} and increased the number of sources per configuration from 3 to 32.
The result a15m135XL is obtained from a newly generated ensemble and uses 8 sources per configuration on 1000 configurations.

\subsection{Models considered}
\label{subsec:models}

In this updated $g_A$ calculation, we use both quantitative and qualitative models of the FV corrections that are allowed to be non-monotonic as predicted from $\chi$PT, as well as more agnostic non-$\chi$PT inspired monotonic models.  These are studied in conjunction both with $\chi$PT expressions for the IV corrections as well as simple Taylor expansion expressions.
For the $\chi$PT models, we only consider the expressions from $SU(2)$ HB$\chi$PT($\D{\hskip-0.5em /}$).  This is because our numerical results are not sufficient to constrain the new parameters describing the axial $N\rightarrow\D$ transitions or the axial coupling of the $\D$, and we wish to minimize the influence from phenomenological input.
For the quantitative $\chi$PT model, we use the expressions given in Eqs.~\eqref{eq:NLO_FV} and~\eqref{eq:NNLO_FV}.

The quark-mass dependence of $g_A$ is parameterized using four  models, 
\begin{subequations}
\label{eq:IVall}
\begin{align}
\label{eq:gA_N2LO}
g_{A,\chi} &= g_0 + \d_{\chi}^{(2)} + \d_{\chi}^{(3)}\,,
\\
\label{eq:gA_epi_epi2}
g_{A,\rm pol}^{(1)} &= c_0 + c_1 \e_\pi + c_2 \e_\pi^2 \,,
\\
\label{eq:gA_epi2_epi3}
g_{A,\rm pol}^{(2)} &= c_0 + c_2 \e_\pi^2 + c_3 \e_\pi^3\,,
\\
\label{eq:g_epi2_epi4}
g_{A,\rm pol}^{(3)} &= c_0 + c_2 \e_\pi^2 + c_4 \e_\pi^4\,.
\end{align}
\end{subequations}
In the above we include the HB$\chi$PT($\D{\hskip-0.5em /}$) prediction together with simple Taylor expansions in different powers of $\e_\pi^n$, as was done in Refs.~\cite{Chang:2018uxx,Walker-Loud:2019cif}. We keep the total number of parameters $c_i$ fixed across the polynomial models.
Equation~\eqref{eq:gA_epi2_epi3} was not used in the previous work.  This model can be considered as HB$\chi$PT($\D{\hskip-0.5em /}$) with the chiral logarithms set to zero.

\begin{figure*}[t]
\includegraphics[width=0.98\textwidth]{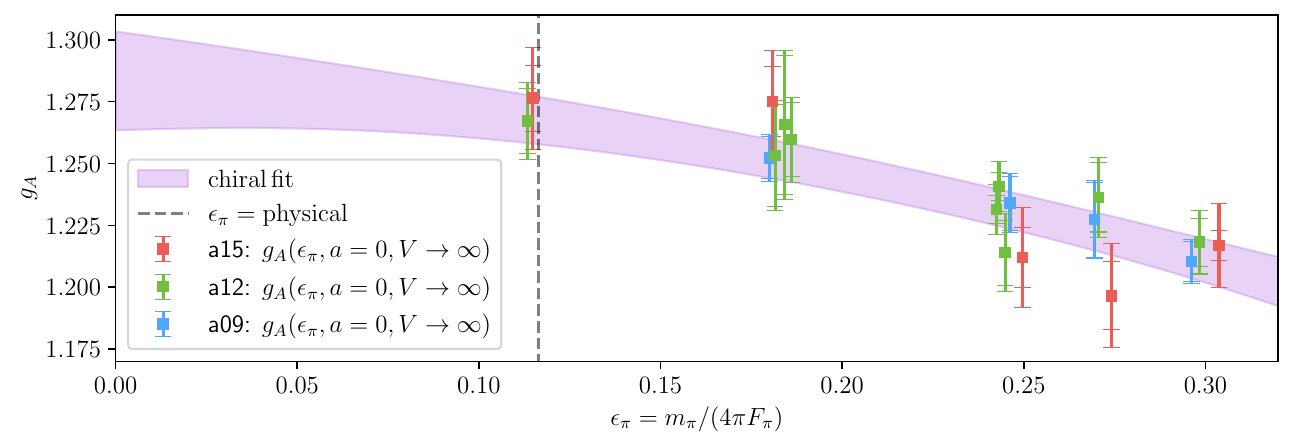}
\caption{\label{fig:modelaveratecurve}
Our final fit for the pion mass dependence of $g_A$ model-averaged over all 64 models considered, as described in Sec.~\ref{subsec:models}, plotted along the values of $g_A$ extrapolated to the continuum and IV limits. The outer error bars correspond to the uncertainty of the extrapolated results, while the inner ones to the uncertainties of the original values listed in Table~\ref{tab:results}. Some of the data has been slightly shifted in the $x$-axis for visibility.}
\end{figure*}

Meanwhile, the non-monotonic FV corrections models used are 
\begin{subequations}
\label{eq:deltaFVnonmon}
\begin{align}
\label{eq:gA_FV_NNLO}
g_{A,\text{FV}}&=\d_{\rm FV}^{(2)} + \d_{\rm FV}^{(3)}\,,
\\
\label{eq:gA_FV_f2}
g_{A,\text{FV}}^{f_2}&=f_2\d_{\rm FV}^{(2)} + \d_{\rm FV}^{(3)}\,,
\\
\label{eq:gA_FV_f3}
g_{A,\text{FV}}^{f_3}&=\d_{\rm FV}^{(2)} + f_3\d_{\rm FV}^{(3)}\,,
\\
\label{eq:gA_FV_f2_f3}
 g_{A,\text{FV}}^{f_2,f_3}&=f_2\d_{\rm FV}^{(2)} + f_3\d_{\rm FV}^{(3)} \,.
\end{align}
\end{subequations}
The above models use the full \nxlo{2} expressions for the FV corrections, and in the case of the later three include different choices of unknown coefficients $f_2$ and $f_3$. The purpose of these latter three models is to use $\chi$PT to guide the analytic form of the IV corrections, but allow for the strength of the FV corrections to become decoupled from the values of the LECs needed to describe the IV pion mass dependence of model $g_{A,\chi}$. We augment this set with four additional $\chi$PT-agnostic models
\begin{subequations}
\label{eq:deltaFVmon}
\begin{align}
\label{eq:gA_FV_mono_1}
g^{\rm{mon}-1}_{A,\text{FV}}&= f_2 \epsilon_{\pi}^2 F_1^{(2)}\,,
\\
\label{eq:gA_FV_mono_2}
g^{\rm{mon}-2}_{A,\text{FV}}&= f_2 \epsilon_{\pi}^2 F_3^{(2)}\,,
\\
\label{eq:gA_FV_mono_3}
g^{\rm{mon}-3}_{A,\text{FV}}&= f_2 \epsilon_{\pi}^2 F_1^{(3)}\,,
 \\
\label{eq:gA_FV_mono_4}
g^{\rm{mon}-4}_{A,\text{FV}}&= f_2 \epsilon_{\pi}^2 F_3^{(3)} \,,
\end{align}
\end{subequations}
which are all monotonic by construction, and have their $m_{\pi} L$ dependence parameterized by the FV functions defined in Eqs.~\eqref{eq:Fn2} and \eqref{eq:F13+F33}. The purpose of these additional models is to diagnose whether statistically significant evidence of the non-monotonic behavior predicted by $\chi$PT, which we observed as a trend in the $a\simeq0.12~$fm data and the RQCD data in the previous section, can be detected in a global analysis of our full dataset.

Finally, the discretization effects are parameterized up to $\Order(a^4)$ in the Symanzik expansion~\cite{Symanzik:1983gh,Symanzik:1983dc}.  For this mixed action with valence fermions that respect chiral symmetry, the leading discretization effects begin at $\Order(a^2)$~\cite{Bar:2005tu}.
We parameterize them as
\begin{subequations}
\label{eq:deltaga}
\begin{align}
\label{eq:deltaga2}
g_{A,a}^{(2)} &=  a_2 \e^{2}_a \,,\\
\label{eq:deltaga4}
g_{A,a}^{(4)} &= a_2 \e^{2}_a + b_4 \e^{2}_a\e_\pi^{2} + a_4\e_a^4 \,,
\end{align}
\end{subequations}
with
\begin{equation}
    \epsilon^2_a = \frac{a^2}{4 w^2_0}\,.
    \label{eq:deltaga}
\end{equation}
The values for $\epsilon_a$ for the ensembles used besides a12m310L  were previously determined in Ref.~\cite{Miller:2020evg}.

We consider these two models for the discretization dependence together with all combinations of the pion mass and FV dependence listed above, resulting in 64 different models. We fit the parameters of each model by $\chi^2$ minimization using the {\tt lsqfit} package, and include a discussion on the choice of Bayesian priors in App.~\ref{app:priors}. We use the posterior distributions to interpolate to the physical point.  We take the FLAG definition of the isospin symmetric point~\cite{FlavourLatticeAveragingGroupFLAG:2024oxs}
\begin{align}
m_\pi^{\rm iso,phys} &= 135~{\rm MeV}\, ,
\notag\\
F_\pi^{\rm iso,phys} &= \frac{130.5}{\sqrt{2}} = 92.3~{\rm MeV}\, ,
\notag\\
\e_\pi^{\rm iso,phys} &= 0.1164\,.
\end{align}
The resulting central values and errors for $g_A$ are model-averaged weighted by the Bayes factors of the corresponding model fit as was done previously~\cite{Chang:2018uxx,Walker-Loud:2019cif}.

\subsection{Final results and comparison with other works}

The final result we obtain for $g_A = 1.2674(96)$, with a percentage uncertainty of $0.76\%$,
\begin{align}\label{eq:gA_update}
g_A &=1.2674(79)^{\text{s}}(28)^{\chi}(05)^{\text{FV}}(38)^{a}(26)^{\text{M}}
\nonumber\\&= 1.2674(96)\,.
\end{align}
The uncertainty breakdown into the statistical (s), chiral extrapolation ($\chi$), FV, and discretization ($a$) is done by differentiating the final posterior prediction at the physical point with respect to the relevant priors.  The statistical uncertainty is with respect to the $g_A$ data and $\e_\pi$, while the FV uncertainty arises from the $f_2$ and $f_3$ coefficients when relevant, for example.  The final model-selection uncertainty arises from the variance of the models with respect to the Bayes Model Average value, see Refs.~\cite{Chang:2018uxx,Miller:2020xhy} for further details.

The model-averaged chiral fit to $g_A$ is plotted in the continuum and IV limits as a function of $\epsilon_{\pi}$ in Fig.~\ref{fig:modelaveratecurve}, along with the individual ensemble values extrapolated to the continuum and to IV. The individual model results including the priors and posteriors for all parameters along with the corresponding Bayes factors are listed in Tables~\ref{tab:priors} and~\ref{tab:postpriors} of App.~\ref{app:models}.

\begin{figure}[t]
\includegraphics[width=\columnwidth,valign=t]{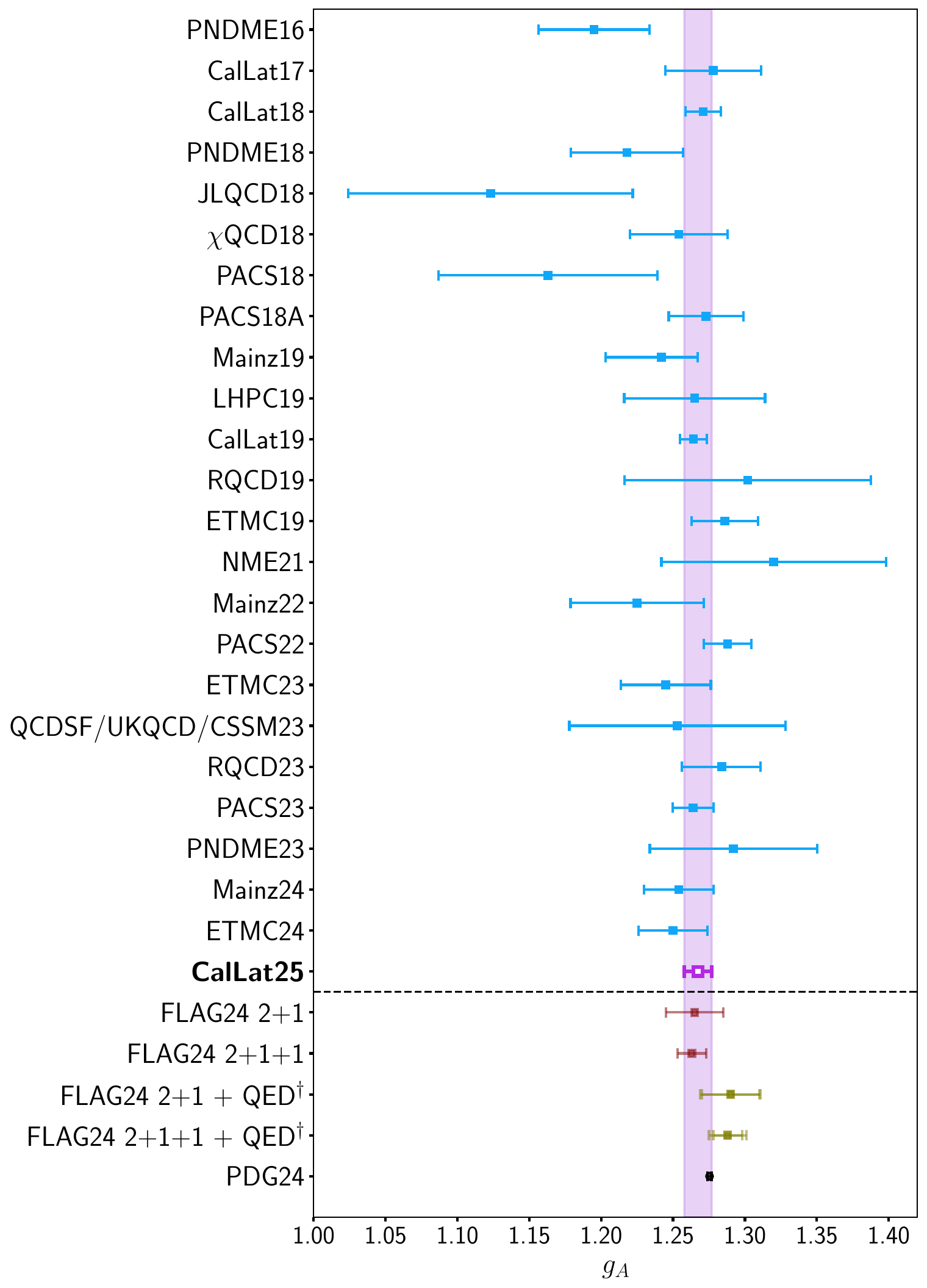}
\caption{\label{fig:summaryplot}
Comparison of our final value for $g_A$ against several LQCD collaborations~\cite{Bhattacharya:2016zcn,Berkowitz:2017gql,Chang:2018uxx,Gupta:2018qil,Yamanaka:2018uud,Liang:2018pis,Ishikawa:2018rew,Shintani:2018ozy,Harris:2019bih,Hasan:2019noy,Walker-Loud:2019cif,RQCD:2019jai,Alexandrou:2019brg,Park:2021ypf,Djukanovic:2022wru,Tsuji:2022ric,Alexandrou:2023qbg,QCDSFUKQCDCSSM:2023qlx,Bali:2023sdi,Tsuji:2023llh,Jang:2023zts,Djukanovic:2024krw,Alexandrou:2024ozj}, the FLAG average~\cite{FlavourLatticeAveragingGroupFLAG:2024oxs}, the FLAG average modified to include an estimate of the QED corrections~\cite{Cirigliano:2022hob}, and the PDG average~\cite{ParticleDataGroup:2024cfk}. }
\end{figure}

In Fig.~\ref{fig:summaryplot}, we plot the updated $g_A$ and compare it against other LQCD calculations. On the lower part of the plot, we include the latest FLAG~\cite{FlavourLatticeAveragingGroupFLAG:2024oxs} averages which, just like all LQCD results, correspond to QCD-only predictions. In Ref.~\cite{Cirigliano:2022hob}, it was estimated that pion-induced radiative corrections to $g_A$ are 
\begin{equation}
g_A^{\rm PDG}=g_A^{\rm QCD}(1+\delta_{\rm RC})\,,\qquad \delta_{\rm RC} = 2.0(6)\times10^{-2}\, .
\end{equation} 
We propagate these estimates to the FLAG $N_f=2+1$ and $N_f=2+1+1$ predictions, and plot the resulting values, denoted as FLAG + QED$^\dagger$, on the bottom of the plot, just above the PDG value $g_A=1.2754(13)$~\cite{ParticleDataGroup:2024cfk}. 

At our current LQCD precision, the FV dependence of our data is small and modeling these corrections contributes the least to the final uncertainty. The result of a global analysis using just the $\chi$PT and $\chi$PT-inspired non-monotonic models of Eq.~\eqref{eq:deltaFVnonmon} is $g_A^{\text{non-mon}}=1.2672(96)$, which is consistent with the value $g_A^{\text{mon}}=1.2674(96)$ obtained by the monotonic FV models of Eq.~\eqref{eq:deltaFVmon}. 

\subsection{\texorpdfstring{$\boldsymbol{SU(2)}$}{} HB\texorpdfstring{$\boldsymbol{\chi}$}{}PT(\texorpdfstring{$\boldsymbol{\D{\hskip-0.5em /}}$}{}) extrapolations}

It is interesting to examine the $SU(2)$ HB$\chi$PT($\D{\hskip-0.5em /}$) extrapolations in more detail.
This is both because this theory serves as a foundation for many modern applications of chiral EFT to predicting the spectrum and properties of nuclei, see Ref.~\cite{Epelbaum:2019kcf} for a recent review.  It also enables us to more stringently test our low-energy understanding of nuclear physics by examining the convergence properties of the theory as well as comparing the LECs determined from comparisons with LQCD versus experimental data.

The first interesting observation is that the $SU(2)$ HB$\chi$PT($\D{\hskip-0.5em /}$) extrapolations have the lowest likelihood of any of the extrapolation models, as can be seen from the logGBF values in Table~\ref{tab:postpriors}.  In fact, the strict $SU(2)$ HB$\chi$PT($\D{\hskip-0.5em /}$) model without the $f_2$ or $f_3$ FV coefficients (those using \eqnref{eq:gA_N2LO} for $m_\pi$ dependence and \eqnref{eq:gA_FV_NNLO} for FV corrections) is the least favored of all fits.  At the same time, it should be noted that the final extrapolated value is perfectly consistent with all other models.  What likely drives these fits to have a lower likelihood than the Taylor expansion fits is that the $\e_\pi^2 \ln \e_\pi^2$ have large coefficients, and thus the counterterm contributions and higher-order corrections must cancel against each other in order to produce such a mild pion mass dependence.  This is contrasted with the Taylor models which all have $\Order(1)$ or smaller coefficients in the expansion, and thus provide a rapidly converging expansion to describe the numerical results.
It will be interesting to see if an extrapolation with explicit $\D$s, that exhibits milder pion mass dependence, is more favored.  Such an extrapolation with only LQCD input will require calculations of $N\rightarrow\D$ and $\D\rightarrow\D$ axial matrix elements.  See Ref.~\cite{Barca:2022uhi} for promissing first steps in this direction.

The second interesting observation concerns the combination of LECs $2c_4-c_3$ that appears in \eqnref{eq:gA_inf}.
A recent phenomenological determination from \nxlo{3} analysis of pion--nucleon ($\pi N$) scattering data~\cite{Hoferichter:2015tha,Hoferichter:2015hva} found the following values for the LECs $c_3$ and $c_4$\footnote{The counting of chiral orders refers to the non-trivial ones, and since chiral corrections to $g_A$ start at $\Order(\e_\pi^2)$, the \nxlo{3} results from Refs.~\cite{Hoferichter:2015tha,Hoferichter:2015hva} are consistent with the \nxlo{2} corrections to $g_A$. This distinction matters since the indicated errors, propagated from the $\pi N$ input, are subleading compared to the differences between determinations at different chiral orders.}
\begin{align}
c_3 &= -5.61(6)~\textrm{GeV}^{-1}\, ,
\nonumber\\
c_4 &= \phantom{-}4.26(4)~\textrm{GeV}^{-1}\, ,
\nonumber\\
2c_4 - c_3 &= \phantom{-}14.1(1)~\textrm{GeV}^{-1}\, .
\end{align}
In our analysis using \eqnref{eq:gA_N2LO} to describe the pion mass dependence of $g_A$, we utilized the dimensionless versions of these LECs, \eqnref{eq:dimensionless_c_d}.
From Table~\ref{tab:postpriors}, one sees that the posterior values of $\tilde{c}_{34}$ for the $SU(2)$ HB$\chi$PT($\D{\hskip-0.5em /}$) fit models range from $0.76$ to $0.81$.  Converting to GeV$^{-1}$ using $F_\pi=92.3$~MeV, the value of $2c_4-c_3$ determined from this analysis becomes
\begin{equation}
    2c_4-c_3 = (0.66\text{--}0.70)\,\textrm{GeV}^{-1}\, ,
\end{equation}
which is 20 times smaller than the phenomenological estimate above.  
The small priors given to $\tilde{c}_{34}$ are not the source of this discrepancy.  These values were chosen to optimize the Bayes factor of the fits.  If instead a prior width of 50 is used, the posterior value of $\tilde{c}_{34}=0.85(25)$ or similar is found.

This discrepancy for the LECs observed here is a particularly extreme case, but similar effects have been encountered in other examples, including the determination from $\pi N$ scattering in the first place~\cite{Hoferichter:2015tha,Hoferichter:2015hva,Siemens:2016jwj}. That is, the large values of $c_3$ and $c_4$ entering at subleading loop level generate sizable corrections that can differ from case to case, and therefore do not necessarily drop out when comparing observables. In the $\pi N$ case, this implies that the  HB$\chi$PT($\D{\hskip-0.5em /}$) theory is not even accurate enough to relate consistently the expansion of the amplitude around subthreshold and threshold kinematics.  The situation improves appreciably when the $\Delta$ resonance is included as an explicit degree of freedom~\cite{Siemens:2016jwj}, as the large loop corrections are reduced, but, for reasons currently not understood, also when resumming higher orders in the $1/M_0$ expansion into a covariant formulation. Understanding this convergence behavior is important, as the LECs are used extensively to predict two-body nuclear currents and three-body nuclear forces~\cite{Hebeler:2020ocj,Krebs:2020pii,Tews:2022yfb}, see, e.g., Ref.~\cite{Hoferichter:2020osn} for a discussion how to account for $c_i$ chiral uncertainties in the axial current. 
The present works suggests that $g_A$ is another quantity particularly sensitive to this issue. Future work that includes the $\Delta$ resonance as an explicit degree of freedom, and/or employs covariant formulations of baryon $\chi$PT, could thus provide valuable new perspectives.    
Understanding the convergence of the chiral expansion is also important for the radiative corrections to  $g_A$, which, at NLO, depend on the combination $c_4 - c_3$ \cite{Cirigliano:2022hob}. Also in this case, it will be important to extend the calculation to include the $\Delta$ resonance and to thoroughly study the convergence.

\section{Discussion and Conclusion \label{sec:discussion}}
A theoretical prediction of $g_A$ with a sub-percent precision can contribute to tests of the SM, as well as put constraints on BSM right-handed currents~\cite{Alioli:2017ces} that offer a possible explanation for the first-row CKM unitarity tension~\cite{Cirigliano:2023nol}.
At this level of precision, 
systematic uncertainties from FV and discretization corrections must be scrutinized more carefully to ensure they remain as small as or smaller than the statistical precision.  
LQCD calculations of $g_A$ all have small discretization errors that are currently compatible with a zero slope with respect to powers of the lattice spacing~\cite{Bhattacharya:2016zcn,Chang:2018uxx,Gupta:2018qil,Alexandrou:2019brg,Bali:2023sdi,Djukanovic:2024krw,Alexandrou:2024ozj}, while some of them display statistically significant FV corrections.  
Therefore, in this work, we have investigated FV corrections in greater detail than has been previously explored in the literature.

Our previous results~\cite{Chang:2018uxx} on three volumes at $m_\pi\simeq220$~MeV showed a trend, albeit not statistically significant, that is consistent with the predicted FV correction from $SU(2)$ HB$\chi$PT($\D{\hskip-0.5em /}$), \eqnref{eq:NLO_FV}.  Subsequently, we generated new results at different volumes at $m_\pi\simeq310$~MeV which showed the opposite trend. Further, the RQCD results at $m_\pi\simeq285$~MeV also showed this opposite trend with statistical significance, see \figref{fig:a12m220_RQCD}, hinting at the possibility of non-monotonic FV corrections.
As discussed in \secref{sec:XPT_largeN}, non-monotonic FV corrections to $g_A$ would not be surprising as there are significant cancellations from radiative corrections to $g_A$ between virtual contributions involving nucleons and $\D$-resonance states.  These cancellations are predicted from large-$N_c$ considerations and the cancellations also manifest in the predicted FV corrections.

In \secref{sec:callat_update}, we performed a thorough analysis of our results, Table~\ref{tab:results}, using various models for the pion mass dependence and various models for the FV corrections.  We report an updated value of $g_A$, \eqnref{eq:gA_update}
, with a $0.78\%$ relative uncertainty.
One interesting finding from our analysis using $SU(2)$ HB$\chi$PT($\D{\hskip-0.5em /}$) to describe the pion mass dependence of $g_A$ is that the value of the LEC $2c_4 - c_3$ we determine is an order of magnitude smaller than the state-of-the-art phenomenological determinations~\cite{Hoferichter:2015hva}. Going forward, it is important to understand this discrepancy as the phenomenological LECs are utilized to make predictions for properties of nuclei through modern applications of chiral EFT~\cite{Epelbaum:2019kcf}. As in similar cases~\cite{Siemens:2016jwj}, studies including the $\D$ resonance as an explicit degree of freedom and/or using covariant formulations could shed light on the convergence properties of the chiral expansion in the baryon sector. 

In the FV correction, we included models that were monotonic by construction and models that allow for non-monotonic behavior. With our current level of precision, our numerical results cannot distinguish between these two types of FV corrections.  In \secref{sec:cautionarytale}, we studied the potential sensitivity of using monotonic versus non-monotonic FV models under the assumption that the numerical results are sufficiently precise to constrain the FV corrections, finding that as the total uncertainty is reduced below a percent, the FV systematic can become a larger uncertainty than the statistical uncertainty.  We conclude that going forward, for sub-percent determinations of $g_A$, it is important to utilize models of FV corrections that support non-monotonic as well as monotonic behavior, at least if the final result depends upon heavier than physical pion mass results to perform all extrapolations.

As noted in \secref{sec:intro}, LQCD results for $g_A$
must not be compared to the PDG value at their current quoted precision
because of unaccounted for QED corrections,
estimated to be 2\% using $SU(2)$ HB$\chi$PT($\D{\hskip-0.5em /}$)
and phenomenological values of some of the LECs~\cite{Cirigliano:2022hob}.
The estimate relied upon Na\"ive Dimensional Analysis~\cite{Manohar:1983md} as one of the LECs has not been constrained.  Therefore, to take advantage of the present and future precision of LQCD results of $g_A$, it is necessary to carry out a determination of these QED corrections with LQCD+QED calculations, as suggested for example in Ref.~\cite{Seng:2024ker,Cirigliano:2024nfi}.

\begin{acknowledgments}
We thank Jacobo Ruiz de Elvira for ongoing collaboration on the chiral extrapolation of $g_A$~\cite{gA_inprep}.
The LQCD calculations were carried out on Summit at the Oak Ridge Leadership Computing Facility at the Oak Ridge National Laboratory, which is supported by the Office of Science of the U.S. Department of Energy under Contract No. DE-AC05-00OR22725, through the Innovative and Novel Computational Impact on Theory and Experiment (INCITE) program; 
on Sierra at Lawrence Livermore National Laboratory (LLNL) through an Early Science Award;
and on Lassen at LLNL through the LLNL Computing Grand Challenge Program.

The computations were performed with \texttt{LALIBE}~\cite{lalibe} which utilizes the \texttt{Chroma} software suite~\cite{Edwards:2004sx} with \texttt{QUDA} solvers~\cite{Clark:2009wm,Babich:2011np} and HDF5~\cite{hdf5} for I/O~\cite{Kurth:2015mqa}.  The calculations were efficiently managed with \texttt{METAQ}~\cite{Berkowitz:2017vcp,Berkowitz:2017xna}.
The HMC was performed with the MILC Code~\cite{milc:code}, and for the ensembles new in this work, running on GPUs using \texttt{QUDA}.  The final extrapolation analysis utilized \texttt{gvar}~\cite{gvar:11.2} and \texttt{lsqfit}~\cite{lsqfit:11.5.1}.

This work was supported by the NVIDIA Corporation (MAC);
by the U.S. National Science Foundation (NSF) through the Mathematical and Physical Sciences Ascending Postdoctoral Research Fellowship under award No. 2402482 (ZH), under NSF Award PHY-2209167 (CM), by the NSF through cooperative agreement
2020275 (TRR), and under the NSF Faculty Early Career Development Program (CAREER) under award PHY-2047185 (AN,ZH);
by the U.S. Department of Energy (DOE), Office of Science, Office of Nuclear Physics, under grant contract numbers
DE-SC0004658 (DAP),
No. 89233218CNA000001 (EM),
DE-AC52-07NA27344 (ASM, PV),
DE-AC02-05CH11231 (ZH, RB, AWL),
DE-AC05-00OR22725 (HJM),
by the Office of Science Graduate Student Research Program (ZH),
the Neutrino Theory Network Program Grant DE-AC02-07CHI11359 and U.S. Department of Energy DE-SC0020250 (ASM),
through the ``Nuclear Theory for New Physics'' Topical Collaboration award No. DE-SC0023663 (TRR, ZH, AWL, EM),
and
the Swiss National Science Foundation, Project No.\ TMCG-2\_213690 (MH).

\end{acknowledgments}

\appendix

\begin{widetext}

\section{NLO \texorpdfstring{$\boldsymbol{\chi}$}{}PT formula for \texorpdfstring{$\boldsymbol{g_A^{(\D)}}$}{} \label{app:gA_NLO_Delta}}

The FV corrections to $g_A$ in $SU(2)$ HB$\chi$PT$(\D)$ that arise at NLO were derived in Ref.~\cite{Beane:2004rf}.  The IV formulae were also provided and, as written there, given by
\begin{align}
\d^{(2)}_\chi&= -i \frac{2}{3F^2}\bigg[
	\frac{3}{2}g_0 R_1(m_\pi,\mu)
	+4g_0^3 J_1(m_\pi, 0, \mu)
\bigg]\, ,\notag\\
\d^{(2,\D)}_\chi  &= -i \frac{2}{3F^2}\bigg[
	4\left(g_{N\D}^2g_0 +\frac{25}{81}g_{N\D}^2 g_{\D}\right)J_1(m_\pi,\D,\mu)-\frac{32}{9}g_{N\D}^2 g_0 N_1(m_\pi,\D,\mu)
\bigg]\, ,
\label{eq:gA_NLO_BS}
\end{align}
where the non-analytic functions are defined as
\begin{align}
iR_1(m,\mu) &= \frac{m^2}{(4\pi)^2} \ln \frac{m^2}{\mu^2}\, ,
\notag\\
iJ_1(m,\D,\mu) &= \frac{3}{4}\frac{\D^2}{(4\pi)^2}\left[
    \left(\frac{m^2}{\D^2} -2\right)\ln \frac{m^2}{\mu^2}+
    2\sqrt{1-\frac{m^2}{\D^2}}
    \ln \frac{1-\sqrt{1-\frac{m^2}{\D^2}}}{1+\sqrt{1-\frac{m^2}{\D^2}}}
\right]\, ,\notag\\
iN(m,\D,\mu) &= \frac{3}{4}\frac{\D^2}{(4\pi)^2}\left[
    \left(\frac{m^2}{\D^2}-\frac{2}{3}\right) \ln \frac{m^2}{\mu^2}
    +\frac{2\pi}{3}\frac{m^3}{\D^3}
+\frac{2}{3}\left(1-\frac{m^2}{\D^2}\right)^{3/2}
    \ln \frac{1-\sqrt{1-\frac{m^2}{\D^2}}}{1+\sqrt{1-\frac{m^2}{\D^2}}}
\right]\, ,
\end{align}
and one can show the latter two functions have the limit
\begin{align}
iN_1(m,0,\mu)=iJ_1(m,0,\mu) = \frac{3}{4} \frac{m^2}{(4\pi)^2} \ln \frac{m^2}{\mu^2}\, .
\end{align}
It is convenient to define subtracted functions for the purpose of chiral extrapolations:
\begin{equation}
\bar{J}_1(m,\D,\mu) = J_1(m,\D,\mu) - J_1(m,0,\mu)\,,\qquad
\bar{N}_1(m,\D,\mu) = N_1(m,\D,\mu) - N_1(m,0,\mu)\,,
\end{equation}
such that the parameter $g_0$ in the extrapolation functions is $g_0=\lim_{m_\pi\rightarrow0}g_A$.
This amounts to a renormalization of $g_0$ in the Lagrangian, which we call $\mathring{g}_0$,
\begin{align}
g_0 &= \mathring{g}_0(\mu)
    +\frac{4\D^2}{(4\pi F)^2}\ln\frac{\D^2}{\mu^2}
    \frac{\mathring{g}_{N\D}^2}{81}
    \left(57\mathring{g}_0+25\mathring{g}_\D\right)\, .
\end{align}
The NLO loop contribution to $g_A$ can then be expressed as
\begin{align}
\d_\chi^{(2)}+\d_\chi^{(2,\D)}\bigg|_\text{loop}  &= -(g_0+2g_0^3)\e_\pi^2 \ln \frac{m_\pi^2}{\mu^2}
-\frac{8}{3}\left(g_{N\D}^2g_0 +\frac{25}{81}g_{N\D}^2 g_{\D}\right)
    \frac{i\bar{J}_1(m_\pi,\D,\mu)}{\L_\chi^2}
+\frac{64}{27}g_0g_{N\D}^2
    \frac{i\bar{N}_1(m_\pi,\D,\mu)}{\L_\chi^2}\, ,
\end{align}
where 
\begin{align}
\frac{i\bar{J}_1(m_\pi,\D,\mu)}{\L_\chi^2} &= \frac{3}{4}\e_\D^2 \left[
    \frac{m_\pi^2}{\D^2}\ln \frac{m_\pi^2}{\mu^2}
    +2R\left(\frac{m_\pi^2}{\D^2}\right)
\right]\, ,\notag
\\
\frac{i\bar{N}_1(m_\pi,\D,\mu)}{\L_\chi^2} &= \frac{3}{4}\e_\pi^2
    \ln \frac{m_\pi^2}{\mu^2}
    +\frac{\pi}{2}\e_\pi^2\frac{m_\pi}{\D}
    +\frac{1}{2}\e_\pi^2\ln\frac{4\D^2}{m_\pi^2}
+\frac{1}{2}\left(\e_\D^2-\e_\pi^2\right)R\left(\frac{m_\pi^2}{\D^2}\right)\, ,
\end{align}
and $R(z)$ is defined in \eqnref{eq:R}.
Finally, we redefine $\tilde{d}_{16}^{r,\D}$ such that the decoupling of the $\D$-resonance is explicit in the limit $\D\rightarrow\infty$ 
and so that our definition agrees with Refs.~\cite{Kambor:1998pi,Becher:2001hv,Gasser:2002am}.  Inserting these into the previous expression and subtracting all corrections proportional to $g_{N\D}^2$ in the heavy-$\D$ limit yields \eqnref{eq:gA_SU2Delta}.

\section{\nxlo{2} FV functions with explicit \texorpdfstring{$\boldsymbol{\D}$}{}s \label{app:FV_n2lo_dleta}}

The full \nxlo{2} result including $\Delta$ degrees of freedom and FV corrections can be expressed as
\begin{align}\label{eq:gA_inf_N2LO}
g_A &= g_0 
    + \d_\chi^{(2)} 
    +\d_\chi^{(3)}
    +\d_{\rm FV}^{(2)}
    +\d_{\rm FV}^{(3)}+ \d_\chi^{(2,\D)} 
    +\d_\chi^{(3,\D)}
    +\d_{\rm FV}^{(2,\D)}
    +\d_{\rm FV}^{(3,\D)}\, ,
\end{align}
where the terms without ``$\Delta$'' superscript refer to the IV chiral and FV corrections in $SU(2)$ HB$\chi$PT$(\D{\hskip-0.5em/})$ as indicated, while the last four terms refer to the additional effects in HB$\chi$PT$(\D)$.
The individual corrections are given by
\begin{align}
\d_\chi^{(2)}&=\e_\pi^2 \left[
    -g_0(1+2g_0^2)\ln \e_\pi^2
    +4\tilde{d}_{16}^r
    -g_0^3
    \right]\,,\notag\\
\d_\chi^{(3)}&=\e_\pi^3 g_0 \frac{2\pi}{3}\left[
    3(1+g_0^2) \frac{4\pi F}{M_0}
    +4(2\tilde{c}_4 -\tilde{c}_3)
    \right]\,,\notag\\
\d_{\rm FV}^{(2)}&=\frac{8}{3}\e_\pi^2\sum_{\vec{n}\neq \vec{0}}\left[g_0^3K_0(x_n)-\frac{g_0(3+2g_0^2)}{2x_n}K_1(x_n)\right]\,,\qquad x_n=m_\pi L |\vec{n}|\,,\notag\\
\d_{\rm FV}^{(3)}&=-\e_\pi^3 g_0\frac{2\pi}{3}\sum_{\vec{n}\neq \vec{0}}\left[\frac{4\pi F}{M_0}\big(3+g_0^2(2-x_n)\big)+4\big(2\tilde c_4-\tilde c_3\big)\right]\frac{e^{-x_n}}{x_n}\,,\notag\\
\d_\chi^{(2,\D)}&=-2\e_\pi^2 \ln \e_\pi^2 \left[
    \frac{1}{9}g_0 g_{N\D}^2 +\frac{25}{81}g_{\D}g_{N\D}^2
    \right]
-R\left(\frac{m_\pi^2}{\D^2}\right) g_{N\D}^2\left[
        \e_\pi^2\frac{32}{27}g_0
        +\e_\D^2 \left(\frac{76}{27}g_0 + \frac{100}{81}g_{\D}\right)
    \right]
\nonumber\\&\phantom{=}
    +\frac{32}{27}g_0 g_{N\D}^2 \e_\pi^2 \left[
    1+\ln \frac{4\e_\D^2}{\e_\pi^2}
    +\frac{\pi\e_\pi}{\e_\D}
    \right]
    +\frac{2}{81}g_{N\D}^2\e_\pi^2\big(9g_0+25g_\D\big)\Big[1+\ln(4\e_\D^2)\Big]\,,\notag\\
\d_\chi^{(3,\D)}&=-\frac{2g_{N\D}^2\e_\D}{243}\frac{4\pi F}{M_0}\left[\big(63g_0-25g_\D\big)\e_\pi^2+20\big(9g_0+5g_\D\big)\e_\D^2+\frac{36\sqrt{2}}{g_{N\D}}\frac{M_0}{4\pi F}\big(13\tilde b_4+12\tilde b_5\big)\big(\e_\D^2-\e_\pi^2\big)\right]R\left(\frac{m_\pi^2}{\D^2}\right)\notag\\
&+\frac{4g_{N\D}^2\e_\pi^2\e_\D}{243}\frac{4\pi F}{M_0}\left[5\big(9g_0+5g_\D\big)+\frac{9\sqrt{2}}{g_{N\D}}\frac{M_0}{4\pi F}\big(13\tilde b_4+12\tilde b_5\big)\right]\left[1+\ln\frac{4\e_\D^2}{\e_\pi^2}\right]\,,\notag\\
\d_{\rm FV}^{(2,\D)}&=\sum_{\vec{n}\neq \vec{0}}\bigg[-\frac{\e_\pi^3}{\e_\D}g_0g_{N\D}^2\frac{32\pi}{27}\frac{e^{-x_n}}{x_n}
-\e_\pi^2 g_{N\D}^2\frac{8}{243}\big(81g_0+25g_\D\big)\int_0^\infty d\lambda\,\lambda\Big[3K_0(y_n)-y_n K_1(y_n)\Big]\notag\\
&+\frac{\e_\pi^3}{\e_\D}g_0 g_{N\D}^2\frac{64}{27}\int_0^\infty d\lambda\,\beta\Big[K_0(y_n)-\frac{1}{y_n}K_1(y_n)\Big]\bigg]\,,\qquad 
y_n=x_n\sqrt{\beta}\,,\qquad \beta = 1+\lambda^2+2\lambda \xi\,,\qquad \xi=\frac{\Delta}{m_\pi}\,,\notag\\
\d_{\rm FV}^{(3,\D)}&=\sum_{\vec{n}\neq \vec{0}}\bigg[\e_\pi^3 g_{N\D}^2\big(81g_0+25g_\D\big)\frac{4}{243}\frac{4\pi F}{M_0}\int_0^\infty d\lambda\,\lambda\Big[2\xi K_0(y_n)-\frac{\lambda y_n}{\beta}\big(3\beta-2\xi^2\big)K_1(y_n)\notag\\
&+\lambda\Big(\frac{y_n}{\beta}\Big)^2\big(\beta^2-\xi^3\lambda+\beta\lambda(\lambda+3\xi)\big)K_2(y_n)\Big]\notag\\
&+\e_\pi^3 g_0 g_{N\D}^2\frac{32}{27}\frac{4\pi F}{M_0}\int_0^\infty d\lambda\,\Big[\big(\beta-4\xi\lambda-3\lambda^2\big)K_0(y_n)+\bigg(\frac{2\beta}{y_n}-\frac{y_n(\beta+\xi\lambda)^2}{\beta}\bigg)K_1(y_n)\Big]\notag\\
&-\e_\pi^3 g_{N\D}\frac{16\sqrt{2}}{27}\big(13\tilde b_4+12\tilde b_5\big)
\int_0^\infty d\lambda\,\beta\Big[K_0(y_n)-\frac{1}{y_n}K_1(y_n)\Big]
\bigg]\,,
\end{align}
with $\tilde b_i=(4\pi F) b_i$ and $b_i$ in the conventions of Ref.~\cite{Siemens:2016jwj}, where large-$N_c$ estimates are derived, leading to $|b_4\pm b_5|\leq 5\,\text{GeV}^{-1}$.  $\d_{\rm FV}^{(2)}$ and $\d_{\rm FV}^{(2,\D)}$ agree with Eq.~\eqref{deltaNLO}, $\d_{\chi}^{(2)}$ and $\d_{\chi}^{(2,\D)}$ combine to Eq.~\eqref{eq:gA_SU2Delta}. 
The \nxlo{2} FV corrections including $\Delta$s were derived in the context of Refs.~\cite{Gupta:2021ahb,Gupta:2022aba}. For the \nxlo{2} IV expressions, see also Refs.~\cite{Siemens:2020vop,gA_inprep}.


\section{Choice of Bayesian priors in global fits}
\label{app:priors}

\begin{figure}
\includegraphics[width=0.7\columnwidth]{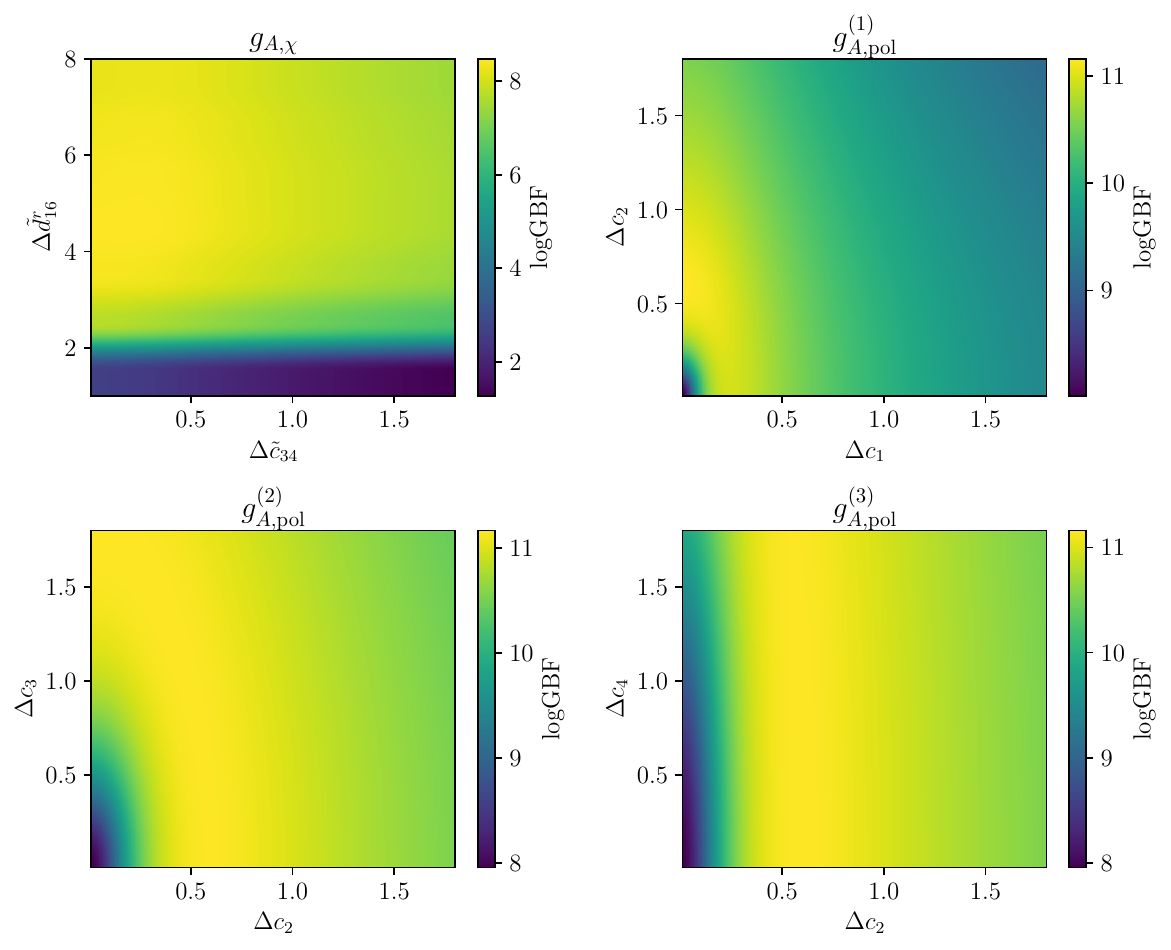}
\caption{\label{fig:IVpriors}
The log of the Bayes factors obtained by fitting the largest volume $a12$ data using the models of
Eqs.~\eqref{eq:IVall}
plotted as functions of the prior width of their free parameters.}
\end{figure}

The global fits for $g_A$ are performed with functional forms consisting of three terms, parameterizing the quark mass, FV, and discretization dependence. We include four possible choices for the quark mass (Eqs.~\eqref{eq:IVall}), eight for the FV (Eqs.~\eqref{eq:deltaFVnonmon} and \eqref{eq:deltaFVmon}), and two for the discretization dependence (Eqs.~\eqref{eq:deltaga}). The large number of parameters for each of the 64 models considered motivates us to follow a Bayesian approach, and to carefully consider the effect of the choice of priors to the fit results.
Our procedure for selecting the priors is described below, while the summary of the final model results for $g_A$, along with the priors, model weights, and posteriors are listed in Appendix~\ref{app:models}.

Our knowledge about the physical value of $g_A$ motivates us to set the prior of the parameters $g_0$ and $c_0$ of
Eq.~\eqref{eq:IVall}
to $1.25(10)$. For the remaining parameters, we set the central values of the priors to $0$, and follow the steps below, motivated by the Empirical Bayes Criterion, to choose the corresponding prior widths. We also note that for all parameters, varying our choices of the prior widths by $\Order(1)$ factors does not affect our final results for $g_A$.
\begin{itemize}
    \item \textbf{Chiral dependence}: We first isolate the parameters of the mass-dependence models of Eq.~\eqref{eq:IVall} by considering fits to the $a12$ data only, choosing the largest volume at the pion masses for which ensembles at multiple volumes are available. We vary the prior widths of the pairs of parameters,  ($\tilde{d}_{16}^r$, 
    $\tilde{c}_{34}$) for $g_{A,\chi}$, ($c_1$, $c_2$) for $g_{A,\text{pol}}^{(1)}$, ($c_2$, $c_3$) for $g_{A,\text{pol}}^{(2)}$, and ($c_2$, $c_4$) for $g_{A,\text{pol}}^{(3)}$, and study the resulting Bayes factors. For $g_{A,\chi}$, there is a maximum in the region we explore for $(\Delta\tilde{d}_{16}^r$, 
    $\Delta\tilde{c}_{34})$. For the three Taylor models, there is a maximum with respect to $\Delta c_2$, while the dependence on the second parameter of each is small and thus we conservatively set the corresponding priors to those chosen for $\Delta c_2$. Our final prior widths choices are well within the regions where the largest Bayes factors are observed, as seen in the plots of Fig.~\ref{fig:IVpriors}, $(\Delta\tilde{d}_{16}^r$, 
    $\Delta\tilde{c}_{34}) = (5,1)$, 
    $(\Delta c_1$, 
    $\Delta c_2) = (0.5,0.5)$, $(\Delta c_2$, 
    $\Delta c_3) = (1,1)$, and $(\Delta c_2$, 
    $\Delta c_4) = (1,1)$. 
    \item \textbf{Discretization dependence}:
    Next, the priors widths for the mass-dependence
    parameters are frozen and prior widths for parameters
    with lattice spacing dependence are varied.
    Our dataset
    exhibits
    minimal dependence on the finite lattice spacing, making it impossible to choose optimal prior widths of the parameters $a_2$, $a_4$, and $b_4$ defined in Eqs.~\eqref{eq:deltaga} using the Empirical Bayes Criterion. However, we observe that the central value and error of our final model-averaged estimate for $g_A$ is identical for any choice of $\Delta a_2 \gtrsim 0.5$, including those larger by several orders of magnitude. As $\Delta a_2$ decreases, the final error becomes smaller and the fit output Bayes factor grows, but simultaneously the probability of the model shrinks by an amount that cannot be quantified. Therefore we choose to set $\Delta a_2 = 0.5$. The dependence of the Bayes factor on the width of the higher-order parameter $b_4$ is much milder, as seen in on the right side of Fig.~\ref{fig:discpriors}, therefore we conservatively set the prior $\Delta b_4 =  \Delta a_2 = 0.5$. The left side of Fig.~\ref{fig:discpriors}shows the dependence of the Bayes factor on the prior width of $a_4$ and $a_2$ at fixed $b_2$, and it is clear that smaller values of $\Delta a_4$ are preferred compared to $\Delta b_4$, so we set its prior to half of $\Delta b_4$, $\Delta a_4 = 0.25$.
    \item \textbf{FV dependence}: After having chosen prior widths for the chiral and discretization part of the models as described above, the only parameters that remain are those that are introduced solely by the FV models of Eqs.~\eqref{eq:deltaFVnonmon} and~\eqref{eq:deltaFVmon}, $f_2$ and $f_3$. We again employ the Empirical Bayes Criterion and perform fits to the full dataset using the 48 out of the 64 models that depend on $f_2$ and/or $f_3$, and varying the corresponding prior widths. In this procedure, we set the priors to the remaining parameters, $g_0$, $\tilde{c}_{34}$, $\tilde{d}_{16}^r$, $c_0$, $c_1$, $c_2$, $c_3$, $c_4$, $a_2$, $a_4$, and $b_4$ to the values described in the previous steps, with the exception of the $g_0$ that shows up in the FV part of the polynomial models, which is set to $1.2(5)$ since, in contrast to the $\chi$PT models, it is not tightly constrained by the chiral dependence of the model. Based on the results for the Bayes factor shown in Fig.~\ref{fig:discFV}, we choose to set $\Delta f_2=1$ for the non-monotonic models, $\Delta f_2=2$ for the monotonic models, $\Delta f_3=6$ for the $g_{A,\chi}+g_{A,\text{FV}}^{f_3}$ model, and $\Delta f_3=1$ for the rest of the models. We only include the results for the discretization model $g_{A,a}^{(2)}$ in the figures, but visually identical results are obtained when using the other discretization model of Eq.~\eqref{eq:deltaga}, $g_{A,a}^{(4)}$.
\end{itemize}

\begin{figure}[]
\captionsetup[subfloat]{captionskip=-3pt}
\centering
\subfloat[\centering  ]
{{\includegraphics[width=0.49\columnwidth]{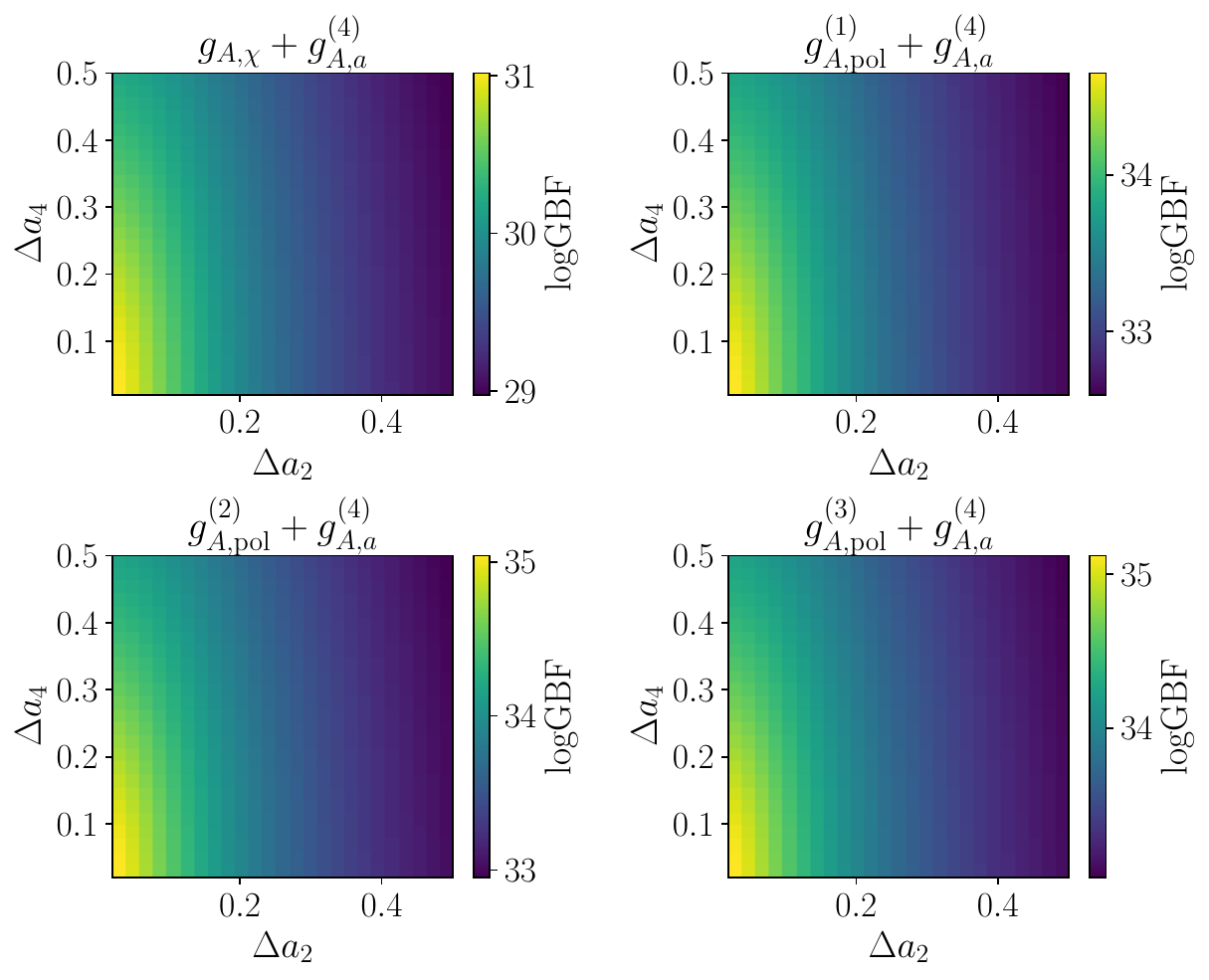} }}
\!
\subfloat[\centering  ]
{{\includegraphics[width=0.49\columnwidth]{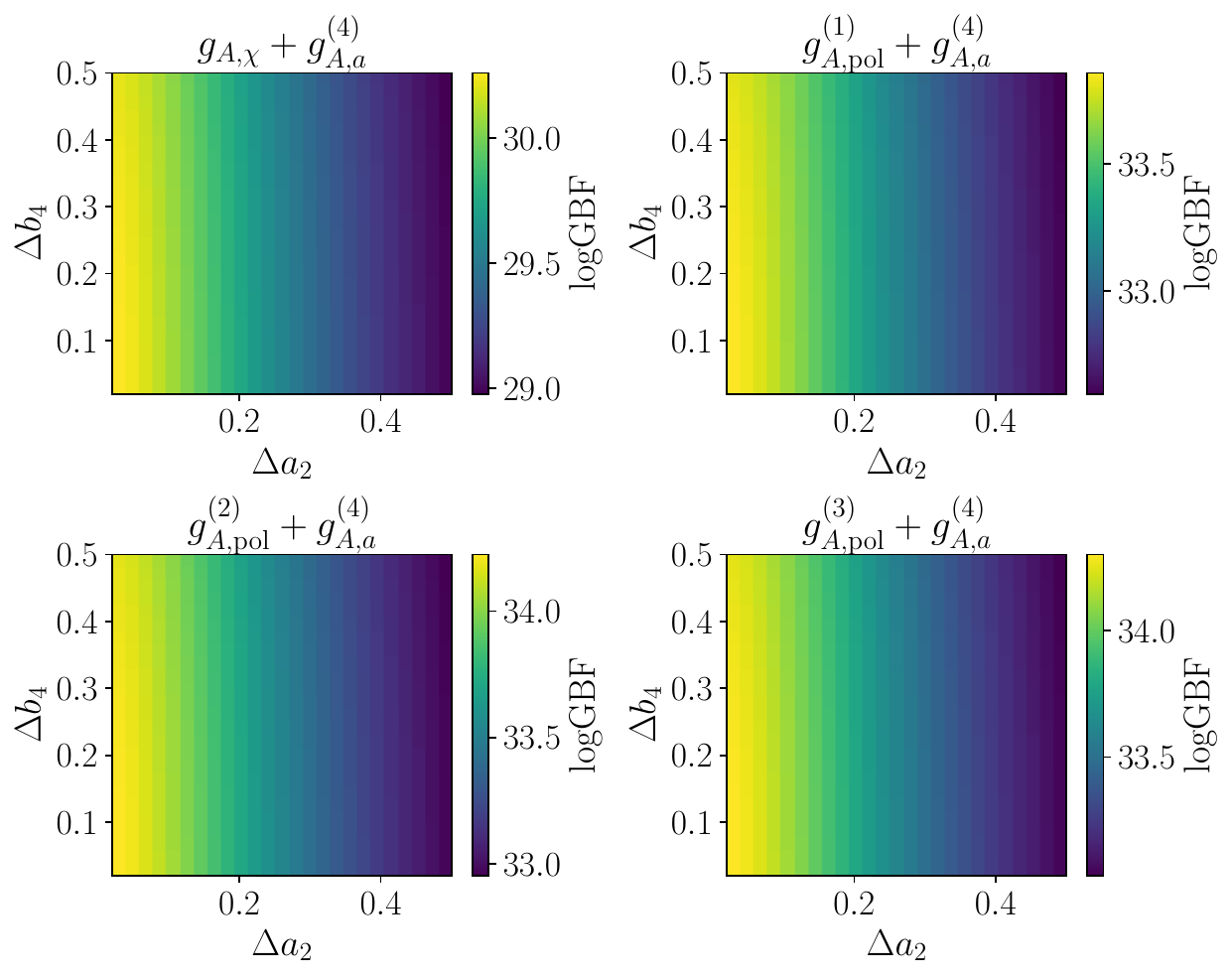} }}
\caption{The log of the Bayes factors obtained by fitting the data at all lattice spacings (keeping the largest volume ensemble for the pion masses for which multiple volumes are available) using the models of Eqs.~\eqref{eq:IVall}
plus the discretization dependence model $g_{A,a}^{(4)}$ defined in Eq.~\eqref{eq:deltaga4}. In (a) we fix $\Delta b_4=0.5$ and vary $a_2$ and $a_4$, while in (b) we fix $\Delta a_4 = 0.5$ and vary $a_2$ and $b_4$.}
\label{fig:discpriors}
\end{figure}

\begin{figure}[]
\begin{tabular}{c}
\includegraphics[width=0.7\columnwidth]{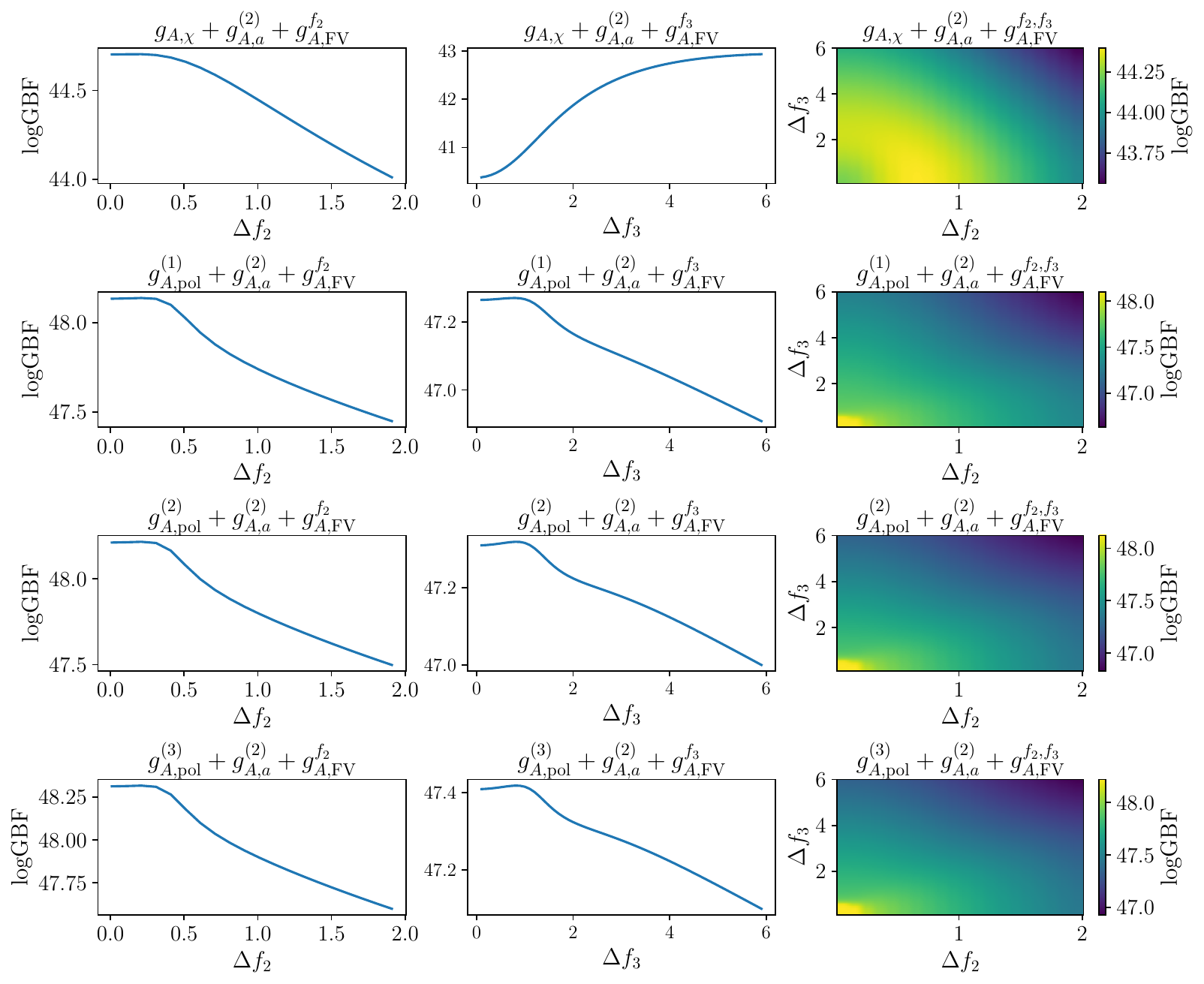}
\\
\includegraphics[width=0.85\columnwidth]{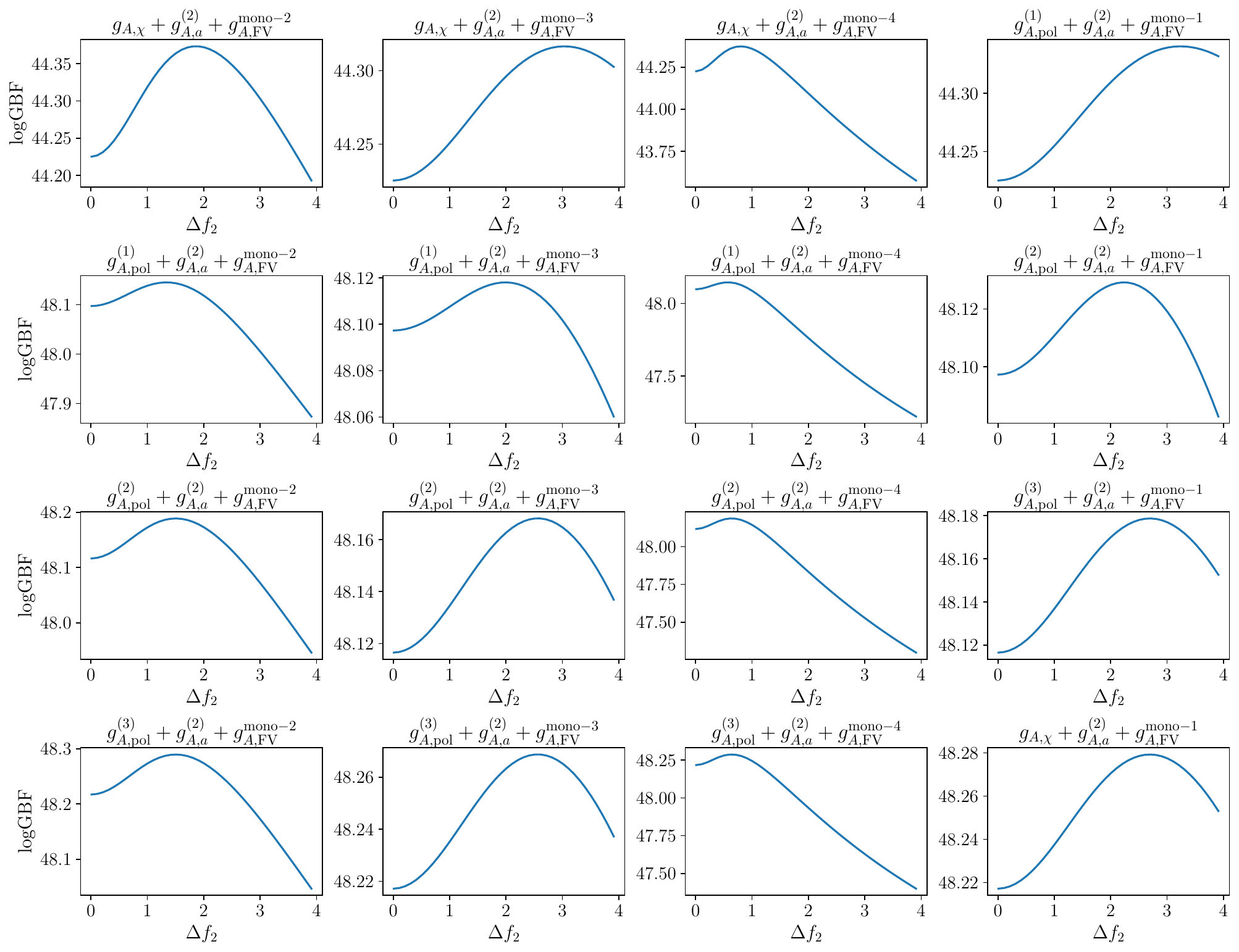}
\end{tabular}
\caption{\label{fig:discFV}
The log of the Bayes factors obtained by fitting the data at all lattice spacings (keeping the largest volume ensemble for the pion masses for which multiple volumes are available) using the models of 
Eqs.~\eqref{eq:IVall}
and Eq.~\eqref{eq:deltaga2} plus the FV models of Eqs.~\eqref{eq:deltaFVnonmon} (upper) and Eqs.~\eqref{eq:deltaFVmon} (lower) that depend on $f_2$ or/and $f_3$, the parameters whose prior width we vary.}
\end{figure}

\newpage
\section{Global fit model results}
\label{app:models}

In this appendix, tables are provided summarizing
 the final choices for fit results.
\tabref{tab:bma_results}
 provides the log Gaussian Bayes Factors
 and model weights along with the posterior
 distributions for $g_A$.
\tabref{tab:priors}
 provides the list of priors for the fits
 used in this manuscript.
\tabref{tab:postpriors}
 gives the final posterior distributions
 for the LECs that have been fit in each model.

 \squeezetable
\begin{table}[]
\caption{\label{tab:bma_results}The extrapolated $g_A$ values of the 64 models included in the analysis along with the log Gaussian Bayes Factors representing the log-likelihood distribution, logGBF, and the normalized model weights, $P(M_k|D)$.
Models are sorted by descending weight.}
\begin{tabular}{lcccccc}
\hline
\hline
Model &Chiral & FV & Disc.  &  logGBF &  $P(M_k|D)$ &   $P(g_A|M_k)$ \\
\hline
poly3\_F21\_a2&\eqref{eq:g_epi2_epi4}&\eqref{eq:gA_FV_mono_1}&\eqref{eq:deltaga2}&48.27 &3.12$\times10^{-2}$ &1.2655(79)\\
poly3\_F33\_a2&\eqref{eq:g_epi2_epi4}&\eqref{eq:gA_FV_mono_4}&\eqref{eq:deltaga2}&48.27 &3.11$\times10^{-2}$ &1.2660(79)\\
poly3\_F23\_a2&\eqref{eq:g_epi2_epi4}&\eqref{eq:gA_FV_mono_2}&\eqref{eq:deltaga2}&48.26 &3.09$\times10^{-2}$ &1.2661(79)\\
poly3\_F21\_a4&\eqref{eq:g_epi2_epi4}&\eqref{eq:gA_FV_mono_1}&\eqref{eq:deltaga4}&48.21 &2.91$\times10^{-2}$ &1.2657(97)\\
poly3\_F33\_a4&\eqref{eq:g_epi2_epi4}&\eqref{eq:gA_FV_mono_4}&\eqref{eq:deltaga4}&48.20 &2.89$\times10^{-2}$ &1.2659(96)\\
poly3\_F23\_a4&\eqref{eq:g_epi2_epi4}&\eqref{eq:gA_FV_mono_2}&\eqref{eq:deltaga4}&48.19 &2.87$\times10^{-2}$ &1.2659(96)\\
poly2\_F21\_a2&\eqref{eq:gA_epi2_epi3}&\eqref{eq:gA_FV_mono_1}&\eqref{eq:deltaga2}&48.17 &2.82$\times10^{-2}$ &1.2655(82)\\
poly2\_F33\_a2&\eqref{eq:gA_epi2_epi3}&\eqref{eq:gA_FV_mono_4}&\eqref{eq:deltaga2}&48.17 &2.81$\times10^{-2}$ &1.2660(81)\\
poly2\_F23\_a2&\eqref{eq:gA_epi2_epi3}&\eqref{eq:gA_FV_mono_2}&\eqref{eq:deltaga2}&48.16 &2.79$\times10^{-2}$ &1.2660(81)\\
poly1\_F33\_a2&\eqref{eq:gA_epi_epi2}&\eqref{eq:gA_FV_mono_4}&\eqref{eq:deltaga2}&48.13 &2.70$\times10^{-2}$ &1.2718(90)\\
poly1\_F21\_a2&\eqref{eq:gA_epi_epi2}&\eqref{eq:gA_FV_mono_1}&\eqref{eq:deltaga2}&48.12 &2.67$\times10^{-2}$ &1.2713(91)\\
poly1\_F23\_a2&\eqref{eq:gA_epi_epi2}&\eqref{eq:gA_FV_mono_2}&\eqref{eq:deltaga2}&48.12 &2.67$\times10^{-2}$ &1.2718(90)\\
poly2\_F21\_a4&\eqref{eq:gA_epi2_epi3}&\eqref{eq:gA_FV_mono_1}&\eqref{eq:deltaga4}&48.11 &2.64$\times10^{-2}$ &1.2657(98)\\
poly2\_F33\_a4&\eqref{eq:gA_epi2_epi3}&\eqref{eq:gA_FV_mono_4}&\eqref{eq:deltaga4}&48.10 &2.62$\times10^{-2}$ &1.2659(98)\\
poly2\_F23\_a4&\eqref{eq:gA_epi2_epi3}&\eqref{eq:gA_FV_mono_2}&\eqref{eq:deltaga4}&48.09 &2.60$\times10^{-2}$ &1.2659(98)\\
poly1\_F33\_a4&\eqref{eq:gA_epi_epi2}&\eqref{eq:gA_FV_mono_4}&\eqref{eq:deltaga4}&48.05 &2.51$\times10^{-2}$ &1.271(11)\\
poly1\_F21\_a4&\eqref{eq:gA_epi_epi2}&\eqref{eq:gA_FV_mono_1}&\eqref{eq:deltaga4}&48.05 &2.49$\times10^{-2}$ &1.271(11)\\
poly1\_F23\_a4&\eqref{eq:gA_epi_epi2}&\eqref{eq:gA_FV_mono_2}&\eqref{eq:deltaga4}&48.04 &2.48$\times10^{-2}$ &1.271(11)\\
poly3\_F31\_a2&\eqref{eq:g_epi2_epi4}&\eqref{eq:gA_FV_mono_3}&\eqref{eq:deltaga2}&47.93 &2.22$\times10^{-2}$ &1.2651(79)\\
poly3\_NNLOf2\_a2&\eqref{eq:g_epi2_epi4}&\eqref{eq:gA_FV_f2}&\eqref{eq:deltaga2}&47.90 &2.15$\times10^{-2}$ &1.2654(81)\\
poly3\_F31\_a4&\eqref{eq:g_epi2_epi4}&\eqref{eq:gA_FV_mono_3}&\eqref{eq:deltaga4}&47.87 &2.09$\times10^{-2}$ &1.2655(97)\\
poly3\_NNLOf2\_a4&\eqref{eq:g_epi2_epi4}&\eqref{eq:gA_FV_f2}&\eqref{eq:deltaga4}&47.84 &2.01$\times10^{-2}$ &1.2658(97)\\
poly2\_F31\_a2&\eqref{eq:gA_epi2_epi3}&\eqref{eq:gA_FV_mono_3}&\eqref{eq:deltaga2}&47.83 &2.01$\times10^{-2}$ &1.2650(82)\\
poly2\_NNLOf2\_a2&\eqref{eq:gA_epi2_epi3}&\eqref{eq:gA_FV_f2}&\eqref{eq:deltaga2}&47.80 &1.95$\times10^{-2}$ &1.2654(83)\\
poly2\_F31\_a4&\eqref{eq:gA_epi2_epi3}&\eqref{eq:gA_FV_mono_3}&\eqref{eq:deltaga4}&47.77 &1.90$\times10^{-2}$ &1.2655(98)\\
poly1\_F31\_a2&\eqref{eq:gA_epi_epi2}&\eqref{eq:gA_FV_mono_3}&\eqref{eq:deltaga2}&47.76 &1.87$\times10^{-2}$ &1.2707(91)\\
poly1\_NNLOf2\_a2&\eqref{eq:gA_epi_epi2}&\eqref{eq:gA_FV_f2}&\eqref{eq:deltaga2}&47.74 &1.83$\times10^{-2}$ &1.2710(92)\\
poly2\_NNLOf2\_a4&\eqref{eq:gA_epi2_epi3}&\eqref{eq:gA_FV_f2}&\eqref{eq:deltaga4}&47.74 &1.82$\times10^{-2}$ &1.2657(99)\\
poly3\_NNLO\_a2&\eqref{eq:g_epi2_epi4}&\eqref{eq:gA_FV_NNLO}&\eqref{eq:deltaga2}&47.73 &1.80$\times10^{-2}$ &1.2656(80)\\
poly3\_NNLOf2f3\_a2&\eqref{eq:g_epi2_epi4}&\eqref{eq:gA_FV_f2_f3}&\eqref{eq:deltaga2}&47.71 &1.78$\times10^{-2}$ &1.2649(80)\\
poly1\_F31\_a4&\eqref{eq:gA_epi_epi2}&\eqref{eq:gA_FV_mono_3}&\eqref{eq:deltaga4}&47.69 &1.75$\times10^{-2}$ &1.271(11)\\
poly1\_NNLOf2\_a4&\eqref{eq:gA_epi_epi2}&\eqref{eq:gA_FV_f2}&\eqref{eq:deltaga4}&47.67 &1.71$\times10^{-2}$ &1.271(11)\\
poly3\_NNLO\_a4&\eqref{eq:g_epi2_epi4}&\eqref{eq:gA_FV_NNLO}&\eqref{eq:deltaga4}&47.66 &1.68$\times10^{-2}$ &1.2654(97)\\
poly3\_NNLOf2f3\_a4&\eqref{eq:g_epi2_epi4}&\eqref{eq:gA_FV_f2_f3}&\eqref{eq:deltaga4}&47.65 &1.67$\times10^{-2}$ &1.2654(97)\\
poly2\_NNLO\_a2&\eqref{eq:gA_epi2_epi3}&\eqref{eq:gA_FV_NNLO}&\eqref{eq:deltaga2}&47.62 &1.63$\times10^{-2}$ &1.2656(82)\\
poly2\_NNLOf2f3\_a2&\eqref{eq:gA_epi2_epi3}&\eqref{eq:gA_FV_f2_f3}&\eqref{eq:deltaga2}&47.61 &1.61$\times10^{-2}$ &1.2648(83)\\
poly1\_NNLOf2f3\_a2&\eqref{eq:gA_epi_epi2}&\eqref{eq:gA_FV_f2_f3}&\eqref{eq:deltaga2}&47.56 &1.54$\times10^{-2}$ &1.2704(92)\\
poly2\_NNLO\_a4&\eqref{eq:gA_epi2_epi3}&\eqref{eq:gA_FV_NNLO}&\eqref{eq:deltaga4}&47.56 &1.52$\times10^{-2}$ &1.2654(98)\\
poly2\_NNLOf2f3\_a4&\eqref{eq:gA_epi2_epi3}&\eqref{eq:gA_FV_f2_f3}&\eqref{eq:deltaga4}&47.55 &1.51$\times10^{-2}$ &1.2654(99)\\
poly1\_NNLO\_a2&\eqref{eq:gA_epi_epi2}&\eqref{eq:gA_FV_NNLO}&\eqref{eq:deltaga2}&47.55 &1.51$\times10^{-2}$ &1.2713(91)\\
poly1\_NNLOf2f3\_a4&\eqref{eq:gA_epi_epi2}&\eqref{eq:gA_FV_f2_f3}&\eqref{eq:deltaga4}&47.50 &1.43$\times10^{-2}$ &1.271(11)\\
poly1\_NNLO\_a4&\eqref{eq:gA_epi_epi2}&\eqref{eq:gA_FV_NNLO}&\eqref{eq:deltaga4}&47.48 &1.40$\times10^{-2}$ &1.271(11)\\
poly3\_NNLOf3\_a2&\eqref{eq:g_epi2_epi4}&\eqref{eq:gA_FV_f3}&\eqref{eq:deltaga2}&47.42 &1.32$\times10^{-2}$ &1.2662(79)\\
poly3\_NNLOf3\_a4&\eqref{eq:g_epi2_epi4}&\eqref{eq:gA_FV_f3}&\eqref{eq:deltaga4}&47.35 &1.23$\times10^{-2}$ &1.2658(97)\\
poly2\_NNLOf3\_a2&\eqref{eq:gA_epi2_epi3}&\eqref{eq:gA_FV_f3}&\eqref{eq:deltaga2}&47.32 &1.20$\times10^{-2}$ &1.2661(82)\\
poly1\_NNLOf3\_a2&\eqref{eq:gA_epi_epi2}&\eqref{eq:gA_FV_f3}&\eqref{eq:deltaga2}&47.27 &1.14$\times10^{-2}$ &1.2720(91)\\
poly2\_NNLOf3\_a4&\eqref{eq:gA_epi2_epi3}&\eqref{eq:gA_FV_f3}&\eqref{eq:deltaga4}&47.25 &1.12$\times10^{-2}$ &1.2658(98)\\
poly1\_NNLOf3\_a4&\eqref{eq:gA_epi_epi2}&\eqref{eq:gA_FV_f3}&\eqref{eq:deltaga4}&47.20 &1.07$\times10^{-2}$ &1.271(11)\\
chPT\_NNLOf2\_a2&\eqref{eq:gA_N2LO}&\eqref{eq:gA_FV_f2}&\eqref{eq:deltaga2}&44.45 &6.80$\times10^{-4}$ &1.266(10)\\
chPT\_NNLOf2\_a4&\eqref{eq:gA_N2LO}&\eqref{eq:gA_FV_f2}&\eqref{eq:deltaga4}&44.39 &6.41$\times10^{-4}$ &1.266(11)\\
chPT\_F21\_a2&\eqref{eq:gA_N2LO}&\eqref{eq:gA_FV_mono_1}&\eqref{eq:deltaga2}&44.37 &6.31$\times10^{-4}$ &1.267(11)\\
chPT\_F21\_a4&\eqref{eq:gA_N2LO}&\eqref{eq:gA_FV_mono_1}&\eqref{eq:deltaga4}&44.31 &5.93$\times10^{-4}$ &1.267(12)\\
chPT\_F33\_a2&\eqref{eq:gA_N2LO}&\eqref{eq:gA_FV_mono_4}&\eqref{eq:deltaga2}&44.31 &5.93$\times10^{-4}$ &1.268(11)\\
chPT\_NNLOf2f3\_a2&\eqref{eq:gA_N2LO}&\eqref{eq:gA_FV_f2_f3}&\eqref{eq:deltaga2}&44.30 &5.88$\times10^{-4}$ &1.267(11)\\
chPT\_F23\_a2&\eqref{eq:gA_N2LO}&\eqref{eq:gA_FV_mono_2}&\eqref{eq:deltaga2}&44.30 &5.85$\times10^{-4}$ &1.268(11)\\
chPT\_F33\_a4&\eqref{eq:gA_N2LO}&\eqref{eq:gA_FV_mono_4}&\eqref{eq:deltaga4}&44.25 &5.59$\times10^{-4}$ &1.267(12)\\
chPT\_NNLOf2f3\_a4&\eqref{eq:gA_N2LO}&\eqref{eq:gA_FV_f2_f3}&\eqref{eq:deltaga4}&44.24 &5.54$\times10^{-4}$ &1.267(12)\\
chPT\_F23\_a4&\eqref{eq:gA_N2LO}&\eqref{eq:gA_FV_mono_2}&\eqref{eq:deltaga4}&44.24 &5.52$\times10^{-4}$ &1.267(12)\\
chPT\_F31\_a2&\eqref{eq:gA_N2LO}&\eqref{eq:gA_FV_mono_3}&\eqref{eq:deltaga2}&44.09 &4.78$\times10^{-4}$ &1.266(11)\\
chPT\_F31\_a4&\eqref{eq:gA_N2LO}&\eqref{eq:gA_FV_mono_3}&\eqref{eq:deltaga4}&44.03 &4.50$\times10^{-4}$ &1.266(12)\\
chPT\_NNLOf3\_a2&\eqref{eq:gA_N2LO}&\eqref{eq:gA_FV_f3}&\eqref{eq:deltaga2}&42.94 &1.51$\times10^{-4}$ &1.267(10)\\
chPT\_NNLOf3\_a4&\eqref{eq:gA_N2LO}&\eqref{eq:gA_FV_f3}&\eqref{eq:deltaga4}&42.91 &1.46$\times10^{-4}$ &1.266(12)\\
chPT\_NNLO\_a4&\eqref{eq:gA_N2LO}&\eqref{eq:gA_FV_NNLO}&\eqref{eq:deltaga4}&41.56 &3.80$\times10^{-5}$ &1.267(12)\\
chPT\_NNLO\_a2&\eqref{eq:gA_N2LO}&\eqref{eq:gA_FV_NNLO}&\eqref{eq:deltaga2}&41.51 &3.61$\times10^{-5}$ &1.270(10)\\
\hline
Bayes Model Average& & & & & & 1.2674(96) \\
\hline
\hline
\end{tabular}
\end{table}

\begin{table}[]
\caption{\label{tab:priors}Priors used in the model fits. The models are ordered according to the order in which the model Eqs.~\eqref{eq:IVall},~\eqref{eq:deltaFVnonmon},~\eqref{eq:deltaFVmon}, and~\eqref{eq:deltaga} appear in the main text.}
\begin{tabular}{lllllllllllllllll}
\hline
\hline
Chiral                                                                            & FV                                                                                & Disc.                                &  & $g_0$         & $\tilde{c}_{34}$ & $\tilde{d}_{16}^r$ & $c_0$        & $c_1$      & $c_2$      & $c_3$    & $c_4$    & $f_2$    & $f_3$    & $a_2$    & $a_4$    & $b_4$    \\ \hline
\multicolumn{1}{l|}{\multirow{16}{*}{\eqref{eq:gA_N2LO}}}       & \multicolumn{1}{l|}{\multirow{2}{*}{\eqref{eq:gA_FV_NNLO}}}    & \eqref{eq:deltaga2} &  & $1.25(10)$    & $0(1)$         & $0(5)$                                                &              &            &            &          &          &          &          & $0(0.5)$ &          &          \\ \cline{3-3}
\multicolumn{1}{l|}{}                                                             & \multicolumn{1}{l|}{}                                                             & \eqref{eq:deltaga4} &  & $1.25(10)$  & $0(1)$         & $0(5)$                                                &              &            &            &          &          &          &          & $0(0.5)$ & $0(0.25)$ & $0(0.5)$ \\ \cline{2-3}
\multicolumn{1}{l|}{}                                                             & \multicolumn{1}{l|}{\multirow{2}{*}{\eqref{eq:gA_FV_f2}}}      & \eqref{eq:deltaga2} &  & $1.25(10)$  & $0(1)$         & $0(5)$                                                &              &            &            &          &          & $0(1)$ &          & $0(0.5)$ &          &          \\ \cline{3-3}
\multicolumn{1}{l|}{}                                                             & \multicolumn{1}{l|}{}                                                             & \eqref{eq:deltaga4} &  & $1.25(10)$  & $0(1)$         & $0(5)$                                                &              &            &            &          &          & $0(1)$ &          & $0(0.5)$ & $0(0.25)$ & $0(0.5)$ \\ \cline{2-3}
\multicolumn{1}{l|}{}                                                             & \multicolumn{1}{l|}{\multirow{2}{*}{\eqref{eq:gA_FV_f3}}}      & \eqref{eq:deltaga2} &  & $1.25(10)$  & $0(1)$         & $0(5)$                                                &              &            &            &          &          &          & $0(6)$ & $0(0.5)$ &          &          \\ \cline{3-3}
\multicolumn{1}{l|}{}                                                             & \multicolumn{1}{l|}{}                                                             & \eqref{eq:deltaga4} &  & $1.25(10)$  & $0(1)$         & $0(5)$                                                &              &            &            &          &          &          & $0(6)$ & $0(0.5)$ & $0(0.25)$ & $0(0.5)$ \\ \cline{2-3}
\multicolumn{1}{l|}{}                                                             & \multicolumn{1}{l|}{\multirow{2}{*}{\eqref{eq:gA_FV_f2_f3}}}  & \eqref{eq:deltaga2} &  & $1.25(10)$  & $0(1)$         & $0(5)$                                                &              &            &            &          &          & $0(1)$ & $0(6)$ & $0(0.5)$ &          &          \\ \cline{3-3}
\multicolumn{1}{l|}{}                                                             & \multicolumn{1}{l|}{}                                                             & \eqref{eq:deltaga4} &  & $1.25(10)$  & $0(1)$         & $0(5)$                                                &              &            &            &          &          & $0(1)$ & $0(6)$ & $0(0.5)$ & $0(0.25)$ & $0(0.5)$ \\ \cline{2-3}
\multicolumn{1}{l|}{}                                                             & \multicolumn{1}{l|}{\multirow{2}{*}{\eqref{eq:gA_FV_mono_1}}} & \eqref{eq:deltaga2} &  & $1.25(10)$   & $0(1)$         & $0(5)$                                                &              &            &            &          &          & $0(2)$ &          & $0(0.5)$ &          &          \\ \cline{3-3}
\multicolumn{1}{l|}{}                                                             & \multicolumn{1}{l|}{}                                                             & \eqref{eq:deltaga4} &  & $1.25(10)$  & $0(1)$         & $0(5)$                                                &              &            &            &          &          & $0(2)$ &          & $0(0.5)$ & $0(0.25)$ & $0(0.5)$ \\ \cline{2-3}
\multicolumn{1}{l|}{}                                                             & \multicolumn{1}{l|}{\multirow{2}{*}{\eqref{eq:gA_FV_mono_2}}} & \eqref{eq:deltaga2} &  & $1.25(10)$  & $0(1)$         & $0(5)$                                                &              &            &            &          &          & $0(2)$ &          & $0(0.5)$ &          &          \\ \cline{3-3}
\multicolumn{1}{l|}{}                                                             & \multicolumn{1}{l|}{}                                                             & \eqref{eq:deltaga4} &  & $1.25(10)$  & $0(1)$         & $0(5)$                                                &              &            &            &          &          & $0(2)$ &          & $0(0.5)$ & $0(0.25)$ & $0(0.5)$   \\ \cline{2-3}
\multicolumn{1}{l|}{}                                                             & \multicolumn{1}{l|}{\multirow{2}{*}{\eqref{eq:gA_FV_mono_3}}} & \eqref{eq:deltaga2} &  & $1.25(10)$  & $0(1)$         & $0(5)$                                                &              &            &            &          &          & $0(2)$ &          & $0(0.5)$ &          &          \\ \cline{3-3}
\multicolumn{1}{l|}{}                                                             & \multicolumn{1}{l|}{}                                                             & \eqref{eq:deltaga4} &  & $1.25(10)$  & $0(1)$         & $0(5)$                                                &              &            &            &          &          & $0(2)$ &          & $0(0.5)$ & $0(0.25)$ & $0(0.5)$ \\ \cline{2-3}
\multicolumn{1}{l|}{}                                                             & \multicolumn{1}{l|}{\multirow{2}{*}{\eqref{eq:gA_FV_mono_4}}} & \eqref{eq:deltaga2} &  & $1.25(10)$  & $0(1)$         & $0(5)$                                                &              &            &            &          &          & $0(2)$ &          & $0(0.5)$ &          &          \\ \cline{3-3}
\multicolumn{1}{l|}{}                                                             & \multicolumn{1}{l|}{}                                                             & \eqref{eq:deltaga4} &  & $1.25(10)$  & $0(1)$         & $0(5)$                                                &              &            &            &          &          & $0(2)$ &          & $0(0.5)$ & $0(0.25)$ & $0(0.5)$ \\ \cline{1-3}
\hline
\multicolumn{1}{l|}{\multirow{16}{*}{\eqref{eq:gA_epi_epi2}}}  & \multicolumn{1}{l|}{\multirow{2}{*}{\eqref{eq:gA_FV_NNLO}}}    & \eqref{eq:deltaga2} &  & $1.25(50)$  & $0(1)$         &                                                         & $1.25(10)$ & $0(0.5)$ & $0(0.5)$ &          &          &          &          & $0(0.5)$ &          &          \\ \cline{3-3}
\multicolumn{1}{l|}{}                                                             & \multicolumn{1}{l|}{}                                                             & \eqref{eq:deltaga4} &  & $1.25(50)$  & $0(1)$         &                                                         & $1.25(10)$ & $0(0.5)$ & $0(0.5)$ &          &          &          &          & $0(0.5)$ & $0(0.25)$ & $0(0.5)$ \\ \cline{2-3}
\multicolumn{1}{l|}{}                                                             & \multicolumn{1}{l|}{\multirow{2}{*}{\eqref{eq:gA_FV_f2}}}      & \eqref{eq:deltaga2} &  & $1.25(50)$  & $0(1)$         &                                                         & $1.25(10)$ & $0(0.5)$ & $0(0.5)$ &          &          & $0(1)$ &          & $0(0.5)$ &          &          \\ \cline{3-3}
\multicolumn{1}{l|}{}                                                             & \multicolumn{1}{l|}{}                                                             & \eqref{eq:deltaga4} &  & $1.25(50)$  & $0(1)$         &                                                         & $1.25(10)$ & $0(0.5)$ & $0(0.5)$ &          &          & $0(1)$ &          & $0(0.5)$ & $0(0.25)$ & $0(0.5)$ \\ \cline{2-3}
\multicolumn{1}{l|}{}                                                             & \multicolumn{1}{l|}{\multirow{2}{*}{\eqref{eq:gA_FV_f3}}}      & \eqref{eq:deltaga2} &  & $1.25(50)$  & $0(1)$         &                                                         & $1.25(10)$ & $0(0.5)$ & $0(0.5)$ &          &          &          & $0(1)$ & $0(0.5)$ &          &          \\ \cline{3-3}
\multicolumn{1}{l|}{}                                                             & \multicolumn{1}{l|}{}                                                             & \eqref{eq:deltaga4} &  & $1.25(50)$  & $0(1)$         &                                                         & $1.25(10)$ & $0(0.5)$ & $0(0.5)$ &          &          &          & $0(1)$ & $0(0.5)$ & $0(0.25)$ & $0(0.5)$ \\ \cline{2-3}
\multicolumn{1}{l|}{}                                                             & \multicolumn{1}{l|}{\multirow{2}{*}{\eqref{eq:gA_FV_f2_f3}}}  & \eqref{eq:deltaga2} &  & $1.25(50)$  & $0(1)$         &                                                         & $1.25(10)$ & $0(0.5)$ & $0(0.5)$ &          &          & $0(1)$ & $0(1)$ & $0(0.5)$ &          &          \\ \cline{3-3}
\multicolumn{1}{l|}{}                                                             & \multicolumn{1}{l|}{}                                                             & \eqref{eq:deltaga4} &  & $1.25(50)$  & $0(1)$         &                                                         & $1.25(10)$ & $0(0.5)$ & $0(0.5)$ &          &          & $0(1)$ & $0(1)$ & $0(0.5)$ & $0(0.25)$ & $0(0.5)$ \\ \cline{2-3}
\multicolumn{1}{l|}{}                                                             & \multicolumn{1}{l|}{\multirow{2}{*}{\eqref{eq:gA_FV_mono_1}}} & \eqref{eq:deltaga2} &  &               &                  &                                                         & $1.25(10)$ & $0(0.5)$ & $0(0.5)$ &          &          & $0(2)$ &          & $0(0.5)$ &          &          \\ \cline{3-3}
\multicolumn{1}{l|}{}                                                             & \multicolumn{1}{l|}{}                                                             & \eqref{eq:deltaga4} &  &               &                  &                                                         & $1.25(10)$ & $0(0.5)$ & $0(0.5)$ &          &          & $0(2)$ &          & $0(0.5)$ & $0(0.25)$ & $0(0.5)$ \\ \cline{2-3}
\multicolumn{1}{l|}{}                                                             & \multicolumn{1}{l|}{\multirow{2}{*}{\eqref{eq:gA_FV_mono_2}}} & \eqref{eq:deltaga2} &  &               &                  &                                                         & $1.25(10)$ & $0(0.5)$ & $0(0.5)$ &          &          & $0(2)$ &          & $0(0.5)$ &          &          \\ \cline{3-3}
\multicolumn{1}{l|}{}                                                             & \multicolumn{1}{l|}{}                                                             & \eqref{eq:deltaga4} &  &               &                  &                                                         & $1.25(10)$ & $0(0.5)$ & $0(0.5)$ &          &          & $0(2)$ &          & $0(0.5)$ & $0(0.25)$ & $0(0.5)$ \\ \cline{2-3}
\multicolumn{1}{l|}{}                                                             & \multicolumn{1}{l|}{\multirow{2}{*}{\eqref{eq:gA_FV_mono_3}}} & \eqref{eq:deltaga2} &  &               &                  &                                                         & $1.25(10)$ & $0(0.5)$ & $0(0.5)$ &          &          & $0(2)$ &          & $0(0.5)$ &          &          \\ \cline{3-3}
\multicolumn{1}{l|}{}                                                             & \multicolumn{1}{l|}{}                                                             & \eqref{eq:deltaga4} &  &               &                  &                                                         & $1.25(10)$ & $0(0.5)$ & $0(0.5)$ &          &          & $0(2)$ &          & $0(0.5)$ & $0(0.25)$ & $0(0.5)$ \\ \cline{2-3}
\multicolumn{1}{l|}{}                                                             & \multicolumn{1}{l|}{\multirow{2}{*}{\eqref{eq:gA_FV_mono_4}}} & \eqref{eq:deltaga2} &  &               &                  &                                                         & $1.25(10)$ & $0(0.5)$ & $0(0.5)$ &          &          & $0(2)$ &          & $0(0.5)$ &          &          \\ \cline{3-3}
\multicolumn{1}{l|}{}                                                             & \multicolumn{1}{l|}{}                                                             & \eqref{eq:deltaga4} &  &               &                  &                                                         & $1.25(10)$ & $0(0.5)$ & $0(0.5)$ &          &          & $0(2)$ &          & $0(0.5)$ & $0(0.25)$ & $0(0.5)$ \\ \cline{1-3}
\hline
\multicolumn{1}{l|}{\multirow{16}{*}{\eqref{eq:gA_epi2_epi3}}} & \multicolumn{1}{l|}{\multirow{2}{*}{\eqref{eq:gA_FV_NNLO}}}    & \eqref{eq:deltaga2} &  & $1.25(50)$  & $0(1)$         &                                                         & $1.25(10)$ &            & $0(1)$   & $0(1)$ &          &          &          & $0(0.5)$ &          &          \\ \cline{3-3}
\multicolumn{1}{l|}{}                                                             & \multicolumn{1}{l|}{}                                                             & \eqref{eq:deltaga4} &  & $1.25(50)$ & $0(1)$         &                                                         & $1.25(10)$ &            & $0(1)$   & $0(1)$ &          &          &          & $0(0.5)$ & $0(0.25)$ & $0(0.5)$ \\ \cline{2-3}
\multicolumn{1}{l|}{}                                                             & \multicolumn{1}{l|}{\multirow{2}{*}{\eqref{eq:gA_FV_f2}}}      & \eqref{eq:deltaga2} &  & $1.25(50)$  & $0(1)$         &                                                         & $1.25(10)$ &            & $0(1)$   & $0(1)$ &          & $0(1)$ &          & $0(0.5)$ &          &          \\ \cline{3-3}
\multicolumn{1}{l|}{}                                                             & \multicolumn{1}{l|}{}                                                             & \eqref{eq:deltaga4} &  & $1.25(50)$  & $0(1)$         &                                                         & $1.25(10)$ &            & $0(1)$   & $0(1)$ &          & $0(1)$ &          & $0(0.5)$ & $0(0.25)$ & $0(0.5)$ \\ \cline{2-3}
\multicolumn{1}{l|}{}                                                             & \multicolumn{1}{l|}{\multirow{2}{*}{\eqref{eq:gA_FV_f3}}}      & \eqref{eq:deltaga2} &  & $1.25(50)$  & $0(1)$         &                                                         & $1.25(10)$ &            & $0(1)$   & $0(1)$ &          &          & $0(1)$ & $0(0.5)$ &          &          \\ \cline{3-3}
\multicolumn{1}{l|}{}                                                             & \multicolumn{1}{l|}{}                                                             & \eqref{eq:deltaga4} &  & $1.25(50)$  & $0(1)$         &                                                         & $1.25(10)$ &            & $0(1)$   & $0(1)$ &          &          & $0(1)$ & $0(0.5)$ & $0(0.25)$ & $0(0.5)$ \\ \cline{2-3}
\multicolumn{1}{l|}{}                                                             & \multicolumn{1}{l|}{\multirow{2}{*}{\eqref{eq:gA_FV_f2_f3}}}  & \eqref{eq:deltaga2} &  & $1.25(50)$  & $0(1)$         &                                                         & $1.25(10)$ &            & $0(1)$   & $0(1)$ &          & $0(1)$ & $0(1)$ & $0(0.5)$ &          &          \\ \cline{3-3}
\multicolumn{1}{l|}{}                                                             & \multicolumn{1}{l|}{}                                                             & \eqref{eq:deltaga4} &  & $1.25(50)$  & $0(1)$         &                                                         & $1.25(10)$ &            & $0(1)$   & $0(1)$ &          & $0(1)$ & $0(1)$ & $0(0.5)$ & $0(0.25)$ & $0(0.5)$ \\ \cline{2-3}
\multicolumn{1}{l|}{}                                                             & \multicolumn{1}{l|}{\multirow{2}{*}{\eqref{eq:gA_FV_mono_1}}} & \eqref{eq:deltaga2} &  &               &                  &                                                         & $1.25(10)$ &            & $0(1)$   & $0(1)$ &          & $0(2)$ &          & $0(0.5)$ &          &          \\ \cline{3-3}
\multicolumn{1}{l|}{}                                                             & \multicolumn{1}{l|}{}                                                             & \eqref{eq:deltaga4} &  &               &                  &                                                         & $1.25(10)$ &            & $0(1)$   & $0(1)$ &          & $0(2)$ &          & $0(0.5)$ & $0(0.25)$ & $0(0.5)$ \\ \cline{2-3}
\multicolumn{1}{l|}{}                                                             & \multicolumn{1}{l|}{\multirow{2}{*}{\eqref{eq:gA_FV_mono_2}}} & \eqref{eq:deltaga2} &  &               &                  &                                                         & $1.25(10)$ &            & $0(1)$   & $0(1)$ &          & $0(2)$ &          & $0(0.5)$ &          &          \\ \cline{3-3}
\multicolumn{1}{l|}{}                                                             & \multicolumn{1}{l|}{}                                                             & \eqref{eq:deltaga4} &  &               &                  &                                                         & $1.25(10)$ &            & $0(1)$   & $0(1)$ &          & $0(2)$ &          & $0(0.5)$ & $0(0.25)$ & $0(0.5)$ \\ \cline{2-3}
\multicolumn{1}{l|}{}                                                             & \multicolumn{1}{l|}{\multirow{2}{*}{\eqref{eq:gA_FV_mono_3}}} & \eqref{eq:deltaga2} &  &               &                  &                                                         & $1.25(10)$ &            & $0(1)$   & $0(1)$ &          & $0(2)$ &          & $0(0.5)$ &          &          \\ \cline{3-3}
\multicolumn{1}{l|}{}                                                             & \multicolumn{1}{l|}{}                                                             & \eqref{eq:deltaga4} &  &               &                  &                                                         & $1.25(10)$ &            & $0(1)$   & $0(1)$ &          & $0(2)$ &          & $0(0.5)$ & $0(0.25)$ & $0(0.5)$ \\ \cline{2-3}
\multicolumn{1}{l|}{}                                                             & \multicolumn{1}{l|}{\multirow{2}{*}{\eqref{eq:gA_FV_mono_4}}} & \eqref{eq:deltaga2} &  &               &                  &                                                         & $1.25(10)$ &            & $0(1)$   & $0(1)$ &          & $0(2)$ &          & $0(0.5)$ &          &          \\ \cline{3-3}
\multicolumn{1}{l|}{}                                                             & \multicolumn{1}{l|}{}                                                             & \eqref{eq:deltaga4} &  &               &                  &                                                         & $1.25(10)$ &            & $0(1)$   & $0(1)$ &          & $0(2)$ &          & $0(0.5)$ & $0(0.25)$ & $0(0.5)$ \\ \cline{1-3}
\hline
\multicolumn{1}{l|}{\multirow{16}{*}{\eqref{eq:g_epi2_epi4}}} & \multicolumn{1}{l|}{\multirow{2}{*}{\eqref{eq:gA_FV_NNLO}}}    & \eqref{eq:deltaga2} &  & $1.25(50)$  & $0(1)$         &                                                         & $1.25(10)$ &            & $0(1)$   &          & $0(1)$ &          &          & $0(0.5)$ &          &          \\ \cline{3-3}
\multicolumn{1}{l|}{}                                                             & \multicolumn{1}{l|}{}                                                             & \eqref{eq:deltaga4} &  & $1.25(50)$  & $0(1)$         &                                                         & $1.25(10)$ &            & $0(1)$   &          & $0(1)$ &          &          & $0(0.5)$ & $0(0.25)$ & $0(0.5)$ \\ \cline{2-3}
\multicolumn{1}{l|}{}                                                             & \multicolumn{1}{l|}{\multirow{2}{*}{\eqref{eq:gA_FV_f2}}}      & \eqref{eq:deltaga2} &  & $1.25(50)$  & $0(1)$         &                                                         & $1.25(10)$ &            & $0(1)$   &          & $0(1)$ & $0(1)$ &          & $0(0.5)$ &          &          \\ \cline{3-3}
\multicolumn{1}{l|}{}                                                             & \multicolumn{1}{l|}{}                                                             & \eqref{eq:deltaga4} &  & $1.25(50)$  & $0(1)$         &                                                         & $1.25(10)$ &            & $0(1)$   &          & $0(1)$ & $0(1)$ &          & $0(0.5)$ & $0(0.25)$ & $0(0.5)$ \\ \cline{2-3}
\multicolumn{1}{l|}{}                                                             & \multicolumn{1}{l|}{\multirow{2}{*}{\eqref{eq:gA_FV_f3}}}      & \eqref{eq:deltaga2} &  & $1.25(50)$  & $0(1)$         &                                                         & $1.25(10)$ &            & $0(1)$   &          & $0(1)$ &          & $0(1)$ & $0(0.5)$ &          &          \\ \cline{3-3}
\multicolumn{1}{l|}{}                                                             & \multicolumn{1}{l|}{}                                                             & \eqref{eq:deltaga4} &  & $1.25(50)$  & $0(1)$         &                                                         & $1.25(10)$ &            & $0(1)$   &          & $0(1)$ &          & $0(1)$ & $0(0.5)$ & $0(0.25)$ & $0(0.5)$ \\ \cline{2-3}
\multicolumn{1}{l|}{}                                                             & \multicolumn{1}{l|}{\multirow{2}{*}{\eqref{eq:gA_FV_f2_f3}}}  & \eqref{eq:deltaga2} &  & $1.25(50)$  & $0(1)$         &                                                         & $1.25(10)$ &            & $0(1)$   &          & $0(1)$ & $0(1)$ & $0(1)$ & $0(0.5)$ &          &          \\ \cline{3-3}
\multicolumn{1}{l|}{}                                                             & \multicolumn{1}{l|}{}                                                             & \eqref{eq:deltaga4} &  & $1.25(50)$  & $0(1)$         &                                                         & $1.25(10)$ &            & $0(1)$   &          & $0(1)$ & $0(1)$ & $0(1)$ &$0(0.5)$ & $0(0.25)$ & $0(0.5)$ \\ \cline{2-3}
\multicolumn{1}{l|}{}                                                             & \multicolumn{1}{l|}{\multirow{2}{*}{\eqref{eq:gA_FV_mono_1}}} & \eqref{eq:deltaga2} &  &               &                  &                                                         & $1.25(10)$ &            & $0(1)$   &          & $0(1)$ & $0(2)$ &          & $0(0.5)$ &          &          \\ \cline{3-3}
\multicolumn{1}{l|}{}                                                             & \multicolumn{1}{l|}{}                                                             & \eqref{eq:deltaga4} &  &               &                  &                                                         & $1.25(10)$ &            & $0(1)$   &          & $0(1)$ & $0(2)$ &          & $0(0.5)$ & $0(0.25)$ & $0(0.5)$ \\ \cline{2-3}
\multicolumn{1}{l|}{}                                                             & \multicolumn{1}{l|}{\multirow{2}{*}{\eqref{eq:gA_FV_mono_2}}} & \eqref{eq:deltaga2} &  &               &                  &                                                         & $1.25(10)$ &            & $0(1)$   &          & $0(1)$ & $0(2)$ &          & $0(0.5)$ &          &          \\ \cline{3-3}
\multicolumn{1}{l|}{}                                                             & \multicolumn{1}{l|}{}                                                             & \eqref{eq:deltaga4} &  &               &                  &                                                         & $1.25(10)$ &            & $0(1)$   &          & $0(1)$ & $0(2)$ &          & $0(0.5)$ & $0(0.25)$ & $0(0.5)$ \\ \cline{2-3}
\multicolumn{1}{l|}{}                                                             & \multicolumn{1}{l|}{\multirow{2}{*}{\eqref{eq:gA_FV_mono_3}}} & \eqref{eq:deltaga2} &  &               &                  &                                                         & $1.25(10)$ &            & $0(1)$   &          & $0(1)$ & $0(2)$ &          & $0(0.5)$ &          &          \\ \cline{3-3}
\multicolumn{1}{l|}{}                                                             & \multicolumn{1}{l|}{}                                                             & \eqref{eq:deltaga4} &  &               &                  &                                                         & $1.25(10)$ &            & $0(1)$   &          & $0(1)$ & $0(2)$ &          & $0(0.5)$ & $0(0.25)$ & $0(0.5)$ \\ \cline{2-3}
\multicolumn{1}{l|}{}                                                             & \multicolumn{1}{l|}{\multirow{2}{*}{\eqref{eq:gA_FV_mono_4}}} & \eqref{eq:deltaga2} &  &               &                  &                                                         & $1.25(10)$ &            & $0(1)$   &          & $0(1)$ & $0(2)$ &          & $0(0.5)$ &          &          \\ \cline{3-3}
\multicolumn{1}{l|}{}                                                             & \multicolumn{1}{l|}{}                                                             & \eqref{eq:deltaga4} &  &               &                  &                                                         & $1.25(10)$ &            & $0(1)$   &          & $0(1)$ & $0(2)$ &          & $0(0.5)$ & $0(0.25)$ & $0(0.5)$ \\ 
\hline
\hline
\end{tabular}
\end{table}

\begin{table}[] 
\caption{\label{tab:postpriors}Posterior values of the parameters from each model fit. The models are ordered according to the order in which the model Eqs.~\eqref{eq:IVall},~\eqref{eq:deltaFVnonmon},~\eqref{eq:deltaFVmon}, and~\eqref{eq:deltaga} appear in the main text.}
\begin{sideways}
\begin{tabular}{llllllllllllllllll} 
\hline
\hline
Chiral  & FV   & Disc.  &  & $g_0$ & $\tilde{c}_{34}$ & $\tilde{d}_{16}^r$ & $c_0$  & $c_1$ & $c_2$  & $c_3$  & $c_4$  & $f_2$  & $f_3$  & $a_2$ & $a_4$ & $b_4$ &    \\ \hline
\multicolumn{1}{l|}{\multirow{16}{*}{\eqref{eq:gA_N2LO}}} &\multicolumn{1}{l|}{\multirow{2}{*}{\eqref{eq:gA_FV_NNLO}}} &\eqref{eq:deltaga2}& &$1.226(15)$ &$0.81(24)$ &$-4.95(33)$ & & & & & & & &$-0.025(50)$ & & &\\ \cline{3-3}
\multicolumn{1}{l|}{}&\multicolumn{1}{l|}{}&\eqref{eq:deltaga4}& &$1.212(20)$ &$0.80(24)$ &$-4.77(37)$ & & & & & & & &$0.10(14)$ &$-0.08(24)$ &$-1.5(1.7)$ &\\ \cline{2-3}
\multicolumn{1}{l|}{}&\multicolumn{1}{l|}{\multirow{2}{*}{\eqref{eq:gA_FV_f2}}} &\eqref{eq:deltaga2}& &$1.221(15)$ &$0.76(24)$ &$-4.85(33)$ & & & & & &$-0.47(52)$ & &$0.033(54)$ & & &\\ \cline{3-3}
\multicolumn{1}{l|}{}&\multicolumn{1}{l|}{}&\eqref{eq:deltaga4}& &$1.217(20)$ &$0.76(24)$ &$-4.80(37)$ & & & & & &$-0.45(53)$ & &$0.06(14)$ &$0.008(244)$ &$-0.5(1.7)$ &\\ \cline{2-3}
\multicolumn{1}{l|}{}&\multicolumn{1}{l|}{\multirow{2}{*}{\eqref{eq:gA_FV_f3}}} &\eqref{eq:deltaga2}& &$1.223(15)$ &$0.79(25)$ &$-4.92(33)$ & & & & & & &$5.6(2.9)$ &$0.022(54)$ & & &\\ \cline{3-3}
\multicolumn{1}{l|}{}&\multicolumn{1}{l|}{}&\eqref{eq:deltaga4}& &$1.212(21)$ &$0.77(25)$ &$-4.76(39)$ & & & & & & &$5.5(2.9)$ &$0.11(14)$ &$-0.03(24)$ &$-1.2(1.7)$ &\\ \cline{2-3}
\multicolumn{1}{l|}{}&\multicolumn{1}{l|}{\multirow{2}{*}{\eqref{eq:gA_FV_f2_f3}}} &\eqref{eq:deltaga2}& &$1.222(15)$ &$0.77(24)$ &$-4.87(33)$ & & & & & &$-0.60(54)$ &$0.16(97)$ &$0.030(54)$ & & &\\ \cline{3-3}
\multicolumn{1}{l|}{}&\multicolumn{1}{l|}{}&\eqref{eq:deltaga4}& &$1.218(21)$ &$0.77(24)$ &$-4.82(38)$ & & & & & &$-0.59(55)$ &$0.17(97)$ &$0.06(14)$ &$0.008(244)$ &$-0.4(1.7)$ &\\ \cline{2-3}
\multicolumn{1}{l|}{}&\multicolumn{1}{l|}{\multirow{2}{*}{\eqref{eq:gA_FV_mono_1}}} &\eqref{eq:deltaga2}& &$1.222(15)$ &$0.76(24)$ &$-4.86(33)$ & & & & & &$-1.4(1.4)$ & &$0.022(53)$ & & &\\ \cline{3-3}
\multicolumn{1}{l|}{}&\multicolumn{1}{l|}{}&\eqref{eq:deltaga4}& &$1.221(16)$ &$0.77(24)$ &$-4.86(34)$ & & & & & &$-1.4(1.4)$ & &$0.028(89)$ &$-0.009(243)$ &$-0.04(48)$ &\\ \cline{2-3}
\multicolumn{1}{l|}{}&\multicolumn{1}{l|}{\multirow{2}{*}{\eqref{eq:gA_FV_mono_2}}} &\eqref{eq:deltaga2}& &$1.222(15)$ &$0.77(24)$ &$-4.87(33)$ & & & & & &$1.1(1.7)$ & &$0.012(51)$ & & &\\ \cline{3-3}
\multicolumn{1}{l|}{}&\multicolumn{1}{l|}{}&\eqref{eq:deltaga4}& &$1.222(16)$ &$0.77(24)$ &$-4.86(34)$ & & & & & &$1.1(1.7)$ & &$0.022(88)$ &$-0.02(24)$ &$-0.05(48)$ &\\ \cline{2-3}
\multicolumn{1}{l|}{}&\multicolumn{1}{l|}{\multirow{2}{*}{\eqref{eq:gA_FV_mono_3}}} &\eqref{eq:deltaga2}& &$1.221(15)$ &$0.76(24)$ &$-4.85(33)$ & & & & & &$-0.98(75)$ & &$0.033(55)$ & & &\\ \cline{3-3}
\multicolumn{1}{l|}{}&\multicolumn{1}{l|}{}&\eqref{eq:deltaga4}& &$1.221(16)$ &$0.76(24)$ &$-4.85(34)$ & & & & & &$-0.98(76)$ & &$0.034(89)$ &$0.007(244)$ &$-0.03(48)$ &\\ \cline{2-3}
\multicolumn{1}{l|}{}&\multicolumn{1}{l|}{\multirow{2}{*}{\eqref{eq:gA_FV_mono_4}}} &\eqref{eq:deltaga2}& &$1.222(15)$ &$0.77(24)$ &$-4.87(33)$ & & & & & &$-1.2(1.7)$ & &$0.012(51)$ & & &\\ \cline{3-3}
\multicolumn{1}{l|}{}&\multicolumn{1}{l|}{}&\eqref{eq:deltaga4}& &$1.222(16)$ &$0.77(24)$ &$-4.86(34)$ & & & & & &$-1.1(1.7)$ & &$0.022(88)$ &$-0.02(24)$ &$-0.05(48)$ &\\ \cline{2-3}
\hline
\multicolumn{1}{l|}{\multirow{16}{*}{\eqref{eq:gA_epi_epi2}}} &\multicolumn{1}{l|}{\multirow{2}{*}{\eqref{eq:gA_FV_NNLO}}} &\eqref{eq:deltaga2}& &$0.79(38)$ &$0.50(95)$ & &$1.302(22)$ &$-0.24(17)$ &$-0.20(38)$ & & & & &$0.009(52)$ & & &\\ \cline{3-3}
\multicolumn{1}{l|}{}&\multicolumn{1}{l|}{}&\eqref{eq:deltaga4}& &$0.80(38)$ &$0.49(95)$ & &$1.296(31)$ &$-0.22(18)$ &$-0.17(39)$ & & & & &$0.04(14)$ &$-0.007(243)$ &$-0.4(1.7)$ &\\ \cline{2-3}
\multicolumn{1}{l|}{}&\multicolumn{1}{l|}{\multirow{2}{*}{\eqref{eq:gA_FV_f2}}} &\eqref{eq:deltaga2}& &$1.27(46)$ &$0.30(95)$ & &$1.301(22)$ &$-0.24(17)$ &$-0.21(38)$ & & &$-0.51(61)$ & &$0.024(53)$ & & &\\ \cline{3-3}
\multicolumn{1}{l|}{}&\multicolumn{1}{l|}{}&\eqref{eq:deltaga4}& &$1.27(46)$ &$0.30(95)$ & &$1.297(31)$ &$-0.22(18)$ &$-0.19(40)$ & & &$-0.50(61)$ & &$0.04(14)$ &$0.01(24)$ &$-0.3(1.7)$ &\\ \cline{2-3}
\multicolumn{1}{l|}{}&\multicolumn{1}{l|}{\multirow{2}{*}{\eqref{eq:gA_FV_f3}}} &\eqref{eq:deltaga2}& &$0.75(39)$ &$0.23(99)$ & &$1.303(22)$ &$-0.24(17)$ &$-0.19(38)$ & & & &$0.41(97)$ &$0.003(51)$ & & &\\ \cline{3-3}
\multicolumn{1}{l|}{}&\multicolumn{1}{l|}{}&\eqref{eq:deltaga4}& &$0.76(39)$ &$0.22(99)$ & &$1.296(31)$ &$-0.23(18)$ &$-0.16(39)$ & & & &$0.40(97)$ &$0.04(14)$ &$-0.01(24)$ &$-0.5(1.7)$ &\\ \cline{2-3}
\multicolumn{1}{l|}{}&\multicolumn{1}{l|}{\multirow{2}{*}{\eqref{eq:gA_FV_f2_f3}}} &\eqref{eq:deltaga2}& &$1.37(44)$ &$-0.04(1.00)$ & &$1.301(22)$ &$-0.24(17)$ &$-0.18(38)$ & & &$-0.40(59)$ &$-0.14(98)$ &$0.028(54)$ & & &\\ \cline{3-3}
\multicolumn{1}{l|}{}&\multicolumn{1}{l|}{}&\eqref{eq:deltaga4}& &$1.37(44)$ &$-0.04(1.00)$ & &$1.299(31)$ &$-0.23(18)$ &$-0.18(40)$ & & &$-0.40(59)$ &$-0.14(98)$ &$0.03(14)$ &$0.02(24)$ &$-0.1(1.7)$ &\\ \cline{2-3}
\multicolumn{1}{l|}{}&\multicolumn{1}{l|}{\multirow{2}{*}{\eqref{eq:gA_FV_mono_1}}} &\eqref{eq:deltaga2}& & & & &$1.302(22)$ &$-0.24(17)$ &$-0.19(38)$ & & &$-1.2(1.4)$ & &$0.016(51)$ & & &\\ \cline{3-3}
\multicolumn{1}{l|}{}&\multicolumn{1}{l|}{}&\eqref{eq:deltaga4}& & & & &$1.302(23)$ &$-0.24(17)$ &$-0.19(39)$ & & &$-1.2(1.4)$ & &$0.018(89)$ &$0.001(242)$ &$-0.03(48)$ &\\ \cline{2-3}
\multicolumn{1}{l|}{}&\multicolumn{1}{l|}{\multirow{2}{*}{\eqref{eq:gA_FV_mono_2}}} &\eqref{eq:deltaga2}& & & & &$1.303(22)$ &$-0.24(17)$ &$-0.19(38)$ & & &$1.0(1.7)$ & &$0.007(50)$ & & &\\ \cline{3-3}
\multicolumn{1}{l|}{}&\multicolumn{1}{l|}{}&\eqref{eq:deltaga4}& & & & &$1.302(23)$ &$-0.24(17)$ &$-0.19(39)$ & & &$1.0(1.7)$ & &$0.013(88)$ &$-0.01(24)$ &$-0.03(48)$ &\\ \cline{2-3}
\multicolumn{1}{l|}{}&\multicolumn{1}{l|}{\multirow{2}{*}{\eqref{eq:gA_FV_mono_3}}} &\eqref{eq:deltaga2}& & & & &$1.301(22)$ &$-0.24(17)$ &$-0.19(38)$ & & &$-0.85(75)$ & &$0.026(53)$ & & &\\ \cline{3-3}
\multicolumn{1}{l|}{}&\multicolumn{1}{l|}{}&\eqref{eq:deltaga4}& & & & &$1.301(23)$ &$-0.24(17)$ &$-0.19(39)$ & & &$-0.86(76)$ & &$0.023(89)$ &$0.02(24)$ &$-0.02(48)$ &\\ \cline{2-3}
\multicolumn{1}{l|}{}&\multicolumn{1}{l|}{\multirow{2}{*}{\eqref{eq:gA_FV_mono_4}}} &\eqref{eq:deltaga2}& & & & &$1.303(22)$ &$-0.24(17)$ &$-0.19(38)$ & & &$-1.0(1.7)$ & &$0.007(50)$ & & &\\ \cline{3-3}
\multicolumn{1}{l|}{}&\multicolumn{1}{l|}{}&\eqref{eq:deltaga4}& & & & &$1.302(23)$ &$-0.24(17)$ &$-0.18(39)$ & & &$-1.0(1.7)$ & &$0.013(88)$ &$-0.01(24)$ &$-0.03(48)$ &\\ \cline{2-3}
\hline
\multicolumn{1}{l|}{\multirow{16}{*}{\eqref{eq:gA_epi2_epi3}}} &\multicolumn{1}{l|}{\multirow{2}{*}{\eqref{eq:gA_FV_NNLO}}} &\eqref{eq:deltaga2}& &$0.79(38)$ &$0.52(95)$ & &$1.276(10)$ & &$-0.76(32)$ &$0.06(88)$ & & & &$0.019(52)$ & & &\\ \cline{3-3}
\multicolumn{1}{l|}{}&\multicolumn{1}{l|}{}&\eqref{eq:deltaga4}& &$0.80(38)$ &$0.51(95)$ & &$1.272(19)$ & &$-0.70(39)$ &$0.08(88)$ & & & &$0.05(14)$ &$0.01(24)$ &$-0.5(1.7)$ &\\ \cline{2-3}
\multicolumn{1}{l|}{}&\multicolumn{1}{l|}{\multirow{2}{*}{\eqref{eq:gA_FV_f2}}} &\eqref{eq:deltaga2}& &$1.29(46)$ &$0.32(96)$ & &$1.275(10)$ & &$-0.75(32)$ &$0.05(88)$ & &$-0.51(59)$ & &$0.035(53)$ & & &\\ \cline{3-3}
\multicolumn{1}{l|}{}&\multicolumn{1}{l|}{}&\eqref{eq:deltaga4}& &$1.29(46)$ &$0.32(96)$ & &$1.273(19)$ & &$-0.72(39)$ &$0.06(88)$ & &$-0.50(59)$ & &$0.05(14)$ &$0.03(24)$ &$-0.3(1.7)$ &\\ \cline{2-3}
\multicolumn{1}{l|}{}&\multicolumn{1}{l|}{\multirow{2}{*}{\eqref{eq:gA_FV_f3}}} &\eqref{eq:deltaga2}& &$0.75(39)$ &$0.27(99)$ & &$1.276(10)$ & &$-0.76(32)$ &$0.07(88)$ & & &$0.46(97)$ &$0.013(51)$ & & &\\ \cline{3-3}
\multicolumn{1}{l|}{}&\multicolumn{1}{l|}{}&\eqref{eq:deltaga4}& &$0.76(39)$ &$0.26(99)$ & &$1.272(19)$ & &$-0.70(39)$ &$0.09(88)$ & & &$0.44(97)$ &$0.05(14)$ &$0.004(242)$ &$-0.5(1.7)$ &\\ \cline{2-3}
\multicolumn{1}{l|}{}&\multicolumn{1}{l|}{\multirow{2}{*}{\eqref{eq:gA_FV_f2_f3}}} &\eqref{eq:deltaga2}& &$1.38(44)$ &$-0.03(1.00)$ & &$1.275(10)$ & &$-0.75(32)$ &$0.08(88)$ & &$-0.41(58)$ &$-0.09(98)$ &$0.038(54)$ & & &\\ \cline{3-3}
\multicolumn{1}{l|}{}&\multicolumn{1}{l|}{}&\eqref{eq:deltaga4}& &$1.38(44)$ &$-0.02(1.00)$ & &$1.274(19)$ & &$-0.72(39)$ &$0.09(88)$ & &$-0.41(59)$ &$-0.08(98)$ &$0.04(14)$ &$0.04(24)$ &$-0.2(1.7)$ &\\ \cline{2-3}
\multicolumn{1}{l|}{}&\multicolumn{1}{l|}{\multirow{2}{*}{\eqref{eq:gA_FV_mono_1}}} &\eqref{eq:deltaga2}& & & & &$1.276(10)$ & &$-0.76(32)$ &$0.08(88)$ & &$-1.3(1.4)$ & &$0.027(51)$ & & &\\ \cline{3-3}
\multicolumn{1}{l|}{}&\multicolumn{1}{l|}{}&\eqref{eq:deltaga4}& & & & &$1.276(12)$ & &$-0.75(33)$ &$0.08(88)$ & &$-1.3(1.4)$ & &$0.023(88)$ &$0.02(24)$ &$-0.03(48)$ &\\ \cline{2-3}
\multicolumn{1}{l|}{}&\multicolumn{1}{l|}{\multirow{2}{*}{\eqref{eq:gA_FV_mono_2}}} &\eqref{eq:deltaga2}& & & & &$1.276(10)$ & &$-0.76(32)$ &$0.08(88)$ & &$1.1(1.7)$ & &$0.018(50)$ & & &\\ \cline{3-3}
\multicolumn{1}{l|}{}&\multicolumn{1}{l|}{}&\eqref{eq:deltaga4}& & & & &$1.276(12)$ & &$-0.76(33)$ &$0.08(88)$ & &$1.1(1.7)$ & &$0.018(88)$ &$0.006(241)$ &$-0.04(48)$ &\\ \cline{2-3}
\multicolumn{1}{l|}{}&\multicolumn{1}{l|}{\multirow{2}{*}{\eqref{eq:gA_FV_mono_3}}} &\eqref{eq:deltaga2}& & & & &$1.275(10)$ & &$-0.75(32)$ &$0.08(88)$ & &$-0.89(75)$ & &$0.037(53)$ & & &\\ \cline{3-3}
\multicolumn{1}{l|}{}&\multicolumn{1}{l|}{}&\eqref{eq:deltaga4}& & & & &$1.275(12)$ & &$-0.75(33)$ &$0.07(88)$ & &$-0.90(76)$ & &$0.029(89)$ &$0.03(24)$ &$-0.02(48)$ &\\ \cline{2-3}
\multicolumn{1}{l|}{}&\multicolumn{1}{l|}{\multirow{2}{*}{\eqref{eq:gA_FV_mono_4}}} &\eqref{eq:deltaga2}& & & & &$1.276(10)$ & &$-0.76(32)$ &$0.08(88)$ & &$-1.1(1.7)$ & &$0.018(50)$ & & &\\ \cline{3-3}
\multicolumn{1}{l|}{}&\multicolumn{1}{l|}{}&\eqref{eq:deltaga4}& & & & &$1.276(12)$ & &$-0.76(33)$ &$0.08(88)$ & &$-1.1(1.7)$ & &$0.019(88)$ &$0.007(241)$ &$-0.04(48)$ &\\ \cline{2-3}
\hline
\multicolumn{1}{l|}{\multirow{16}{*}{\eqref{eq:g_epi2_epi4}}} &\multicolumn{1}{l|}{\multirow{2}{*}{\eqref{eq:gA_FV_NNLO}}} &\eqref{eq:deltaga2}& &$0.79(38)$ &$0.52(95)$ & &$1.2758(91)$ & &$-0.75(15)$ & &$0.13(96)$ & & &$0.019(51)$ & & &\\ \cline{3-3}
\multicolumn{1}{l|}{}&\multicolumn{1}{l|}{}&\eqref{eq:deltaga4}& &$0.80(38)$ &$0.51(95)$ & &$1.272(18)$ & &$-0.68(28)$ & &$0.14(97)$ & & &$0.05(14)$ &$0.01(24)$ &$-0.5(1.7)$ &\\ \cline{2-3}
\multicolumn{1}{l|}{}&\multicolumn{1}{l|}{\multirow{2}{*}{\eqref{eq:gA_FV_f2}}} &\eqref{eq:deltaga2}& &$1.29(46)$ &$0.32(96)$ & &$1.2756(92)$ & &$-0.75(15)$ & &$0.13(96)$ &$-0.51(59)$ & &$0.035(53)$ & & &\\ \cline{3-3}
\multicolumn{1}{l|}{}&\multicolumn{1}{l|}{}&\eqref{eq:deltaga4}& &$1.29(46)$ &$0.32(96)$ & &$1.273(18)$ & &$-0.71(28)$ & &$0.13(97)$ &$-0.50(59)$ & &$0.05(14)$ &$0.03(24)$ &$-0.3(1.7)$ &\\ \cline{2-3}
\multicolumn{1}{l|}{}&\multicolumn{1}{l|}{\multirow{2}{*}{\eqref{eq:gA_FV_f3}}} &\eqref{eq:deltaga2}& &$0.75(39)$ &$0.27(99)$ & &$1.2764(90)$ & &$-0.75(15)$ & &$0.14(96)$ & &$0.46(97)$ &$0.013(51)$ & & &\\ \cline{3-3}
\multicolumn{1}{l|}{}&\multicolumn{1}{l|}{}&\eqref{eq:deltaga4}& &$0.76(39)$ &$0.26(99)$ & &$1.272(18)$ & &$-0.68(28)$ & &$0.15(97)$ & &$0.44(97)$ &$0.05(14)$ &$0.004(242)$ &$-0.5(1.7)$ &\\ \cline{2-3}
\multicolumn{1}{l|}{}&\multicolumn{1}{l|}{\multirow{2}{*}{\eqref{eq:gA_FV_f2_f3}}} &\eqref{eq:deltaga2}& &$1.38(44)$ &$-0.03(1.00)$ & &$1.2748(91)$ & &$-0.73(15)$ & &$0.15(96)$ &$-0.41(58)$ &$-0.09(98)$ &$0.038(53)$ & & &\\ \cline{3-3}
\multicolumn{1}{l|}{}&\multicolumn{1}{l|}{}&\eqref{eq:deltaga4}& &$1.38(44)$ &$-0.02(1.00)$ & &$1.274(18)$ & &$-0.71(28)$ & &$0.15(97)$ &$-0.41(59)$ &$-0.08(98)$ &$0.04(14)$ &$0.04(24)$ &$-0.2(1.7)$ &\\ \cline{2-3}
\multicolumn{1}{l|}{}&\multicolumn{1}{l|}{\multirow{2}{*}{\eqref{eq:gA_FV_mono_1}}} &\eqref{eq:deltaga2}& & & & &$1.2756(90)$ & &$-0.74(15)$ & &$0.14(96)$ &$-1.3(1.4)$ & &$0.027(51)$ & & &\\ \cline{3-3}
\multicolumn{1}{l|}{}&\multicolumn{1}{l|}{}&\eqref{eq:deltaga4}& & & & &$1.276(11)$ & &$-0.74(16)$ & &$0.14(96)$ &$-1.3(1.4)$ & &$0.023(88)$ &$0.02(24)$ &$-0.03(48)$ &\\ \cline{2-3}
\multicolumn{1}{l|}{}&\multicolumn{1}{l|}{\multirow{2}{*}{\eqref{eq:gA_FV_mono_2}}} &\eqref{eq:deltaga2}& & & & &$1.2762(90)$ & &$-0.75(15)$ & &$0.14(96)$ &$1.1(1.7)$ & &$0.018(50)$ & & &\\ \cline{3-3}
\multicolumn{1}{l|}{}&\multicolumn{1}{l|}{}&\eqref{eq:deltaga4}& & & & &$1.276(11)$ & &$-0.75(16)$ & &$0.14(96)$ &$1.1(1.7)$ & &$0.018(88)$ &$0.006(241)$ &$-0.04(48)$ &\\ \cline{2-3}
\multicolumn{1}{l|}{}&\multicolumn{1}{l|}{\multirow{2}{*}{\eqref{eq:gA_FV_mono_3}}} &\eqref{eq:deltaga2}& & & & &$1.2750(90)$ & &$-0.74(15)$ & &$0.14(96)$ &$-0.89(75)$ & &$0.037(53)$ & & &\\ \cline{3-3}
\multicolumn{1}{l|}{}&\multicolumn{1}{l|}{}&\eqref{eq:deltaga4}& & & & &$1.275(11)$ & &$-0.74(16)$ & &$0.14(96)$ &$-0.90(76)$ & &$0.029(89)$ &$0.03(24)$ &$-0.02(48)$ &\\ \cline{2-3}
\multicolumn{1}{l|}{}&\multicolumn{1}{l|}{\multirow{2}{*}{\eqref{eq:gA_FV_mono_4}}} &\eqref{eq:deltaga2}& & & & &$1.2762(90)$ & &$-0.75(15)$ & &$0.15(96)$ &$-1.1(1.7)$ & &$0.018(50)$ & & &\\ \cline{3-3}
\multicolumn{1}{l|}{}&\multicolumn{1}{l|}{}&\eqref{eq:deltaga4}& & & & &$1.276(11)$ & &$-0.75(16)$ & &$0.15(96)$ &$-1.1(1.7)$ & &$0.019(88)$ &$0.007(241)$ &$-0.04(48)$ &\\ \cline{2-3}
\hline
\hline
\end{tabular}
\end{sideways}
\end{table}

\end{widetext}

\bibliography{gA}

\end{document}

%% file: preamble.tex
\usepackage[export]{adjustbox} 
\usepackage{dcolumn}   
\usepackage{bm}        
\usepackage{amssymb}   
\usepackage{standalone}
\usepackage{enumitem}
\usepackage[pdftex]{color}
\usepackage{xcolor}
\usepackage{slashed}
\usepackage{booktabs}
\usepackage{multirow}
\usepackage{amsmath}
\usepackage{stackrel}
\usepackage{rotating}
\usepackage{CJKutf8}
\usepackage{pifont}
\usepackage{mathtools}
\usepackage{hyperref}
\hypersetup{
    colorlinks=true,       
    linkcolor=blue,          
    citecolor=blue,        
    filecolor=blue,      
    urlcolor=blue           
}
\usepackage{simplewick}
\usepackage{braket}

\usepackage{rotating}
\usepackage[caption=false]{subfig}

\allowdisplaybreaks

\AtBeginDocument{%
    \newwrite\bibnotes
    \def\bibnotesext{Notes.bib}
    \immediate\openout\bibnotes=\jobname\bibnotesext
    \immediate\write\bibnotes{@CONTROL{REVTEX41Control}}
    \immediate\write\bibnotes{@CONTROL{%
    apsrev41Control,author="08",editor="1",pages="1",title="0",year="1"}}
     \if@filesw
     \immediate\write\@auxout{\string\citation{apsrev41Control}}%
    \fi
}%

%% file: command_list.tex
\newcommand{\asmadd}[1]{{\color{cyan} #1}}

\newcommand{\eqnref}[1]{Eq.~(\ref{#1})}
\newcommand{\secref}[1]{Sec.~\ref{#1}}
\newcommand{\tabref}[1]{Table~\ref{#1}}
\newcommand{\figref}[1]{Fig.~\ref{#1}}
\def\beq{\begin{equation}}
\def\eeq{\end{equation}}

\newcommand{\nxlo}[1]{N${}^{#1}$LO}

\definecolor{kngrey}{HTML}{A6AAA9}
\definecolor{knred}{HTML}{EC5D57}
\definecolor{knorange}{HTML}{F39019}
\definecolor{knyellow}{HTML}{F5D328}
\definecolor{kngreen}{HTML}{70BF41}
\definecolor{knblue}{HTML}{51A7F9}
\definecolor{knpurple}{HTML}{B36AE2}

\newcommand{\Order}{{\mathcal O}}

\def\d{{\delta}}
\def\D{{\Delta}}
\def\t{\tau}
\def\e{{\varepsilon}}

\def\L{{\Lambda}}

\def\s{{\sigma}}

\def\ln{{\textrm{ln}}}

\definecolor{gray}{rgb}{0.6,0.6,0.6}
\newcommand{\old}[1]{{\color{gray}{#1}}}

%% file: def.tex
\def\d{{\delta}}
\def\D{{\Delta}}
\def\e{{\epsilon}}

\def\L{{\Lambda}}

\def\s{{\sigma}}
\def\t{{\tau}}



%% file: affiliations.tex
\newcommand{\bern}{
    Albert Einstein Center for Fundamental Physics, 
    Institute for Theoretical Physics, 
    University of Bern, Sidlerstrasse 5, 3012 Bern, Switzerland
}
\newcommand{\cmu}{
    Department of Physics, Carnegie Mellon University,
    Pittsburgh, Pennsylvania 15213, USA
}
\newcommand{\lanl}{
    Theoretical Division, 
    Los Alamos National Laboratory, 
    Los Alamos, NM 87545, USA
}
\newcommand{\lblnsd}{
    Nuclear Science Division,
    Lawrence Berkeley National Laboratory,
	Berkeley, CA 94720, USA
	}

\newcommand{\llnl}{
	Lawrence Livermore National Laboratory,
	Livermore, CA 94550, USA
	}

\newcommand{\nvidia}{
    NVIDIA Corporation,
    2701 San Tomas Expressway, Santa Clara, CA 95050, USA
    }
\newcommand{\ornl}{
    National Center For Computational Sciences 
    Oak Ridge National Laboratory, 
    Oak Ridge, Tennessee, USA
}
\newcommand{\ucb}{
	Department of Physics,
	University of California,
	Berkeley, CA 94720, USA
	}
\newcommand{\unc}{
	Department of Physics and Astronomy,
	University of North Carolina,
	Chapel Hill, NC 27516-3255, USA
	}

%% file: gA_fv_mdwf_hisq.bbl
\begin{thebibliography}{151}%
\makeatletter
\providecommand \@ifxundefined [1]{%
 \@ifx{#1\undefined}
}%
\providecommand \@ifnum [1]{%
 \ifnum #1\expandafter \@firstoftwo
 \else \expandafter \@secondoftwo
 \fi
}%
\providecommand \@ifx [1]{%
 \ifx #1\expandafter \@firstoftwo
 \else \expandafter \@secondoftwo
 \fi
}%
\providecommand \natexlab [1]{#1}%
\providecommand \enquote  [1]{``#1''}%
\providecommand \bibnamefont  [1]{#1}%
\providecommand \bibfnamefont [1]{#1}%
\providecommand \citenamefont [1]{#1}%
\providecommand \href@noop [0]{\@secondoftwo}%
\providecommand \href [0]{\begingroup \@sanitize@url \@href}%
\providecommand \@href[1]{\@@startlink{#1}\@@href}%
\providecommand \@@href[1]{\endgroup#1\@@endlink}%
\providecommand \@sanitize@url [0]{\catcode `\\12\catcode `\$12\catcode
  `\&12\catcode `\#12\catcode `\^12\catcode `\_12\catcode `\%12\relax}%
\providecommand \@@startlink[1]{}%
\providecommand \@@endlink[0]{}%
\providecommand \url  [0]{\begingroup\@sanitize@url \@url }%
\providecommand \@url [1]{\endgroup\@href {#1}{\urlprefix }}%
\providecommand \urlprefix  [0]{URL }%
\providecommand \Eprint [0]{\href }%
\providecommand \doibase [0]{http://dx.doi.org/}%
\providecommand \selectlanguage [0]{\@gobble}%
\providecommand \bibinfo  [0]{\@secondoftwo}%
\providecommand \bibfield  [0]{\@secondoftwo}%
\providecommand \translation [1]{[#1]}%
\providecommand \BibitemOpen [0]{}%
\providecommand \bibitemStop [0]{}%
\providecommand \bibitemNoStop [0]{.\EOS\space}%
\providecommand \EOS [0]{\spacefactor3000\relax}%
\providecommand \BibitemShut  [1]{\csname bibitem#1\endcsname}%
\let\auto@bib@innerbib\@empty
\bibitem [{\citenamefont {Fields}\ \emph {et~al.}(2020)\citenamefont {Fields},
  \citenamefont {Olive}, \citenamefont {Yeh},\ and\ \citenamefont
  {Young}}]{Fields:2019pfx}%
  \BibitemOpen
  \bibfield  {author} {\bibinfo {author} {\bibfnamefont {B.~D.}\ \bibnamefont
  {Fields}}, \bibinfo {author} {\bibfnamefont {K.~A.}\ \bibnamefont {Olive}},
  \bibinfo {author} {\bibfnamefont {T.-H.}\ \bibnamefont {Yeh}}, \ and\
  \bibinfo {author} {\bibfnamefont {C.}~\bibnamefont {Young}},\ }\bibfield
  {title} {\enquote {\bibinfo {title} {{Big-Bang Nucleosynthesis after
  Planck}},}\ }\href {\doibase 10.1088/1475-7516/2020/03/010} {\bibfield
  {journal} {\bibinfo  {journal} {JCAP}\ }\textbf {\bibinfo {volume} {03}},\
  \bibinfo {pages} {010} (\bibinfo {year} {2020})},\ \bibinfo {note} {[Erratum:
  JCAP {\bf 11}, E02 (2020)]},\ \Eprint {http://arxiv.org/abs/1912.01132}
  {arXiv:1912.01132 [astro-ph.CO]} \BibitemShut {NoStop}%
\bibitem [{\citenamefont {Acharya}\ \emph {et~al.}(2024)\citenamefont {Acharya}
  \emph {et~al.}}]{Acharya:2024lke}%
  \BibitemOpen
  \bibfield  {author} {\bibinfo {author} {\bibfnamefont {B.}~\bibnamefont
  {Acharya}} \emph {et~al.},\ }\bibfield  {title} {\enquote {\bibinfo {title}
  {{Solar fusion III: New data and theory for hydrogen-burning stars}},}\
  }\href@noop {} {\  (\bibinfo {year} {2024})},\ \Eprint
  {http://arxiv.org/abs/2405.06470} {arXiv:2405.06470 [astro-ph.SR]}
  \BibitemShut {NoStop}%
\bibitem [{\citenamefont {Hayen}(2024)}]{Hayen:2024xjf}%
  \BibitemOpen
  \bibfield  {author} {\bibinfo {author} {\bibfnamefont {L.}~\bibnamefont
  {Hayen}},\ }\bibfield  {title} {\enquote {\bibinfo {title} {{Opportunities
  and open questions in modern $\beta$ decay}},}\ }\href {\doibase
  10.1146/annurev-nucl-121423-100730} {\bibfield  {journal} {\bibinfo
  {journal} {Ann. Rev. Nucl. Part. Sci.}\ }\textbf {\bibinfo {volume} {74}},\
  \bibinfo {pages} {497--528} (\bibinfo {year} {2024})},\ \Eprint
  {http://arxiv.org/abs/2403.08485} {arXiv:2403.08485 [nucl-th]} \BibitemShut
  {NoStop}%
\bibitem [{\citenamefont {Bopp}\ \emph {et~al.}(1986)\citenamefont {Bopp},
  \citenamefont {Dubbers}, \citenamefont {Hornig}, \citenamefont {Klemt},
  \citenamefont {Last}, \citenamefont {Schutze}, \citenamefont {Freedman},\
  and\ \citenamefont {Scharpf}}]{Bopp:1986rt}%
  \BibitemOpen
  \bibfield  {author} {\bibinfo {author} {\bibfnamefont {P.}~\bibnamefont
  {Bopp}}, \bibinfo {author} {\bibfnamefont {D.}~\bibnamefont {Dubbers}},
  \bibinfo {author} {\bibfnamefont {L.}~\bibnamefont {Hornig}}, \bibinfo
  {author} {\bibfnamefont {E.}~\bibnamefont {Klemt}}, \bibinfo {author}
  {\bibfnamefont {J.}~\bibnamefont {Last}}, \bibinfo {author} {\bibfnamefont
  {H.}~\bibnamefont {Schutze}}, \bibinfo {author} {\bibfnamefont {S.~J.}\
  \bibnamefont {Freedman}}, \ and\ \bibinfo {author} {\bibfnamefont
  {O.}~\bibnamefont {Scharpf}},\ }\bibfield  {title} {\enquote {\bibinfo
  {title} {{The Beta Decay Asymmetry of the Neutron and $g_A/g_V$}},}\ }\href
  {\doibase 10.1103/PhysRevLett.56.919} {\bibfield  {journal} {\bibinfo
  {journal} {Phys. Rev. Lett.}\ }\textbf {\bibinfo {volume} {56}},\ \bibinfo
  {pages} {919} (\bibinfo {year} {1986})},\ \bibinfo {note} {[Erratum: Phys.
  Rev. Lett. {\bf 57}, 1192 (1986)]}\BibitemShut {NoStop}%
\bibitem [{\citenamefont {Erozolimsky}\ \emph {et~al.}(1991)\citenamefont
  {Erozolimsky}, \citenamefont {Kuznetsov}, \citenamefont {Kujda},
  \citenamefont {Mostovoi},\ and\ \citenamefont
  {Stepanenko}}]{Erozolimsky:1990ui}%
  \BibitemOpen
  \bibfield  {author} {\bibinfo {author} {\bibfnamefont {B.~G.}\ \bibnamefont
  {Erozolimsky}}, \bibinfo {author} {\bibfnamefont {I.~A.}\ \bibnamefont
  {Kuznetsov}}, \bibinfo {author} {\bibfnamefont {I.~A.}\ \bibnamefont
  {Kujda}}, \bibinfo {author} {\bibfnamefont {Y.~A.}\ \bibnamefont {Mostovoi}},
  \ and\ \bibinfo {author} {\bibfnamefont {I.~V.}\ \bibnamefont {Stepanenko}},\
  }\bibfield  {title} {\enquote {\bibinfo {title} {{New measurements of the
  electron - neutron spin asymmetry}},}\ }\href {\doibase
  10.1016/0370-2693(91)91703-X} {\bibfield  {journal} {\bibinfo  {journal}
  {Phys. Lett. B}\ }\textbf {\bibinfo {volume} {263}},\ \bibinfo {pages}
  {33--38} (\bibinfo {year} {1991})},\ \bibinfo {note} {[Erratum: Phys. Lett. B
  {\bf 412}, 240 (1997)]}\BibitemShut {NoStop}%
\bibitem [{\citenamefont {Liaud}\ \emph {et~al.}(1997)\citenamefont {Liaud},
  \citenamefont {Schreckenbach}, \citenamefont {Kossakowski}, \citenamefont
  {Nastoll}, \citenamefont {Bussiere}, \citenamefont {Guillaud},\ and\
  \citenamefont {Beck}}]{Liaud:1997vu}%
  \BibitemOpen
  \bibfield  {author} {\bibinfo {author} {\bibfnamefont {P.}~\bibnamefont
  {Liaud}}, \bibinfo {author} {\bibfnamefont {K.}~\bibnamefont
  {Schreckenbach}}, \bibinfo {author} {\bibfnamefont {R.}~\bibnamefont
  {Kossakowski}}, \bibinfo {author} {\bibfnamefont {H.}~\bibnamefont
  {Nastoll}}, \bibinfo {author} {\bibfnamefont {A.}~\bibnamefont {Bussiere}},
  \bibinfo {author} {\bibfnamefont {J.~P.}\ \bibnamefont {Guillaud}}, \ and\
  \bibinfo {author} {\bibfnamefont {L.}~\bibnamefont {Beck}},\ }\bibfield
  {title} {\enquote {\bibinfo {title} {{The measurement of the beta asymmetry
  in the decay of polarized neutrons}},}\ }\href {\doibase
  10.1016/S0375-9474(96)00325-9} {\bibfield  {journal} {\bibinfo  {journal}
  {Nucl. Phys. A}\ }\textbf {\bibinfo {volume} {612}},\ \bibinfo {pages}
  {53--81} (\bibinfo {year} {1997})}\BibitemShut {NoStop}%
\bibitem [{\citenamefont {Mostovoi}\ \emph {et~al.}(2001)\citenamefont
  {Mostovoi} \emph {et~al.}}]{Mostovoi:2001ye}%
  \BibitemOpen
  \bibfield  {author} {\bibinfo {author} {\bibfnamefont {Y.~A.}\ \bibnamefont
  {Mostovoi}} \emph {et~al.},\ }\bibfield  {title} {\enquote {\bibinfo {title}
  {{Experimental value of $G_A/G_V$ from a measurement of both $P$-odd
  correlations in free-neutron decay}},}\ }\href {\doibase 10.1134/1.1423745}
  {\bibfield  {journal} {\bibinfo  {journal} {Phys. Atom. Nucl.}\ }\textbf
  {\bibinfo {volume} {64}},\ \bibinfo {pages} {1955--1960} (\bibinfo {year}
  {2001})}\BibitemShut {NoStop}%
\bibitem [{\citenamefont {Schumann}\ \emph {et~al.}(2008)\citenamefont
  {Schumann}, \citenamefont {Kreuz}, \citenamefont {Deissenroth}, \citenamefont
  {Gl{\"u}ck}, \citenamefont {Krempel}, \citenamefont {M{\"a}rkisch},
  \citenamefont {Mund}, \citenamefont {Petoukhov}, \citenamefont {Soldner},\
  and\ \citenamefont {Abele}}]{Schumann:2007hz}%
  \BibitemOpen
  \bibfield  {author} {\bibinfo {author} {\bibfnamefont {M.}~\bibnamefont
  {Schumann}}, \bibinfo {author} {\bibfnamefont {M.}~\bibnamefont {Kreuz}},
  \bibinfo {author} {\bibfnamefont {M.}~\bibnamefont {Deissenroth}}, \bibinfo
  {author} {\bibfnamefont {F.}~\bibnamefont {Gl{\"u}ck}}, \bibinfo {author}
  {\bibfnamefont {J.}~\bibnamefont {Krempel}}, \bibinfo {author} {\bibfnamefont
  {B.}~\bibnamefont {M{\"a}rkisch}}, \bibinfo {author} {\bibfnamefont
  {D.}~\bibnamefont {Mund}}, \bibinfo {author} {\bibfnamefont {A.}~\bibnamefont
  {Petoukhov}}, \bibinfo {author} {\bibfnamefont {T.}~\bibnamefont {Soldner}},
  \ and\ \bibinfo {author} {\bibfnamefont {H.}~\bibnamefont {Abele}},\
  }\bibfield  {title} {\enquote {\bibinfo {title} {{Measurement of the Proton
  Asymmetry Parameter $C$ in Neutron Beta Decay}},}\ }\href {\doibase
  10.1103/PhysRevLett.100.151801} {\bibfield  {journal} {\bibinfo  {journal}
  {Phys. Rev. Lett.}\ }\textbf {\bibinfo {volume} {100}},\ \bibinfo {pages}
  {151801} (\bibinfo {year} {2008})},\ \Eprint {http://arxiv.org/abs/0712.2442}
  {arXiv:0712.2442 [hep-ph]} \BibitemShut {NoStop}%
\bibitem [{\citenamefont {Mund}\ \emph {et~al.}(2013)\citenamefont {Mund},
  \citenamefont {M{\"a}rkisch}, \citenamefont {Deissenroth}, \citenamefont
  {Krempel}, \citenamefont {Schumann}, \citenamefont {Abele}, \citenamefont
  {Petoukhov},\ and\ \citenamefont {Soldner}}]{Mund:2012fq}%
  \BibitemOpen
  \bibfield  {author} {\bibinfo {author} {\bibfnamefont {D.}~\bibnamefont
  {Mund}}, \bibinfo {author} {\bibfnamefont {B.}~\bibnamefont {M{\"a}rkisch}},
  \bibinfo {author} {\bibfnamefont {M.}~\bibnamefont {Deissenroth}}, \bibinfo
  {author} {\bibfnamefont {J.}~\bibnamefont {Krempel}}, \bibinfo {author}
  {\bibfnamefont {M.}~\bibnamefont {Schumann}}, \bibinfo {author}
  {\bibfnamefont {H.}~\bibnamefont {Abele}}, \bibinfo {author} {\bibfnamefont
  {A.}~\bibnamefont {Petoukhov}}, \ and\ \bibinfo {author} {\bibfnamefont
  {T.}~\bibnamefont {Soldner}},\ }\bibfield  {title} {\enquote {\bibinfo
  {title} {{Determination of the Weak Axial Vector Coupling $\lambda=g_A/g_V$
  from a Measurement of the $\beta$-Asymmetry Parameter $A$ in Neutron Beta
  Decay}},}\ }\href {\doibase 10.1103/PhysRevLett.110.172502} {\bibfield
  {journal} {\bibinfo  {journal} {Phys. Rev. Lett.}\ }\textbf {\bibinfo
  {volume} {110}},\ \bibinfo {pages} {172502} (\bibinfo {year} {2013})},\
  \Eprint {http://arxiv.org/abs/1204.0013} {arXiv:1204.0013 [hep-ex]}
  \BibitemShut {NoStop}%
\bibitem [{\citenamefont {Brown}\ \emph {et~al.}(2018)\citenamefont {Brown}
  \emph {et~al.}}]{UCNA:2017obv}%
  \BibitemOpen
  \bibfield  {author} {\bibinfo {author} {\bibfnamefont {M.~A.~P.}\
  \bibnamefont {Brown}} \emph {et~al.} (\bibinfo {collaboration} {UCNA}),\
  }\bibfield  {title} {\enquote {\bibinfo {title} {{New result for the neutron
  $\beta$-asymmetry parameter $A_0$ from UCNA}},}\ }\href {\doibase
  10.1103/PhysRevC.97.035505} {\bibfield  {journal} {\bibinfo  {journal} {Phys.
  Rev. C}\ }\textbf {\bibinfo {volume} {97}},\ \bibinfo {pages} {035505}
  (\bibinfo {year} {2018})},\ \Eprint {http://arxiv.org/abs/1712.00884}
  {arXiv:1712.00884 [nucl-ex]} \BibitemShut {NoStop}%
\bibitem [{\citenamefont {M\"arkisch}\ \emph {et~al.}(2019)\citenamefont
  {M\"arkisch} \emph {et~al.}}]{Markisch:2018ndu}%
  \BibitemOpen
  \bibfield  {author} {\bibinfo {author} {\bibfnamefont {B.}~\bibnamefont
  {M\"arkisch}} \emph {et~al.},\ }\bibfield  {title} {\enquote {\bibinfo
  {title} {{Measurement of the Weak Axial-Vector Coupling Constant in the Decay
  of Free Neutrons Using a Pulsed Cold Neutron Beam}},}\ }\href {\doibase
  10.1103/PhysRevLett.122.242501} {\bibfield  {journal} {\bibinfo  {journal}
  {Phys. Rev. Lett.}\ }\textbf {\bibinfo {volume} {122}},\ \bibinfo {pages}
  {242501} (\bibinfo {year} {2019})},\ \Eprint
  {http://arxiv.org/abs/1812.04666} {arXiv:1812.04666 [nucl-ex]} \BibitemShut
  {NoStop}%
\bibitem [{\citenamefont {Beck}\ \emph {et~al.}(2020)\citenamefont {Beck} \emph
  {et~al.}}]{Beck:2019xye}%
  \BibitemOpen
  \bibfield  {author} {\bibinfo {author} {\bibfnamefont {M.}~\bibnamefont
  {Beck}} \emph {et~al.},\ }\bibfield  {title} {\enquote {\bibinfo {title}
  {{Improved determination of the $\beta$-$\overline{\nu}_e$ angular
  correlation coefficient $a$ in free neutron decay with the $a$SPECT
  spectrometer}},}\ }\href {\doibase 10.1103/PhysRevC.101.055506} {\bibfield
  {journal} {\bibinfo  {journal} {Phys. Rev. C}\ }\textbf {\bibinfo {volume}
  {101}},\ \bibinfo {pages} {055506} (\bibinfo {year} {2020})},\ \Eprint
  {http://arxiv.org/abs/1908.04785} {arXiv:1908.04785 [nucl-ex]} \BibitemShut
  {NoStop}%
\bibitem [{\citenamefont {Hassan}\ \emph {et~al.}(2021)\citenamefont {Hassan}
  \emph {et~al.}}]{Hassan:2020hrj}%
  \BibitemOpen
  \bibfield  {author} {\bibinfo {author} {\bibfnamefont {M.~T.}\ \bibnamefont
  {Hassan}} \emph {et~al.},\ }\bibfield  {title} {\enquote {\bibinfo {title}
  {{Measurement of the neutron decay electron-antineutrino angular correlation
  by the aCORN experiment}},}\ }\href {\doibase 10.1103/PhysRevC.103.045502}
  {\bibfield  {journal} {\bibinfo  {journal} {Phys. Rev. C}\ }\textbf {\bibinfo
  {volume} {103}},\ \bibinfo {pages} {045502} (\bibinfo {year} {2021})},\
  \Eprint {http://arxiv.org/abs/2012.14379} {arXiv:2012.14379 [nucl-ex]}
  \BibitemShut {NoStop}%
\bibitem [{\citenamefont {Navas}\ \emph {et~al.}(2024)\citenamefont {Navas}
  \emph {et~al.}}]{ParticleDataGroup:2024cfk}%
  \BibitemOpen
  \bibfield  {author} {\bibinfo {author} {\bibfnamefont {S.}~\bibnamefont
  {Navas}} \emph {et~al.} (\bibinfo {collaboration} {Particle Data Group}),\
  }\bibfield  {title} {\enquote {\bibinfo {title} {{Review of particle
  physics}},}\ }\href {\doibase 10.1103/PhysRevD.110.030001} {\bibfield
  {journal} {\bibinfo  {journal} {Phys. Rev. D}\ }\textbf {\bibinfo {volume}
  {110}},\ \bibinfo {pages} {030001} (\bibinfo {year} {2024})}\BibitemShut
  {NoStop}%
\bibitem [{\citenamefont {Kossakowski}\ \emph {et~al.}(1989)\citenamefont
  {Kossakowski}, \citenamefont {Liaud}, \citenamefont {Azuelos}, \citenamefont
  {Grivot},\ and\ \citenamefont {Schreckenbach}}]{Kossakowski:1989dc}%
  \BibitemOpen
  \bibfield  {author} {\bibinfo {author} {\bibfnamefont {R.}~\bibnamefont
  {Kossakowski}}, \bibinfo {author} {\bibfnamefont {P.}~\bibnamefont {Liaud}},
  \bibinfo {author} {\bibfnamefont {G.}~\bibnamefont {Azuelos}}, \bibinfo
  {author} {\bibfnamefont {P.}~\bibnamefont {Grivot}}, \ and\ \bibinfo {author}
  {\bibfnamefont {K.}~\bibnamefont {Schreckenbach}},\ }\bibfield  {title}
  {\enquote {\bibinfo {title} {{Neutron Lifetime Measurement With a Helium
  Filled Time Projection Chamber}},}\ }\href {\doibase
  10.1016/0375-9474(89)90246-7} {\bibfield  {journal} {\bibinfo  {journal}
  {Nucl. Phys. A}\ }\textbf {\bibinfo {volume} {503}},\ \bibinfo {pages}
  {473--500} (\bibinfo {year} {1989})}\BibitemShut {NoStop}%
\bibitem [{\citenamefont {Byrne}\ and\ \citenamefont
  {Dawber}(1996)}]{Byrne:1996zz}%
  \BibitemOpen
  \bibfield  {author} {\bibinfo {author} {\bibfnamefont {J.}~\bibnamefont
  {Byrne}}\ and\ \bibinfo {author} {\bibfnamefont {P.~G.}\ \bibnamefont
  {Dawber}},\ }\bibfield  {title} {\enquote {\bibinfo {title} {{A Revised Value
  for the Neutron Lifetime Measured Using a Penning Trap}},}\ }\href {\doibase
  10.1209/epl/i1996-00319-x} {\bibfield  {journal} {\bibinfo  {journal} {EPL}\
  }\textbf {\bibinfo {volume} {33}},\ \bibinfo {pages} {187} (\bibinfo {year}
  {1996})}\BibitemShut {NoStop}%
\bibitem [{\citenamefont {Nico}\ \emph {et~al.}(2005)\citenamefont {Nico} \emph
  {et~al.}}]{Nico:2004ie}%
  \BibitemOpen
  \bibfield  {author} {\bibinfo {author} {\bibfnamefont {J.~S.}\ \bibnamefont
  {Nico}} \emph {et~al.},\ }\bibfield  {title} {\enquote {\bibinfo {title}
  {{Measurement of the neutron lifetime by counting trapped protons in a cold
  neutron beam}},}\ }\href {\doibase 10.1103/PhysRevC.71.055502} {\bibfield
  {journal} {\bibinfo  {journal} {Phys. Rev. C}\ }\textbf {\bibinfo {volume}
  {71}},\ \bibinfo {pages} {055502} (\bibinfo {year} {2005})},\ \Eprint
  {http://arxiv.org/abs/nucl-ex/0411041} {arXiv:nucl-ex/0411041} \BibitemShut
  {NoStop}%
\bibitem [{\citenamefont {Yue}\ \emph {et~al.}(2013)\citenamefont {Yue},
  \citenamefont {Dewey}, \citenamefont {Gilliam}, \citenamefont {Greene},
  \citenamefont {Laptev}, \citenamefont {Nico}, \citenamefont {Snow},\ and\
  \citenamefont {Wietfeldt}}]{Yue:2013qrc}%
  \BibitemOpen
  \bibfield  {author} {\bibinfo {author} {\bibfnamefont {A.~T.}\ \bibnamefont
  {Yue}}, \bibinfo {author} {\bibfnamefont {M.~S.}\ \bibnamefont {Dewey}},
  \bibinfo {author} {\bibfnamefont {D.~M.}\ \bibnamefont {Gilliam}}, \bibinfo
  {author} {\bibfnamefont {G.~L.}\ \bibnamefont {Greene}}, \bibinfo {author}
  {\bibfnamefont {A.~B.}\ \bibnamefont {Laptev}}, \bibinfo {author}
  {\bibfnamefont {J.~S.}\ \bibnamefont {Nico}}, \bibinfo {author}
  {\bibfnamefont {W.~M.}\ \bibnamefont {Snow}}, \ and\ \bibinfo {author}
  {\bibfnamefont {F.~E.}\ \bibnamefont {Wietfeldt}},\ }\bibfield  {title}
  {\enquote {\bibinfo {title} {{Improved Determination of the Neutron
  Lifetime}},}\ }\href {\doibase 10.1103/PhysRevLett.111.222501} {\bibfield
  {journal} {\bibinfo  {journal} {Phys. Rev. Lett.}\ }\textbf {\bibinfo
  {volume} {111}},\ \bibinfo {pages} {222501} (\bibinfo {year} {2013})},\
  \Eprint {http://arxiv.org/abs/1309.2623} {arXiv:1309.2623 [nucl-ex]}
  \BibitemShut {NoStop}%
\bibitem [{\citenamefont {Serebrov}\ \emph {et~al.}(2005)\citenamefont
  {Serebrov} \emph {et~al.}}]{Serebrov:2004zf}%
  \BibitemOpen
  \bibfield  {author} {\bibinfo {author} {\bibfnamefont {A.}~\bibnamefont
  {Serebrov}} \emph {et~al.},\ }\bibfield  {title} {\enquote {\bibinfo {title}
  {{Measurement of the neutron lifetime using a gravitational trap and a
  low-temperature Fomblin coating}},}\ }\href {\doibase
  10.1016/j.physletb.2004.11.013} {\bibfield  {journal} {\bibinfo  {journal}
  {Phys. Lett. B}\ }\textbf {\bibinfo {volume} {605}},\ \bibinfo {pages}
  {72--78} (\bibinfo {year} {2005})},\ \Eprint
  {http://arxiv.org/abs/nucl-ex/0408009} {arXiv:nucl-ex/0408009} \BibitemShut
  {NoStop}%
\bibitem [{\citenamefont {Pichlmaier}\ \emph {et~al.}(2010)\citenamefont
  {Pichlmaier}, \citenamefont {Varlamov}, \citenamefont {Schreckenbach},\ and\
  \citenamefont {Geltenbort}}]{Pichlmaier:2010zz}%
  \BibitemOpen
  \bibfield  {author} {\bibinfo {author} {\bibfnamefont {A.}~\bibnamefont
  {Pichlmaier}}, \bibinfo {author} {\bibfnamefont {V.}~\bibnamefont
  {Varlamov}}, \bibinfo {author} {\bibfnamefont {K.}~\bibnamefont
  {Schreckenbach}}, \ and\ \bibinfo {author} {\bibfnamefont {P.}~\bibnamefont
  {Geltenbort}},\ }\bibfield  {title} {\enquote {\bibinfo {title} {{Neutron
  lifetime measurement with the UCN trap-in-trap MAMBO II}},}\ }\href {\doibase
  10.1016/j.physletb.2010.08.032} {\bibfield  {journal} {\bibinfo  {journal}
  {Phys. Lett. B}\ }\textbf {\bibinfo {volume} {693}},\ \bibinfo {pages}
  {221--226} (\bibinfo {year} {2010})}\BibitemShut {NoStop}%
\bibitem [{\citenamefont {Steyerl}\ \emph {et~al.}(2012)\citenamefont
  {Steyerl}, \citenamefont {Pendlebury}, \citenamefont {Kaufman}, \citenamefont
  {Malik},\ and\ \citenamefont {Desai}}]{Steyerl:2012zz}%
  \BibitemOpen
  \bibfield  {author} {\bibinfo {author} {\bibfnamefont {A.}~\bibnamefont
  {Steyerl}}, \bibinfo {author} {\bibfnamefont {J.~M.}\ \bibnamefont
  {Pendlebury}}, \bibinfo {author} {\bibfnamefont {C.}~\bibnamefont {Kaufman}},
  \bibinfo {author} {\bibfnamefont {S.~S.}\ \bibnamefont {Malik}}, \ and\
  \bibinfo {author} {\bibfnamefont {A.~M.}\ \bibnamefont {Desai}},\ }\bibfield
  {title} {\enquote {\bibinfo {title} {{Quasielastic scattering in the
  interaction of ultracold neutrons with a liquid wall and application in a
  reanalysis of the Mambo I neutron-lifetime experiment}},}\ }\href {\doibase
  10.1103/PhysRevC.85.065503} {\bibfield  {journal} {\bibinfo  {journal} {Phys.
  Rev. C}\ }\textbf {\bibinfo {volume} {85}},\ \bibinfo {pages} {065503}
  (\bibinfo {year} {2012})}\BibitemShut {NoStop}%
\bibitem [{\citenamefont {Arzumanov}\ \emph {et~al.}(2015)\citenamefont
  {Arzumanov}, \citenamefont {Bondarenko}, \citenamefont {Chernyavsky},
  \citenamefont {Geltenbort}, \citenamefont {Morozov}, \citenamefont
  {Nesvizhevsky}, \citenamefont {Panin},\ and\ \citenamefont
  {Strepetov}}]{Arzumanov:2015tea}%
  \BibitemOpen
  \bibfield  {author} {\bibinfo {author} {\bibfnamefont {S.}~\bibnamefont
  {Arzumanov}}, \bibinfo {author} {\bibfnamefont {L.}~\bibnamefont
  {Bondarenko}}, \bibinfo {author} {\bibfnamefont {S.}~\bibnamefont
  {Chernyavsky}}, \bibinfo {author} {\bibfnamefont {P.}~\bibnamefont
  {Geltenbort}}, \bibinfo {author} {\bibfnamefont {V.}~\bibnamefont {Morozov}},
  \bibinfo {author} {\bibfnamefont {V.~V.}\ \bibnamefont {Nesvizhevsky}},
  \bibinfo {author} {\bibfnamefont {Y.}~\bibnamefont {Panin}}, \ and\ \bibinfo
  {author} {\bibfnamefont {A.}~\bibnamefont {Strepetov}},\ }\bibfield  {title}
  {\enquote {\bibinfo {title} {{A measurement of the neutron lifetime using the
  method of storage of ultracold neutrons and detection of inelastically
  up-scattered neutrons}},}\ }\href {\doibase 10.1016/j.physletb.2015.04.021}
  {\bibfield  {journal} {\bibinfo  {journal} {Phys. Lett. B}\ }\textbf
  {\bibinfo {volume} {745}},\ \bibinfo {pages} {79--89} (\bibinfo {year}
  {2015})}\BibitemShut {NoStop}%
\bibitem [{\citenamefont {Serebrov}\ \emph {et~al.}(2018)\citenamefont
  {Serebrov} \emph {et~al.}}]{Serebrov:2017bzo}%
  \BibitemOpen
  \bibfield  {author} {\bibinfo {author} {\bibfnamefont {A.~P.}\ \bibnamefont
  {Serebrov}} \emph {et~al.},\ }\bibfield  {title} {\enquote {\bibinfo {title}
  {{Neutron lifetime measurements with a large gravitational trap for ultracold
  neutrons}},}\ }\href {\doibase 10.1103/PhysRevC.97.055503} {\bibfield
  {journal} {\bibinfo  {journal} {Phys. Rev. C}\ }\textbf {\bibinfo {volume}
  {97}},\ \bibinfo {pages} {055503} (\bibinfo {year} {2018})},\ \Eprint
  {http://arxiv.org/abs/1712.05663} {arXiv:1712.05663 [nucl-ex]} \BibitemShut
  {NoStop}%
\bibitem [{\citenamefont {Pattie}\ \emph {et~al.}(2018)\citenamefont {Pattie}
  \emph {et~al.}}]{Pattie:2017vsj}%
  \BibitemOpen
  \bibfield  {author} {\bibinfo {author} {\bibfnamefont {R.~W.}\ \bibnamefont
  {Pattie}, \bibfnamefont {Jr.}} \emph {et~al.},\ }\bibfield  {title} {\enquote
  {\bibinfo {title} {{Measurement of the neutron lifetime using a
  magneto-gravitational trap and in situ detection}},}\ }\href {\doibase
  10.1126/science.aan8895} {\bibfield  {journal} {\bibinfo  {journal}
  {Science}\ }\textbf {\bibinfo {volume} {360}},\ \bibinfo {pages} {627--632}
  (\bibinfo {year} {2018})},\ \Eprint {http://arxiv.org/abs/1707.01817}
  {arXiv:1707.01817 [nucl-ex]} \BibitemShut {NoStop}%
\bibitem [{\citenamefont {Ezhov}\ \emph {et~al.}(2018)\citenamefont {Ezhov}
  \emph {et~al.}}]{Ezhov:2014tna}%
  \BibitemOpen
  \bibfield  {author} {\bibinfo {author} {\bibfnamefont {V.~F.}\ \bibnamefont
  {Ezhov}} \emph {et~al.},\ }\bibfield  {title} {\enquote {\bibinfo {title}
  {{Measurement of the neutron lifetime with ultra-cold neutrons stored in a
  magneto-gravitational trap}},}\ }\href {\doibase 10.1134/S0021364018110024}
  {\bibfield  {journal} {\bibinfo  {journal} {JETP Lett.}\ }\textbf {\bibinfo
  {volume} {107}},\ \bibinfo {pages} {671--675} (\bibinfo {year} {2018})},\
  \Eprint {http://arxiv.org/abs/1412.7434} {arXiv:1412.7434 [nucl-ex]}
  \BibitemShut {NoStop}%
\bibitem [{\citenamefont {Gonzalez}\ \emph {et~al.}(2021)\citenamefont
  {Gonzalez} \emph {et~al.}}]{UCNt:2021pcg}%
  \BibitemOpen
  \bibfield  {author} {\bibinfo {author} {\bibfnamefont {F.~M.}\ \bibnamefont
  {Gonzalez}} \emph {et~al.} (\bibinfo {collaboration}
  {UCN\ensuremath{\tau}}),\ }\bibfield  {title} {\enquote {\bibinfo {title}
  {{Improved neutron lifetime measurement with UCN$\tau$}},}\ }\href {\doibase
  10.1103/PhysRevLett.127.162501} {\bibfield  {journal} {\bibinfo  {journal}
  {Phys. Rev. Lett.}\ }\textbf {\bibinfo {volume} {127}},\ \bibinfo {pages}
  {162501} (\bibinfo {year} {2021})},\ \Eprint
  {http://arxiv.org/abs/2106.10375} {arXiv:2106.10375 [nucl-ex]} \BibitemShut
  {NoStop}%
\bibitem [{\citenamefont {Musedinovic}\ \emph {et~al.}(2024)\citenamefont
  {Musedinovic} \emph {et~al.}}]{Musedinovic:2024gms}%
  \BibitemOpen
  \bibfield  {author} {\bibinfo {author} {\bibfnamefont {R.}~\bibnamefont
  {Musedinovic}} \emph {et~al.},\ }\bibfield  {title} {\enquote {\bibinfo
  {title} {{Measurement of the Free Neutron Lifetime in a Magneto-Gravitational
  Trap with In Situ Detection}},}\ }\href@noop {} {\  (\bibinfo {year}
  {2024})},\ \Eprint {http://arxiv.org/abs/2409.05560} {arXiv:2409.05560
  [nucl-ex]} \BibitemShut {NoStop}%
\bibitem [{\citenamefont {Fornal}\ and\ \citenamefont
  {Grinstein}(2018)}]{Fornal:2018eol}%
  \BibitemOpen
  \bibfield  {author} {\bibinfo {author} {\bibfnamefont {B.}~\bibnamefont
  {Fornal}}\ and\ \bibinfo {author} {\bibfnamefont {B.}~\bibnamefont
  {Grinstein}},\ }\bibfield  {title} {\enquote {\bibinfo {title} {{Dark Matter
  Interpretation of the Neutron Decay Anomaly}},}\ }\href {\doibase
  10.1103/PhysRevLett.120.191801} {\bibfield  {journal} {\bibinfo  {journal}
  {Phys. Rev. Lett.}\ }\textbf {\bibinfo {volume} {120}},\ \bibinfo {pages}
  {191801} (\bibinfo {year} {2018})},\ \bibinfo {note} {[Erratum: Phys. Rev.
  Lett. {\bf 124}, 219901 (2020)]},\ \Eprint {http://arxiv.org/abs/1801.01124}
  {arXiv:1801.01124 [hep-ph]} \BibitemShut {NoStop}%
\bibitem [{\citenamefont {Fuwa}\ \emph {et~al.}(2024)\citenamefont {Fuwa} \emph
  {et~al.}}]{Fuwa:2024cdf}%
  \BibitemOpen
  \bibfield  {author} {\bibinfo {author} {\bibfnamefont {Y.}~\bibnamefont
  {Fuwa}} \emph {et~al.},\ }\bibfield  {title} {\enquote {\bibinfo {title}
  {{Improved measurements of neutron lifetime with cold neutron beam at
  J-PARC}},}\ }\href@noop {} {\  (\bibinfo {year} {2024})},\ \Eprint
  {http://arxiv.org/abs/2412.19519} {arXiv:2412.19519 [nucl-ex]} \BibitemShut
  {NoStop}%
\bibitem [{\citenamefont {Czarnecki}\ \emph {et~al.}(2018)\citenamefont
  {Czarnecki}, \citenamefont {Marciano},\ and\ \citenamefont
  {Sirlin}}]{Czarnecki:2018okw}%
  \BibitemOpen
  \bibfield  {author} {\bibinfo {author} {\bibfnamefont {A.}~\bibnamefont
  {Czarnecki}}, \bibinfo {author} {\bibfnamefont {W.~J.}\ \bibnamefont
  {Marciano}}, \ and\ \bibinfo {author} {\bibfnamefont {A.}~\bibnamefont
  {Sirlin}},\ }\bibfield  {title} {\enquote {\bibinfo {title} {{Neutron
  Lifetime and Axial Coupling Connection}},}\ }\href {\doibase
  10.1103/PhysRevLett.120.202002} {\bibfield  {journal} {\bibinfo  {journal}
  {Phys. Rev. Lett.}\ }\textbf {\bibinfo {volume} {120}},\ \bibinfo {pages}
  {202002} (\bibinfo {year} {2018})},\ \Eprint
  {http://arxiv.org/abs/1802.01804} {arXiv:1802.01804 [hep-ph]} \BibitemShut
  {NoStop}%
\bibitem [{\citenamefont {Cirigliano}\ \emph
  {et~al.}(2023{\natexlab{a}})\citenamefont {Cirigliano}, \citenamefont
  {Dekens}, \citenamefont {Mereghetti},\ and\ \citenamefont
  {Tomalak}}]{Cirigliano:2023fnz}%
  \BibitemOpen
  \bibfield  {author} {\bibinfo {author} {\bibfnamefont {V.}~\bibnamefont
  {Cirigliano}}, \bibinfo {author} {\bibfnamefont {W.}~\bibnamefont {Dekens}},
  \bibinfo {author} {\bibfnamefont {E.}~\bibnamefont {Mereghetti}}, \ and\
  \bibinfo {author} {\bibfnamefont {O.}~\bibnamefont {Tomalak}},\ }\bibfield
  {title} {\enquote {\bibinfo {title} {{Effective field theory for radiative
  corrections to charged-current processes: Vector coupling}},}\ }\href
  {\doibase 10.1103/PhysRevD.108.053003} {\bibfield  {journal} {\bibinfo
  {journal} {Phys. Rev. D}\ }\textbf {\bibinfo {volume} {108}},\ \bibinfo
  {pages} {053003} (\bibinfo {year} {2023}{\natexlab{a}})},\ \Eprint
  {http://arxiv.org/abs/2306.03138} {arXiv:2306.03138 [hep-ph]} \BibitemShut
  {NoStop}%
\bibitem [{\citenamefont {Vander~Griend}\ \emph {et~al.}(2025)\citenamefont
  {Vander~Griend}, \citenamefont {Cao}, \citenamefont {Hill},\ and\
  \citenamefont {Plestid}}]{VanderGriend:2025mdc}%
  \BibitemOpen
  \bibfield  {author} {\bibinfo {author} {\bibfnamefont {P.}~\bibnamefont
  {Vander~Griend}}, \bibinfo {author} {\bibfnamefont {Z.}~\bibnamefont {Cao}},
  \bibinfo {author} {\bibfnamefont {R.}~\bibnamefont {Hill}}, \ and\ \bibinfo
  {author} {\bibfnamefont {R.}~\bibnamefont {Plestid}},\ }\bibfield  {title}
  {\enquote {\bibinfo {title} {{The Fermi function and the neutron's
  lifetime}},}\ }\href@noop {} {\  (\bibinfo {year} {2025})},\ \Eprint
  {http://arxiv.org/abs/2501.17916} {arXiv:2501.17916 [hep-ph]} \BibitemShut
  {NoStop}%
\bibitem [{\citenamefont {Sirlin}(1967)}]{Sirlin:1967zza}%
  \BibitemOpen
  \bibfield  {author} {\bibinfo {author} {\bibfnamefont {A.}~\bibnamefont
  {Sirlin}},\ }\bibfield  {title} {\enquote {\bibinfo {title} {{General
  Properties of the Electromagnetic Corrections to the Beta Decay of a Physical
  Nucleon}},}\ }\href {\doibase 10.1103/PhysRev.164.1767} {\bibfield  {journal}
  {\bibinfo  {journal} {Phys. Rev.}\ }\textbf {\bibinfo {volume} {164}},\
  \bibinfo {pages} {1767--1775} (\bibinfo {year} {1967})}\BibitemShut {NoStop}%
\bibitem [{\citenamefont {Sirlin}(1978)}]{Sirlin:1977sv}%
  \BibitemOpen
  \bibfield  {author} {\bibinfo {author} {\bibfnamefont {A.}~\bibnamefont
  {Sirlin}},\ }\bibfield  {title} {\enquote {\bibinfo {title} {{Current Algebra
  Formulation of Radiative Corrections in Gauge Theories and the Universality
  of the Weak Interactions}},}\ }\href {\doibase 10.1103/RevModPhys.50.573}
  {\bibfield  {journal} {\bibinfo  {journal} {Rev. Mod. Phys.}\ }\textbf
  {\bibinfo {volume} {50}},\ \bibinfo {pages} {573} (\bibinfo {year} {1978})},\
  \bibinfo {note} {[Erratum: Rev. Mod. Phys. {\bf 50}, 905 (1978)]}\BibitemShut
  {NoStop}%
\bibitem [{\citenamefont {Seng}\ \emph {et~al.}(2018)\citenamefont {Seng},
  \citenamefont {Gorchtein}, \citenamefont {Patel},\ and\ \citenamefont
  {Ramsey-Musolf}}]{Seng:2018yzq}%
  \BibitemOpen
  \bibfield  {author} {\bibinfo {author} {\bibfnamefont {C.-Y.}\ \bibnamefont
  {Seng}}, \bibinfo {author} {\bibfnamefont {M.}~\bibnamefont {Gorchtein}},
  \bibinfo {author} {\bibfnamefont {H.~H.}\ \bibnamefont {Patel}}, \ and\
  \bibinfo {author} {\bibfnamefont {M.~J.}\ \bibnamefont {Ramsey-Musolf}},\
  }\bibfield  {title} {\enquote {\bibinfo {title} {{Reduced Hadronic
  Uncertainty in the Determination of $V_{ud}$}},}\ }\href {\doibase
  10.1103/PhysRevLett.121.241804} {\bibfield  {journal} {\bibinfo  {journal}
  {Phys. Rev. Lett.}\ }\textbf {\bibinfo {volume} {121}},\ \bibinfo {pages}
  {241804} (\bibinfo {year} {2018})},\ \Eprint
  {http://arxiv.org/abs/1807.10197} {arXiv:1807.10197 [hep-ph]} \BibitemShut
  {NoStop}%
\bibitem [{\citenamefont {Seng}\ \emph {et~al.}(2019)\citenamefont {Seng},
  \citenamefont {Gorchtein},\ and\ \citenamefont
  {Ramsey-Musolf}}]{Seng:2018qru}%
  \BibitemOpen
  \bibfield  {author} {\bibinfo {author} {\bibfnamefont {C.~Y.}\ \bibnamefont
  {Seng}}, \bibinfo {author} {\bibfnamefont {M.}~\bibnamefont {Gorchtein}}, \
  and\ \bibinfo {author} {\bibfnamefont {M.~J.}\ \bibnamefont
  {Ramsey-Musolf}},\ }\bibfield  {title} {\enquote {\bibinfo {title}
  {{Dispersive evaluation of the inner radiative correction in neutron and
  nuclear $\beta$ decay}},}\ }\href {\doibase 10.1103/PhysRevD.100.013001}
  {\bibfield  {journal} {\bibinfo  {journal} {Phys. Rev. D}\ }\textbf {\bibinfo
  {volume} {100}},\ \bibinfo {pages} {013001} (\bibinfo {year} {2019})},\
  \Eprint {http://arxiv.org/abs/1812.03352} {arXiv:1812.03352 [nucl-th]}
  \BibitemShut {NoStop}%
\bibitem [{\citenamefont {Czarnecki}\ \emph {et~al.}(2019)\citenamefont
  {Czarnecki}, \citenamefont {Marciano},\ and\ \citenamefont
  {Sirlin}}]{Czarnecki:2019mwq}%
  \BibitemOpen
  \bibfield  {author} {\bibinfo {author} {\bibfnamefont {A.}~\bibnamefont
  {Czarnecki}}, \bibinfo {author} {\bibfnamefont {W.~J.}\ \bibnamefont
  {Marciano}}, \ and\ \bibinfo {author} {\bibfnamefont {A.}~\bibnamefont
  {Sirlin}},\ }\bibfield  {title} {\enquote {\bibinfo {title} {{Radiative
  Corrections to Neutron and Nuclear Beta Decays Revisited}},}\ }\href
  {\doibase 10.1103/PhysRevD.100.073008} {\bibfield  {journal} {\bibinfo
  {journal} {Phys. Rev. D}\ }\textbf {\bibinfo {volume} {100}},\ \bibinfo
  {pages} {073008} (\bibinfo {year} {2019})},\ \Eprint
  {http://arxiv.org/abs/1907.06737} {arXiv:1907.06737 [hep-ph]} \BibitemShut
  {NoStop}%
\bibitem [{\citenamefont {Hayen}(2021{\natexlab{a}})}]{Hayen:2020cxh}%
  \BibitemOpen
  \bibfield  {author} {\bibinfo {author} {\bibfnamefont {L.}~\bibnamefont
  {Hayen}},\ }\bibfield  {title} {\enquote {\bibinfo {title} {{Standard model
  $\mathcal{O}(\alpha)$ renormalization of $g_A$ and its impact on new physics
  searches}},}\ }\href {\doibase 10.1103/PhysRevD.103.113001} {\bibfield
  {journal} {\bibinfo  {journal} {Phys. Rev. D}\ }\textbf {\bibinfo {volume}
  {103}},\ \bibinfo {pages} {113001} (\bibinfo {year} {2021}{\natexlab{a}})},\
  \Eprint {http://arxiv.org/abs/2010.07262} {arXiv:2010.07262 [hep-ph]}
  \BibitemShut {NoStop}%
\bibitem [{\citenamefont {Hayen}(2021{\natexlab{b}})}]{Hayen:2021iga}%
  \BibitemOpen
  \bibfield  {author} {\bibinfo {author} {\bibfnamefont {L.}~\bibnamefont
  {Hayen}},\ }\bibfield  {title} {\enquote {\bibinfo {title} {{Radiative
  corrections to nucleon weak charges and Beyond Standard Model impact}},}\
  }\href@noop {} {\  (\bibinfo {year} {2021}{\natexlab{b}})},\ \Eprint
  {http://arxiv.org/abs/2102.03458} {arXiv:2102.03458 [hep-ph]} \BibitemShut
  {NoStop}%
\bibitem [{\citenamefont {Gorchtein}\ and\ \citenamefont
  {Seng}(2023)}]{Gorchtein:2023srs}%
  \BibitemOpen
  \bibfield  {author} {\bibinfo {author} {\bibfnamefont {M.}~\bibnamefont
  {Gorchtein}}\ and\ \bibinfo {author} {\bibfnamefont {C.-Y.}\ \bibnamefont
  {Seng}},\ }\bibfield  {title} {\enquote {\bibinfo {title} {{The Standard
  Model Theory of Neutron Beta Decay}},}\ }\href {\doibase
  10.3390/universe9090422} {\bibfield  {journal} {\bibinfo  {journal}
  {Universe}\ }\textbf {\bibinfo {volume} {9}},\ \bibinfo {pages} {422}
  (\bibinfo {year} {2023})},\ \Eprint {http://arxiv.org/abs/2307.01145}
  {arXiv:2307.01145 [hep-ph]} \BibitemShut {NoStop}%
\bibitem [{\citenamefont {Cirigliano}\ \emph
  {et~al.}(2022{\natexlab{a}})\citenamefont {Cirigliano}, \citenamefont
  {de~Vries}, \citenamefont {Hayen}, \citenamefont {Mereghetti},\ and\
  \citenamefont {Walker-Loud}}]{Cirigliano:2022hob}%
  \BibitemOpen
  \bibfield  {author} {\bibinfo {author} {\bibfnamefont {V.}~\bibnamefont
  {Cirigliano}}, \bibinfo {author} {\bibfnamefont {J.}~\bibnamefont
  {de~Vries}}, \bibinfo {author} {\bibfnamefont {L.}~\bibnamefont {Hayen}},
  \bibinfo {author} {\bibfnamefont {E.}~\bibnamefont {Mereghetti}}, \ and\
  \bibinfo {author} {\bibfnamefont {A.}~\bibnamefont {Walker-Loud}},\
  }\bibfield  {title} {\enquote {\bibinfo {title} {{Pion-Induced Radiative
  Corrections to Neutron \ensuremath{\beta} Decay}},}\ }\href {\doibase
  10.1103/PhysRevLett.129.121801} {\bibfield  {journal} {\bibinfo  {journal}
  {Phys. Rev. Lett.}\ }\textbf {\bibinfo {volume} {129}},\ \bibinfo {pages}
  {121801} (\bibinfo {year} {2022}{\natexlab{a}})},\ \Eprint
  {http://arxiv.org/abs/2202.10439} {arXiv:2202.10439 [nucl-th]} \BibitemShut
  {NoStop}%
\bibitem [{\citenamefont {Jenkins}\ and\ \citenamefont
  {Manohar}(1991{\natexlab{a}})}]{Jenkins:1990jv}%
  \BibitemOpen
  \bibfield  {author} {\bibinfo {author} {\bibfnamefont {E.~E.}\ \bibnamefont
  {Jenkins}}\ and\ \bibinfo {author} {\bibfnamefont {A.~V.}\ \bibnamefont
  {Manohar}},\ }\bibfield  {title} {\enquote {\bibinfo {title} {{Baryon chiral
  perturbation theory using a heavy fermion Lagrangian}},}\ }\href {\doibase
  10.1016/0370-2693(91)90266-S} {\bibfield  {journal} {\bibinfo  {journal}
  {Phys. Lett. B}\ }\textbf {\bibinfo {volume} {255}},\ \bibinfo {pages}
  {558--562} (\bibinfo {year} {1991}{\natexlab{a}})}\BibitemShut {NoStop}%
\bibitem [{\citenamefont {Jenkins}\ and\ \citenamefont
  {Manohar}(1991{\natexlab{b}})}]{Jenkins:1991es}%
  \BibitemOpen
  \bibfield  {author} {\bibinfo {author} {\bibfnamefont {E.~E.}\ \bibnamefont
  {Jenkins}}\ and\ \bibinfo {author} {\bibfnamefont {A.~V.}\ \bibnamefont
  {Manohar}},\ }\bibfield  {title} {\enquote {\bibinfo {title} {{Chiral
  corrections to the baryon axial currents}},}\ }\href {\doibase
  10.1016/0370-2693(91)90840-M} {\bibfield  {journal} {\bibinfo  {journal}
  {Phys. Lett. B}\ }\textbf {\bibinfo {volume} {259}},\ \bibinfo {pages}
  {353--358} (\bibinfo {year} {1991}{\natexlab{b}})}\BibitemShut {NoStop}%
\bibitem [{\citenamefont {Bhattacharya}\ \emph {et~al.}(2016)\citenamefont
  {Bhattacharya}, \citenamefont {Cirigliano}, \citenamefont {Cohen},
  \citenamefont {Gupta}, \citenamefont {Lin},\ and\ \citenamefont
  {Yoon}}]{Bhattacharya:2016zcn}%
  \BibitemOpen
  \bibfield  {author} {\bibinfo {author} {\bibfnamefont {T.}~\bibnamefont
  {Bhattacharya}}, \bibinfo {author} {\bibfnamefont {V.}~\bibnamefont
  {Cirigliano}}, \bibinfo {author} {\bibfnamefont {S.}~\bibnamefont {Cohen}},
  \bibinfo {author} {\bibfnamefont {R.}~\bibnamefont {Gupta}}, \bibinfo
  {author} {\bibfnamefont {H.-W.}\ \bibnamefont {Lin}}, \ and\ \bibinfo
  {author} {\bibfnamefont {B.}~\bibnamefont {Yoon}},\ }\bibfield  {title}
  {\enquote {\bibinfo {title} {{Axial, Scalar and Tensor Charges of the Nucleon
  from 2+1+1-flavor Lattice QCD}},}\ }\href {\doibase
  10.1103/PhysRevD.94.054508} {\bibfield  {journal} {\bibinfo  {journal} {Phys.
  Rev. D}\ }\textbf {\bibinfo {volume} {94}},\ \bibinfo {pages} {054508}
  (\bibinfo {year} {2016})},\ \Eprint {http://arxiv.org/abs/1606.07049}
  {arXiv:1606.07049 [hep-lat]} \BibitemShut {NoStop}%
\bibitem [{\citenamefont {Chang}\ \emph {et~al.}(2018)\citenamefont {Chang},
  \citenamefont {Nicholson}, \citenamefont {Rinaldi}, \citenamefont
  {Berkowitz}, \citenamefont {Garron}, \citenamefont {Brantley}, \citenamefont
  {Monge-Camacho}, \citenamefont {Monahan}, \citenamefont {Bouchard},
  \citenamefont {Clark}, \citenamefont {Jo{\'o}}, \citenamefont {Kurth},
  \citenamefont {Orginos}, \citenamefont {Vranas},\ and\ \citenamefont
  {Walker-Loud}}]{Chang:2018uxx}%
  \BibitemOpen
  \bibfield  {author} {\bibinfo {author} {\bibfnamefont {C.~C.}\ \bibnamefont
  {Chang}}, \bibinfo {author} {\bibfnamefont {A.}~\bibnamefont {Nicholson}},
  \bibinfo {author} {\bibfnamefont {E.}~\bibnamefont {Rinaldi}}, \bibinfo
  {author} {\bibfnamefont {E.}~\bibnamefont {Berkowitz}}, \bibinfo {author}
  {\bibfnamefont {N.}~\bibnamefont {Garron}}, \bibinfo {author} {\bibfnamefont
  {D.~A.}\ \bibnamefont {Brantley}}, \bibinfo {author} {\bibfnamefont
  {H.}~\bibnamefont {Monge-Camacho}}, \bibinfo {author} {\bibfnamefont {C.~J.}\
  \bibnamefont {Monahan}}, \bibinfo {author} {\bibfnamefont {C.}~\bibnamefont
  {Bouchard}}, \bibinfo {author} {\bibfnamefont {M.~A.}\ \bibnamefont {Clark}},
  \bibinfo {author} {\bibfnamefont {B.}~\bibnamefont {Jo{\'o}}}, \bibinfo
  {author} {\bibfnamefont {T.}~\bibnamefont {Kurth}}, \bibinfo {author}
  {\bibfnamefont {K.}~\bibnamefont {Orginos}}, \bibinfo {author} {\bibfnamefont
  {P.}~\bibnamefont {Vranas}}, \ and\ \bibinfo {author} {\bibfnamefont
  {A.}~\bibnamefont {Walker-Loud}},\ }\bibfield  {title} {\enquote {\bibinfo
  {title} {{A per-cent-level determination of the nucleon axial coupling from
  quantum chromodynamics}},}\ }\href {\doibase 10.1038/s41586-018-0161-8}
  {\bibfield  {journal} {\bibinfo  {journal} {Nature}\ }\textbf {\bibinfo
  {volume} {558}},\ \bibinfo {pages} {91--94} (\bibinfo {year} {2018})},\
  \Eprint {http://arxiv.org/abs/1805.12130} {arXiv:1805.12130 [hep-lat]}
  \BibitemShut {NoStop}%
\bibitem [{\citenamefont {Gupta}\ \emph {et~al.}(2018)\citenamefont {Gupta},
  \citenamefont {Jang}, \citenamefont {Yoon}, \citenamefont {Lin},
  \citenamefont {Cirigliano},\ and\ \citenamefont
  {Bhattacharya}}]{Gupta:2018qil}%
  \BibitemOpen
  \bibfield  {author} {\bibinfo {author} {\bibfnamefont {R.}~\bibnamefont
  {Gupta}}, \bibinfo {author} {\bibfnamefont {Y.-C.}\ \bibnamefont {Jang}},
  \bibinfo {author} {\bibfnamefont {B.}~\bibnamefont {Yoon}}, \bibinfo {author}
  {\bibfnamefont {H.-W.}\ \bibnamefont {Lin}}, \bibinfo {author} {\bibfnamefont
  {V.}~\bibnamefont {Cirigliano}}, \ and\ \bibinfo {author} {\bibfnamefont
  {T.}~\bibnamefont {Bhattacharya}},\ }\bibfield  {title} {\enquote {\bibinfo
  {title} {{Isovector Charges of the Nucleon from 2+1+1-flavor Lattice QCD}},}\
  }\href {\doibase 10.1103/PhysRevD.98.034503} {\bibfield  {journal} {\bibinfo
  {journal} {Phys. Rev. D}\ }\textbf {\bibinfo {volume} {98}},\ \bibinfo
  {pages} {034503} (\bibinfo {year} {2018})},\ \Eprint
  {http://arxiv.org/abs/1806.09006} {arXiv:1806.09006 [hep-lat]} \BibitemShut
  {NoStop}%
\bibitem [{\citenamefont {Alexandrou}\ \emph {et~al.}(2020)\citenamefont
  {Alexandrou}, \citenamefont {Bacchio}, \citenamefont {Constantinou},
  \citenamefont {Finkenrath}, \citenamefont {Hadjiyiannakou}, \citenamefont
  {Jansen}, \citenamefont {Koutsou},\ and\ \citenamefont {Vaquero
  Aviles-Casco}}]{Alexandrou:2019brg}%
  \BibitemOpen
  \bibfield  {author} {\bibinfo {author} {\bibfnamefont {C.}~\bibnamefont
  {Alexandrou}}, \bibinfo {author} {\bibfnamefont {S.}~\bibnamefont {Bacchio}},
  \bibinfo {author} {\bibfnamefont {M.}~\bibnamefont {Constantinou}}, \bibinfo
  {author} {\bibfnamefont {J.}~\bibnamefont {Finkenrath}}, \bibinfo {author}
  {\bibfnamefont {K.}~\bibnamefont {Hadjiyiannakou}}, \bibinfo {author}
  {\bibfnamefont {K.}~\bibnamefont {Jansen}}, \bibinfo {author} {\bibfnamefont
  {G.}~\bibnamefont {Koutsou}}, \ and\ \bibinfo {author} {\bibfnamefont
  {A.}~\bibnamefont {Vaquero Aviles-Casco}},\ }\bibfield  {title} {\enquote
  {\bibinfo {title} {{Nucleon axial, tensor, and scalar charges and
  $\sigma$-terms in lattice QCD}},}\ }\href {\doibase
  10.1103/PhysRevD.102.054517} {\bibfield  {journal} {\bibinfo  {journal}
  {Phys. Rev. D}\ }\textbf {\bibinfo {volume} {102}},\ \bibinfo {pages}
  {054517} (\bibinfo {year} {2020})},\ \Eprint
  {http://arxiv.org/abs/1909.00485} {arXiv:1909.00485 [hep-lat]} \BibitemShut
  {NoStop}%
\bibitem [{\citenamefont {Bali}\ \emph {et~al.}(2023)\citenamefont {Bali},
  \citenamefont {Collins}, \citenamefont {Heybrock}, \citenamefont {L\"offler},
  \citenamefont {R\"odl}, \citenamefont {S\"oldner},\ and\ \citenamefont
  {Weish\"aupl}}]{Bali:2023sdi}%
  \BibitemOpen
  \bibfield  {author} {\bibinfo {author} {\bibfnamefont {G.~S.}\ \bibnamefont
  {Bali}}, \bibinfo {author} {\bibfnamefont {S.}~\bibnamefont {Collins}},
  \bibinfo {author} {\bibfnamefont {S.}~\bibnamefont {Heybrock}}, \bibinfo
  {author} {\bibfnamefont {M.}~\bibnamefont {L\"offler}}, \bibinfo {author}
  {\bibfnamefont {R.}~\bibnamefont {R\"odl}}, \bibinfo {author} {\bibfnamefont
  {W.}~\bibnamefont {S\"oldner}}, \ and\ \bibinfo {author} {\bibfnamefont
  {S.}~\bibnamefont {Weish\"aupl}} (\bibinfo {collaboration} {RQCD}),\
  }\bibfield  {title} {\enquote {\bibinfo {title} {{Octet baryon isovector
  charges from $N_f=2+1$ lattice QCD}},}\ }\href {\doibase
  10.1103/PhysRevD.108.034512} {\bibfield  {journal} {\bibinfo  {journal}
  {Phys. Rev. D}\ }\textbf {\bibinfo {volume} {108}},\ \bibinfo {pages}
  {034512} (\bibinfo {year} {2023})},\ \Eprint
  {http://arxiv.org/abs/2305.04717} {arXiv:2305.04717 [hep-lat]} \BibitemShut
  {NoStop}%
\bibitem [{\citenamefont {Djukanovic}\ \emph {et~al.}(2024)\citenamefont
  {Djukanovic}, \citenamefont {von Hippel}, \citenamefont {Meyer},
  \citenamefont {Ottnad},\ and\ \citenamefont {Wittig}}]{Djukanovic:2024krw}%
  \BibitemOpen
  \bibfield  {author} {\bibinfo {author} {\bibfnamefont {D.}~\bibnamefont
  {Djukanovic}}, \bibinfo {author} {\bibfnamefont {G.}~\bibnamefont {von
  Hippel}}, \bibinfo {author} {\bibfnamefont {H.~B.}\ \bibnamefont {Meyer}},
  \bibinfo {author} {\bibfnamefont {K.}~\bibnamefont {Ottnad}}, \ and\ \bibinfo
  {author} {\bibfnamefont {H.}~\bibnamefont {Wittig}},\ }\bibfield  {title}
  {\enquote {\bibinfo {title} {{Improved analysis of isovector nucleon matrix
  elements with $N_f=2+1$ flavors of ${\mathcal O}(a)$ improved Wilson
  fermions}},}\ }\href {\doibase 10.1103/PhysRevD.109.074507} {\bibfield
  {journal} {\bibinfo  {journal} {Phys. Rev. D}\ }\textbf {\bibinfo {volume}
  {109}},\ \bibinfo {pages} {074507} (\bibinfo {year} {2024})},\ \Eprint
  {http://arxiv.org/abs/2402.03024} {arXiv:2402.03024 [hep-lat]} \BibitemShut
  {NoStop}%
\bibitem [{\citenamefont {Alexandrou}\ \emph
  {et~al.}(2024{\natexlab{a}})\citenamefont {Alexandrou}, \citenamefont
  {Bacchio}, \citenamefont {Finkenrath}, \citenamefont {Iona}, \citenamefont
  {Koutsou}, \citenamefont {Li},\ and\ \citenamefont
  {Spanoudes}}]{Alexandrou:2024ozj}%
  \BibitemOpen
  \bibfield  {author} {\bibinfo {author} {\bibfnamefont {C.}~\bibnamefont
  {Alexandrou}}, \bibinfo {author} {\bibfnamefont {S.}~\bibnamefont {Bacchio}},
  \bibinfo {author} {\bibfnamefont {J.}~\bibnamefont {Finkenrath}}, \bibinfo
  {author} {\bibfnamefont {C.}~\bibnamefont {Iona}}, \bibinfo {author}
  {\bibfnamefont {G.}~\bibnamefont {Koutsou}}, \bibinfo {author} {\bibfnamefont
  {Y.}~\bibnamefont {Li}}, \ and\ \bibinfo {author} {\bibfnamefont
  {G.}~\bibnamefont {Spanoudes}},\ }\bibfield  {title} {\enquote {\bibinfo
  {title} {{Nucleon charges and $\sigma$-terms in lattice QCD}},}\ }\href@noop
  {} {\  (\bibinfo {year} {2024}{\natexlab{a}})},\ \Eprint
  {http://arxiv.org/abs/2412.01535} {arXiv:2412.01535 [hep-lat]} \BibitemShut
  {NoStop}%
\bibitem [{\citenamefont {Seng}(2024)}]{Seng:2024ker}%
  \BibitemOpen
  \bibfield  {author} {\bibinfo {author} {\bibfnamefont {C.-Y.}\ \bibnamefont
  {Seng}},\ }\bibfield  {title} {\enquote {\bibinfo {title} {{Hybrid analysis
  of radiative corrections to neutron decay with current algebra and effective
  field theory}},}\ }\href {\doibase 10.1007/JHEP07(2024)175} {\bibfield
  {journal} {\bibinfo  {journal} {JHEP}\ }\textbf {\bibinfo {volume} {07}},\
  \bibinfo {pages} {175} (\bibinfo {year} {2024})},\ \Eprint
  {http://arxiv.org/abs/2403.08976} {arXiv:2403.08976 [hep-ph]} \BibitemShut
  {NoStop}%
\bibitem [{\citenamefont {Cirigliano}\ \emph
  {et~al.}(2024{\natexlab{a}})\citenamefont {Cirigliano}, \citenamefont
  {Dekens}, \citenamefont {Mereghetti},\ and\ \citenamefont
  {Tomalak}}]{Cirigliano:2024nfi}%
  \BibitemOpen
  \bibfield  {author} {\bibinfo {author} {\bibfnamefont {V.}~\bibnamefont
  {Cirigliano}}, \bibinfo {author} {\bibfnamefont {W.}~\bibnamefont {Dekens}},
  \bibinfo {author} {\bibfnamefont {E.}~\bibnamefont {Mereghetti}}, \ and\
  \bibinfo {author} {\bibfnamefont {O.}~\bibnamefont {Tomalak}},\ }\bibfield
  {title} {\enquote {\bibinfo {title} {{Effective field theory for radiative
  corrections to charged-current processes II: Axial-vector coupling}},}\
  }\href@noop {} {\  (\bibinfo {year} {2024}{\natexlab{a}})},\ \Eprint
  {http://arxiv.org/abs/2410.21404} {arXiv:2410.21404 [nucl-th]} \BibitemShut
  {NoStop}%
\bibitem [{\citenamefont {Cirigliano}\ \emph
  {et~al.}(2023{\natexlab{b}})\citenamefont {Cirigliano}, \citenamefont
  {Crivellin}, \citenamefont {Hoferichter},\ and\ \citenamefont
  {Moulson}}]{Cirigliano:2022yyo}%
  \BibitemOpen
  \bibfield  {author} {\bibinfo {author} {\bibfnamefont {V.}~\bibnamefont
  {Cirigliano}}, \bibinfo {author} {\bibfnamefont {A.}~\bibnamefont
  {Crivellin}}, \bibinfo {author} {\bibfnamefont {M.}~\bibnamefont
  {Hoferichter}}, \ and\ \bibinfo {author} {\bibfnamefont {M.}~\bibnamefont
  {Moulson}},\ }\bibfield  {title} {\enquote {\bibinfo {title} {{Scrutinizing
  CKM unitarity with a new measurement of the $K_{\mu3}/K_{\mu2}$ branching
  fraction}},}\ }\href {\doibase 10.1016/j.physletb.2023.137748} {\bibfield
  {journal} {\bibinfo  {journal} {Phys. Lett. B}\ }\textbf {\bibinfo {volume}
  {838}},\ \bibinfo {pages} {137748} (\bibinfo {year} {2023}{\natexlab{b}})},\
  \Eprint {http://arxiv.org/abs/2208.11707} {arXiv:2208.11707 [hep-ph]}
  \BibitemShut {NoStop}%
\bibitem [{\citenamefont {Hardy}\ and\ \citenamefont
  {Towner}(2020)}]{Hardy:2020qwl}%
  \BibitemOpen
  \bibfield  {author} {\bibinfo {author} {\bibfnamefont {J.~C.}\ \bibnamefont
  {Hardy}}\ and\ \bibinfo {author} {\bibfnamefont {I.~S.}\ \bibnamefont
  {Towner}},\ }\bibfield  {title} {\enquote {\bibinfo {title} {{Superallowed
  $0^+ \to 0^+$ nuclear $\beta$ decays: 2020 critical survey, with implications
  for V$_{ud}$ and CKM unitarity}},}\ }\href {\doibase
  10.1103/PhysRevC.102.045501} {\bibfield  {journal} {\bibinfo  {journal}
  {Phys. Rev. C}\ }\textbf {\bibinfo {volume} {102}},\ \bibinfo {pages}
  {045501} (\bibinfo {year} {2020})}\BibitemShut {NoStop}%
\bibitem [{\citenamefont {Cirigliano}\ \emph
  {et~al.}(2024{\natexlab{b}})\citenamefont {Cirigliano}, \citenamefont
  {Dekens}, \citenamefont {de~Vries}, \citenamefont {Gandolfi}, \citenamefont
  {Hoferichter},\ and\ \citenamefont {Mereghetti}}]{Cirigliano:2024rfk}%
  \BibitemOpen
  \bibfield  {author} {\bibinfo {author} {\bibfnamefont {V.}~\bibnamefont
  {Cirigliano}}, \bibinfo {author} {\bibfnamefont {W.}~\bibnamefont {Dekens}},
  \bibinfo {author} {\bibfnamefont {J.}~\bibnamefont {de~Vries}}, \bibinfo
  {author} {\bibfnamefont {S.}~\bibnamefont {Gandolfi}}, \bibinfo {author}
  {\bibfnamefont {M.}~\bibnamefont {Hoferichter}}, \ and\ \bibinfo {author}
  {\bibfnamefont {E.}~\bibnamefont {Mereghetti}},\ }\bibfield  {title}
  {\enquote {\bibinfo {title} {{Radiative Corrections to Superallowed
  \ensuremath{\beta} Decays in Effective Field Theory}},}\ }\href {\doibase
  10.1103/PhysRevLett.133.211801} {\bibfield  {journal} {\bibinfo  {journal}
  {Phys. Rev. Lett.}\ }\textbf {\bibinfo {volume} {133}},\ \bibinfo {pages}
  {211801} (\bibinfo {year} {2024}{\natexlab{b}})},\ \Eprint
  {http://arxiv.org/abs/2405.18469} {arXiv:2405.18469 [hep-ph]} \BibitemShut
  {NoStop}%
\bibitem [{\citenamefont {Cirigliano}\ \emph
  {et~al.}(2024{\natexlab{c}})\citenamefont {Cirigliano}, \citenamefont
  {Dekens}, \citenamefont {de~Vries}, \citenamefont {Gandolfi}, \citenamefont
  {Hoferichter},\ and\ \citenamefont {Mereghetti}}]{Cirigliano:2024msg}%
  \BibitemOpen
  \bibfield  {author} {\bibinfo {author} {\bibfnamefont {V.}~\bibnamefont
  {Cirigliano}}, \bibinfo {author} {\bibfnamefont {W.}~\bibnamefont {Dekens}},
  \bibinfo {author} {\bibfnamefont {J.}~\bibnamefont {de~Vries}}, \bibinfo
  {author} {\bibfnamefont {S.}~\bibnamefont {Gandolfi}}, \bibinfo {author}
  {\bibfnamefont {M.}~\bibnamefont {Hoferichter}}, \ and\ \bibinfo {author}
  {\bibfnamefont {E.}~\bibnamefont {Mereghetti}},\ }\bibfield  {title}
  {\enquote {\bibinfo {title} {{Ab initio electroweak corrections to
  superallowed $\beta$ decays and their impact on $V_{ud}$}},}\ }\href
  {\doibase 10.1103/PhysRevC.110.055502} {\bibfield  {journal} {\bibinfo
  {journal} {Phys. Rev. C}\ }\textbf {\bibinfo {volume} {110}},\ \bibinfo
  {pages} {055502} (\bibinfo {year} {2024}{\natexlab{c}})},\ \Eprint
  {http://arxiv.org/abs/2405.18464} {arXiv:2405.18464 [nucl-th]} \BibitemShut
  {NoStop}%
\bibitem [{\citenamefont {Alioli}\ \emph {et~al.}(2017)\citenamefont {Alioli},
  \citenamefont {Cirigliano}, \citenamefont {Dekens}, \citenamefont
  {de~Vries},\ and\ \citenamefont {Mereghetti}}]{Alioli:2017ces}%
  \BibitemOpen
  \bibfield  {author} {\bibinfo {author} {\bibfnamefont {S.}~\bibnamefont
  {Alioli}}, \bibinfo {author} {\bibfnamefont {V.}~\bibnamefont {Cirigliano}},
  \bibinfo {author} {\bibfnamefont {W.}~\bibnamefont {Dekens}}, \bibinfo
  {author} {\bibfnamefont {J.}~\bibnamefont {de~Vries}}, \ and\ \bibinfo
  {author} {\bibfnamefont {E.}~\bibnamefont {Mereghetti}},\ }\bibfield  {title}
  {\enquote {\bibinfo {title} {{Right-handed charged currents in the era of the
  Large Hadron Collider}},}\ }\href {\doibase 10.1007/JHEP05(2017)086}
  {\bibfield  {journal} {\bibinfo  {journal} {JHEP}\ }\textbf {\bibinfo
  {volume} {05}},\ \bibinfo {pages} {086} (\bibinfo {year} {2017})},\ \Eprint
  {http://arxiv.org/abs/1703.04751} {arXiv:1703.04751 [hep-ph]} \BibitemShut
  {NoStop}%
\bibitem [{\citenamefont {Cirigliano}\ \emph
  {et~al.}(2024{\natexlab{d}})\citenamefont {Cirigliano}, \citenamefont
  {Dekens}, \citenamefont {de~Vries}, \citenamefont {Mereghetti},\ and\
  \citenamefont {Tong}}]{Cirigliano:2023nol}%
  \BibitemOpen
  \bibfield  {author} {\bibinfo {author} {\bibfnamefont {V.}~\bibnamefont
  {Cirigliano}}, \bibinfo {author} {\bibfnamefont {W.}~\bibnamefont {Dekens}},
  \bibinfo {author} {\bibfnamefont {J.}~\bibnamefont {de~Vries}}, \bibinfo
  {author} {\bibfnamefont {E.}~\bibnamefont {Mereghetti}}, \ and\ \bibinfo
  {author} {\bibfnamefont {T.}~\bibnamefont {Tong}},\ }\bibfield  {title}
  {\enquote {\bibinfo {title} {{Anomalies in global SMEFT analyses. A case
  study of first-row CKM unitarity}},}\ }\href {\doibase
  10.1007/JHEP03(2024)033} {\bibfield  {journal} {\bibinfo  {journal} {JHEP}\
  }\textbf {\bibinfo {volume} {03}},\ \bibinfo {pages} {033} (\bibinfo {year}
  {2024}{\natexlab{d}})},\ \Eprint {http://arxiv.org/abs/2311.00021}
  {arXiv:2311.00021 [hep-ph]} \BibitemShut {NoStop}%
\bibitem [{\citenamefont {Cirigliano}\ \emph
  {et~al.}(2022{\natexlab{b}})\citenamefont {Cirigliano}, \citenamefont
  {D\'\i{}az-Calder\'on}, \citenamefont {Falkowski}, \citenamefont
  {Gonz\'alez-Alonso},\ and\ \citenamefont
  {Rodr\'\i{}guez-S\'anchez}}]{Cirigliano:2021yto}%
  \BibitemOpen
  \bibfield  {author} {\bibinfo {author} {\bibfnamefont {V.}~\bibnamefont
  {Cirigliano}}, \bibinfo {author} {\bibfnamefont {D.}~\bibnamefont
  {D\'\i{}az-Calder\'on}}, \bibinfo {author} {\bibfnamefont {A.}~\bibnamefont
  {Falkowski}}, \bibinfo {author} {\bibfnamefont {M.}~\bibnamefont
  {Gonz\'alez-Alonso}}, \ and\ \bibinfo {author} {\bibfnamefont
  {A.}~\bibnamefont {Rodr\'\i{}guez-S\'anchez}},\ }\bibfield  {title} {\enquote
  {\bibinfo {title} {{Semileptonic tau decays beyond the Standard Model}},}\
  }\href {\doibase 10.1007/JHEP04(2022)152} {\bibfield  {journal} {\bibinfo
  {journal} {JHEP}\ }\textbf {\bibinfo {volume} {04}},\ \bibinfo {pages} {152}
  (\bibinfo {year} {2022}{\natexlab{b}})},\ \Eprint
  {http://arxiv.org/abs/2112.02087} {arXiv:2112.02087 [hep-ph]} \BibitemShut
  {NoStop}%
\bibitem [{\citenamefont {Edwards}\ \emph {et~al.}(2006)\citenamefont
  {Edwards}, \citenamefont {Fleming}, \citenamefont {H{\"a}gler}, \citenamefont
  {Negele}, \citenamefont {Orginos}, \citenamefont {Pochinsky}, \citenamefont
  {Renner}, \citenamefont {Richards},\ and\ \citenamefont
  {Schroers}}]{Edwards:2005ym}%
  \BibitemOpen
  \bibfield  {author} {\bibinfo {author} {\bibfnamefont {R.~G.}\ \bibnamefont
  {Edwards}}, \bibinfo {author} {\bibfnamefont {G.~T.}\ \bibnamefont
  {Fleming}}, \bibinfo {author} {\bibfnamefont {P.}~\bibnamefont {H{\"a}gler}},
  \bibinfo {author} {\bibfnamefont {J.~W.}\ \bibnamefont {Negele}}, \bibinfo
  {author} {\bibfnamefont {K.}~\bibnamefont {Orginos}}, \bibinfo {author}
  {\bibfnamefont {A.~V.}\ \bibnamefont {Pochinsky}}, \bibinfo {author}
  {\bibfnamefont {D.~B.}\ \bibnamefont {Renner}}, \bibinfo {author}
  {\bibfnamefont {D.~G.}\ \bibnamefont {Richards}}, \ and\ \bibinfo {author}
  {\bibfnamefont {W.}~\bibnamefont {Schroers}} (\bibinfo {collaboration}
  {LHPC}),\ }\bibfield  {title} {\enquote {\bibinfo {title} {{The Nucleon axial
  charge in full lattice QCD}},}\ }\href {\doibase
  10.1103/PhysRevLett.96.052001} {\bibfield  {journal} {\bibinfo  {journal}
  {Phys. Rev. Lett.}\ }\textbf {\bibinfo {volume} {96}},\ \bibinfo {pages}
  {052001} (\bibinfo {year} {2006})},\ \Eprint
  {http://arxiv.org/abs/hep-lat/0510062} {arXiv:hep-lat/0510062 [hep-lat]}
  \BibitemShut {NoStop}%
\bibitem [{\citenamefont {Drischler}\ \emph {et~al.}(2021)\citenamefont
  {Drischler}, \citenamefont {Haxton}, \citenamefont {McElvain}, \citenamefont
  {Mereghetti}, \citenamefont {Nicholson}, \citenamefont {Vranas},\ and\
  \citenamefont {Walker-Loud}}]{Drischler:2019xuo}%
  \BibitemOpen
  \bibfield  {author} {\bibinfo {author} {\bibfnamefont {C.}~\bibnamefont
  {Drischler}}, \bibinfo {author} {\bibfnamefont {W.}~\bibnamefont {Haxton}},
  \bibinfo {author} {\bibfnamefont {K.}~\bibnamefont {McElvain}}, \bibinfo
  {author} {\bibfnamefont {E.}~\bibnamefont {Mereghetti}}, \bibinfo {author}
  {\bibfnamefont {A.}~\bibnamefont {Nicholson}}, \bibinfo {author}
  {\bibfnamefont {P.}~\bibnamefont {Vranas}}, \ and\ \bibinfo {author}
  {\bibfnamefont {A.}~\bibnamefont {Walker-Loud}},\ }\bibfield  {title}
  {\enquote {\bibinfo {title} {{Towards grounding nuclear physics in QCD}},}\
  }\href {\doibase 10.1016/j.ppnp.2021.103888} {\bibfield  {journal} {\bibinfo
  {journal} {Prog. Part. Nucl. Phys.}\ }\textbf {\bibinfo {volume} {121}},\
  \bibinfo {pages} {103888} (\bibinfo {year} {2021})},\ \Eprint
  {http://arxiv.org/abs/1910.07961} {arXiv:1910.07961 [nucl-th]} \BibitemShut
  {NoStop}%
\bibitem [{\citenamefont {Gasser}\ and\ \citenamefont
  {Leutwyler}(1988)}]{Gasser:1987zq}%
  \BibitemOpen
  \bibfield  {author} {\bibinfo {author} {\bibfnamefont {J.}~\bibnamefont
  {Gasser}}\ and\ \bibinfo {author} {\bibfnamefont {H.}~\bibnamefont
  {Leutwyler}},\ }\bibfield  {title} {\enquote {\bibinfo {title}
  {{Spontaneously Broken Symmetries: Effective Lagrangians at Finite
  Volume}},}\ }\href {\doibase 10.1016/0550-3213(88)90107-1} {\bibfield
  {journal} {\bibinfo  {journal} {Nucl. Phys. B}\ }\textbf {\bibinfo {volume}
  {307}},\ \bibinfo {pages} {763--778} (\bibinfo {year} {1988})}\BibitemShut
  {NoStop}%
\bibitem [{\citenamefont {Jaffe}(2002)}]{Jaffe:2001eb}%
  \BibitemOpen
  \bibfield  {author} {\bibinfo {author} {\bibfnamefont {R.~L.}\ \bibnamefont
  {Jaffe}},\ }\bibfield  {title} {\enquote {\bibinfo {title} {{Delocalization
  of the axial charge in the chiral limit}},}\ }\href {\doibase
  10.1016/S0370-2693(02)01242-X} {\bibfield  {journal} {\bibinfo  {journal}
  {Phys. Lett. B}\ }\textbf {\bibinfo {volume} {529}},\ \bibinfo {pages}
  {105--110} (\bibinfo {year} {2002})},\ \Eprint
  {http://arxiv.org/abs/hep-ph/0108015} {arXiv:hep-ph/0108015 [hep-ph]}
  \BibitemShut {NoStop}%
\bibitem [{\citenamefont {Cohen}(2002)}]{Cohen:2001bg}%
  \BibitemOpen
  \bibfield  {author} {\bibinfo {author} {\bibfnamefont {T.~D.}\ \bibnamefont
  {Cohen}},\ }\bibfield  {title} {\enquote {\bibinfo {title} {{The Extraction
  of $g_A$ from finite volume systems: The Long and short of it}},}\ }\href
  {\doibase 10.1016/S0370-2693(02)01241-8} {\bibfield  {journal} {\bibinfo
  {journal} {Phys. Lett. B}\ }\textbf {\bibinfo {volume} {529}},\ \bibinfo
  {pages} {50--56} (\bibinfo {year} {2002})},\ \Eprint
  {http://arxiv.org/abs/hep-lat/0112014} {arXiv:hep-lat/0112014 [hep-lat]}
  \BibitemShut {NoStop}%
\bibitem [{\citenamefont {Yamazaki}\ \emph {et~al.}(2008)\citenamefont
  {Yamazaki}, \citenamefont {Aoki}, \citenamefont {Blum}, \citenamefont {Lin},
  \citenamefont {Lin}, \citenamefont {Ohta}, \citenamefont {Sasaki},
  \citenamefont {Tweedie},\ and\ \citenamefont {Zanotti}}]{Yamazaki:2008py}%
  \BibitemOpen
  \bibfield  {author} {\bibinfo {author} {\bibfnamefont {T.}~\bibnamefont
  {Yamazaki}}, \bibinfo {author} {\bibfnamefont {Y.}~\bibnamefont {Aoki}},
  \bibinfo {author} {\bibfnamefont {T.}~\bibnamefont {Blum}}, \bibinfo {author}
  {\bibfnamefont {H.~W.}\ \bibnamefont {Lin}}, \bibinfo {author} {\bibfnamefont
  {M.~F.}\ \bibnamefont {Lin}}, \bibinfo {author} {\bibfnamefont
  {S.}~\bibnamefont {Ohta}}, \bibinfo {author} {\bibfnamefont {S.}~\bibnamefont
  {Sasaki}}, \bibinfo {author} {\bibfnamefont {R.~J.}\ \bibnamefont {Tweedie}},
  \ and\ \bibinfo {author} {\bibfnamefont {J.~M.}\ \bibnamefont {Zanotti}}
  (\bibinfo {collaboration} {RBC+UKQCD}),\ }\bibfield  {title} {\enquote
  {\bibinfo {title} {{Nucleon axial charge in 2+1 flavor dynamical lattice QCD
  with domain wall fermions}},}\ }\href {\doibase
  10.1103/PhysRevLett.100.171602} {\bibfield  {journal} {\bibinfo  {journal}
  {Phys. Rev. Lett.}\ }\textbf {\bibinfo {volume} {100}},\ \bibinfo {pages}
  {171602} (\bibinfo {year} {2008})},\ \Eprint {http://arxiv.org/abs/0801.4016}
  {arXiv:0801.4016 [hep-lat]} \BibitemShut {NoStop}%
\bibitem [{\citenamefont {Yamazaki}\ \emph {et~al.}(2009)\citenamefont
  {Yamazaki}, \citenamefont {Aoki}, \citenamefont {Blum}, \citenamefont {Lin},
  \citenamefont {Ohta}, \citenamefont {Sasaki}, \citenamefont {Tweedie},\ and\
  \citenamefont {Zanotti}}]{Yamazaki:2009zq}%
  \BibitemOpen
  \bibfield  {author} {\bibinfo {author} {\bibfnamefont {T.}~\bibnamefont
  {Yamazaki}}, \bibinfo {author} {\bibfnamefont {Y.}~\bibnamefont {Aoki}},
  \bibinfo {author} {\bibfnamefont {T.}~\bibnamefont {Blum}}, \bibinfo {author}
  {\bibfnamefont {H.-W.}\ \bibnamefont {Lin}}, \bibinfo {author} {\bibfnamefont
  {S.}~\bibnamefont {Ohta}}, \bibinfo {author} {\bibfnamefont {S.}~\bibnamefont
  {Sasaki}}, \bibinfo {author} {\bibfnamefont {R.}~\bibnamefont {Tweedie}}, \
  and\ \bibinfo {author} {\bibfnamefont {J.}~\bibnamefont {Zanotti}},\
  }\bibfield  {title} {\enquote {\bibinfo {title} {{Nucleon form factors with
  2+1 flavor dynamical domain-wall fermions}},}\ }\href {\doibase
  10.1103/PhysRevD.79.114505} {\bibfield  {journal} {\bibinfo  {journal} {Phys.
  Rev. D}\ }\textbf {\bibinfo {volume} {79}},\ \bibinfo {pages} {114505}
  (\bibinfo {year} {2009})},\ \Eprint {http://arxiv.org/abs/0904.2039}
  {arXiv:0904.2039 [hep-lat]} \BibitemShut {NoStop}%
\bibitem [{\citenamefont {Capitani}\ \emph {et~al.}(2012)\citenamefont
  {Capitani}, \citenamefont {Della~Morte}, \citenamefont {von Hippel},
  \citenamefont {Jager}, \citenamefont {Juttner}, \citenamefont {Knippschild},
  \citenamefont {Meyer},\ and\ \citenamefont {Wittig}}]{Capitani:2012gj}%
  \BibitemOpen
  \bibfield  {author} {\bibinfo {author} {\bibfnamefont {S.}~\bibnamefont
  {Capitani}}, \bibinfo {author} {\bibfnamefont {M.}~\bibnamefont
  {Della~Morte}}, \bibinfo {author} {\bibfnamefont {G.}~\bibnamefont {von
  Hippel}}, \bibinfo {author} {\bibfnamefont {B.}~\bibnamefont {Jager}},
  \bibinfo {author} {\bibfnamefont {A.}~\bibnamefont {Juttner}}, \bibinfo
  {author} {\bibfnamefont {B.}~\bibnamefont {Knippschild}}, \bibinfo {author}
  {\bibfnamefont {H.~B.}\ \bibnamefont {Meyer}}, \ and\ \bibinfo {author}
  {\bibfnamefont {H.}~\bibnamefont {Wittig}},\ }\bibfield  {title} {\enquote
  {\bibinfo {title} {{The nucleon axial charge from lattice QCD with controlled
  errors}},}\ }\href {\doibase 10.1103/PhysRevD.86.074502} {\bibfield
  {journal} {\bibinfo  {journal} {Phys. Rev. D}\ }\textbf {\bibinfo {volume}
  {86}},\ \bibinfo {pages} {074502} (\bibinfo {year} {2012})},\ \Eprint
  {http://arxiv.org/abs/1205.0180} {arXiv:1205.0180 [hep-lat]} \BibitemShut
  {NoStop}%
\bibitem [{\citenamefont {Walker-Loud}\ \emph {et~al.}(2020)\citenamefont
  {Walker-Loud} \emph {et~al.}}]{Walker-Loud:2019cif}%
  \BibitemOpen
  \bibfield  {author} {\bibinfo {author} {\bibfnamefont {A.}~\bibnamefont
  {Walker-Loud}} \emph {et~al.},\ }\bibfield  {title} {\enquote {\bibinfo
  {title} {{Lattice QCD Determination of $g_A$}},}\ }\href {\doibase
  10.22323/1.317.0020} {\bibfield  {journal} {\bibinfo  {journal} {PoS}\
  }\textbf {\bibinfo {volume} {CD2018}},\ \bibinfo {pages} {020} (\bibinfo
  {year} {2020})},\ \Eprint {http://arxiv.org/abs/1912.08321} {arXiv:1912.08321
  [hep-lat]} \BibitemShut {NoStop}%
\bibitem [{\citenamefont {Beane}\ and\ \citenamefont
  {Savage}(2004)}]{Beane:2004rf}%
  \BibitemOpen
  \bibfield  {author} {\bibinfo {author} {\bibfnamefont {S.~R.}\ \bibnamefont
  {Beane}}\ and\ \bibinfo {author} {\bibfnamefont {M.~J.}\ \bibnamefont
  {Savage}},\ }\bibfield  {title} {\enquote {\bibinfo {title} {{Baryon axial
  charge in a finite volume}},}\ }\href {\doibase 10.1103/PhysRevD.70.074029}
  {\bibfield  {journal} {\bibinfo  {journal} {Phys. Rev. D}\ }\textbf {\bibinfo
  {volume} {70}},\ \bibinfo {pages} {074029} (\bibinfo {year} {2004})},\
  \Eprint {http://arxiv.org/abs/hep-ph/0404131} {arXiv:hep-ph/0404131 [hep-ph]}
  \BibitemShut {NoStop}%
\bibitem [{\citenamefont {Walker-Loud}\ \emph {et~al.}(2009)\citenamefont
  {Walker-Loud} \emph {et~al.}}]{Walker-Loud:2008rui}%
  \BibitemOpen
  \bibfield  {author} {\bibinfo {author} {\bibfnamefont {A.}~\bibnamefont
  {Walker-Loud}} \emph {et~al.},\ }\bibfield  {title} {\enquote {\bibinfo
  {title} {{Light hadron spectroscopy using domain wall valence quarks on an
  Asqtad sea}},}\ }\href {\doibase 10.1103/PhysRevD.79.054502} {\bibfield
  {journal} {\bibinfo  {journal} {Phys. Rev. D}\ }\textbf {\bibinfo {volume}
  {79}},\ \bibinfo {pages} {054502} (\bibinfo {year} {2009})},\ \Eprint
  {http://arxiv.org/abs/0806.4549} {arXiv:0806.4549 [hep-lat]} \BibitemShut
  {NoStop}%
\bibitem [{\citenamefont {Jenkins}(1996)}]{Jenkins:1995gc}%
  \BibitemOpen
  \bibfield  {author} {\bibinfo {author} {\bibfnamefont {E.~E.}\ \bibnamefont
  {Jenkins}},\ }\bibfield  {title} {\enquote {\bibinfo {title} {{Chiral
  Lagrangian for baryons in the $1/N_c$ expansion}},}\ }\href {\doibase
  10.1103/PhysRevD.53.2625} {\bibfield  {journal} {\bibinfo  {journal} {Phys.
  Rev. D}\ }\textbf {\bibinfo {volume} {53}},\ \bibinfo {pages} {2625--2644}
  (\bibinfo {year} {1996})},\ \Eprint {http://arxiv.org/abs/hep-ph/9509433}
  {arXiv:hep-ph/9509433 [hep-ph]} \BibitemShut {NoStop}%
\bibitem [{\citenamefont {Hemmert}\ \emph {et~al.}(1997)\citenamefont
  {Hemmert}, \citenamefont {Holstein},\ and\ \citenamefont
  {Kambor}}]{Hemmert:1996rw}%
  \BibitemOpen
  \bibfield  {author} {\bibinfo {author} {\bibfnamefont {T.~R.}\ \bibnamefont
  {Hemmert}}, \bibinfo {author} {\bibfnamefont {B.~R.}\ \bibnamefont
  {Holstein}}, \ and\ \bibinfo {author} {\bibfnamefont {J.}~\bibnamefont
  {Kambor}},\ }\bibfield  {title} {\enquote {\bibinfo {title} {{$\Delta(1232)$
  and the polarizabilities of the nucleon}},}\ }\href {\doibase
  10.1103/PhysRevD.55.5598} {\bibfield  {journal} {\bibinfo  {journal} {Phys.
  Rev. D}\ }\textbf {\bibinfo {volume} {55}},\ \bibinfo {pages} {5598--5612}
  (\bibinfo {year} {1997})},\ \Eprint {http://arxiv.org/abs/hep-ph/9612374}
  {arXiv:hep-ph/9612374} \BibitemShut {NoStop}%
\bibitem [{\citenamefont {Flores-Mendieta}\ \emph {et~al.}(2000)\citenamefont
  {Flores-Mendieta}, \citenamefont {Hofmann}, \citenamefont {Jenkins},\ and\
  \citenamefont {Manohar}}]{Flores-Mendieta:2000ljq}%
  \BibitemOpen
  \bibfield  {author} {\bibinfo {author} {\bibfnamefont {R.}~\bibnamefont
  {Flores-Mendieta}}, \bibinfo {author} {\bibfnamefont {C.~P.}\ \bibnamefont
  {Hofmann}}, \bibinfo {author} {\bibfnamefont {E.~E.}\ \bibnamefont
  {Jenkins}}, \ and\ \bibinfo {author} {\bibfnamefont {A.~V.}\ \bibnamefont
  {Manohar}},\ }\bibfield  {title} {\enquote {\bibinfo {title} {{On the
  structure of large $N_c$ cancellations in baryon chiral perturbation
  theory}},}\ }\href {\doibase 10.1103/PhysRevD.62.034001} {\bibfield
  {journal} {\bibinfo  {journal} {Phys. Rev. D}\ }\textbf {\bibinfo {volume}
  {62}},\ \bibinfo {pages} {034001} (\bibinfo {year} {2000})},\ \Eprint
  {http://arxiv.org/abs/hep-ph/0001218} {arXiv:hep-ph/0001218} \BibitemShut
  {NoStop}%
\bibitem [{\citenamefont {Jenkins}\ \emph {et~al.}(2002)\citenamefont
  {Jenkins}, \citenamefont {Ji},\ and\ \citenamefont
  {Manohar}}]{Jenkins:2002rj}%
  \BibitemOpen
  \bibfield  {author} {\bibinfo {author} {\bibfnamefont {E.~E.}\ \bibnamefont
  {Jenkins}}, \bibinfo {author} {\bibfnamefont {X.-d.}\ \bibnamefont {Ji}}, \
  and\ \bibinfo {author} {\bibfnamefont {A.~V.}\ \bibnamefont {Manohar}},\
  }\bibfield  {title} {\enquote {\bibinfo {title} {{$\Delta\to N\gamma$ in
  large-$N_c$ QCD}},}\ }\href {\doibase 10.1103/PhysRevLett.89.242001}
  {\bibfield  {journal} {\bibinfo  {journal} {Phys. Rev. Lett.}\ }\textbf
  {\bibinfo {volume} {89}},\ \bibinfo {pages} {242001} (\bibinfo {year}
  {2002})},\ \Eprint {http://arxiv.org/abs/hep-ph/0207092}
  {arXiv:hep-ph/0207092} \BibitemShut {NoStop}%
\bibitem [{\citenamefont {Siemens}\ \emph {et~al.}(2017)\citenamefont
  {Siemens}, \citenamefont {Ruiz~de Elvira}, \citenamefont {Epelbaum},
  \citenamefont {Hoferichter}, \citenamefont {Krebs}, \citenamefont {Kubis},\
  and\ \citenamefont {Mei\ss{}ner}}]{Siemens:2016jwj}%
  \BibitemOpen
  \bibfield  {author} {\bibinfo {author} {\bibfnamefont {D.}~\bibnamefont
  {Siemens}}, \bibinfo {author} {\bibfnamefont {J.}~\bibnamefont {Ruiz~de
  Elvira}}, \bibinfo {author} {\bibfnamefont {E.}~\bibnamefont {Epelbaum}},
  \bibinfo {author} {\bibfnamefont {M.}~\bibnamefont {Hoferichter}}, \bibinfo
  {author} {\bibfnamefont {H.}~\bibnamefont {Krebs}}, \bibinfo {author}
  {\bibfnamefont {B.}~\bibnamefont {Kubis}}, \ and\ \bibinfo {author}
  {\bibfnamefont {U.-G.}\ \bibnamefont {Mei\ss{}ner}},\ }\bibfield  {title}
  {\enquote {\bibinfo {title} {{Reconciling threshold and subthreshold
  expansions for pion\textendash{}nucleon scattering}},}\ }\href {\doibase
  10.1016/j.physletb.2017.04.039} {\bibfield  {journal} {\bibinfo  {journal}
  {Phys. Lett. B}\ }\textbf {\bibinfo {volume} {770}},\ \bibinfo {pages}
  {27--34} (\bibinfo {year} {2017})},\ \Eprint
  {http://arxiv.org/abs/1610.08978} {arXiv:1610.08978 [nucl-th]} \BibitemShut
  {NoStop}%
\bibitem [{\citenamefont {Hemmert}\ \emph {et~al.}(1998)\citenamefont
  {Hemmert}, \citenamefont {Holstein},\ and\ \citenamefont
  {Kambor}}]{Hemmert:1997ye}%
  \BibitemOpen
  \bibfield  {author} {\bibinfo {author} {\bibfnamefont {T.~R.}\ \bibnamefont
  {Hemmert}}, \bibinfo {author} {\bibfnamefont {B.~R.}\ \bibnamefont
  {Holstein}}, \ and\ \bibinfo {author} {\bibfnamefont {J.}~\bibnamefont
  {Kambor}},\ }\bibfield  {title} {\enquote {\bibinfo {title} {{Chiral
  Lagrangians and $\Delta(1232)$ interactions: Formalism}},}\ }\href {\doibase
  10.1088/0954-3899/24/10/003} {\bibfield  {journal} {\bibinfo  {journal} {J.
  Phys. G}\ }\textbf {\bibinfo {volume} {24}},\ \bibinfo {pages} {1831--1859}
  (\bibinfo {year} {1998})},\ \Eprint {http://arxiv.org/abs/hep-ph/9712496}
  {arXiv:hep-ph/9712496} \BibitemShut {NoStop}%
\bibitem [{\citenamefont {Dashen}\ and\ \citenamefont
  {Manohar}(1993{\natexlab{a}})}]{Dashen:1993as}%
  \BibitemOpen
  \bibfield  {author} {\bibinfo {author} {\bibfnamefont {R.~F.}\ \bibnamefont
  {Dashen}}\ and\ \bibinfo {author} {\bibfnamefont {A.~V.}\ \bibnamefont
  {Manohar}},\ }\bibfield  {title} {\enquote {\bibinfo {title} {{Baryon - pion
  couplings from large-$N_c$ QCD}},}\ }\href {\doibase
  10.1016/0370-2693(93)91635-Z} {\bibfield  {journal} {\bibinfo  {journal}
  {Phys. Lett. B}\ }\textbf {\bibinfo {volume} {315}},\ \bibinfo {pages}
  {425--430} (\bibinfo {year} {1993}{\natexlab{a}})},\ \Eprint
  {http://arxiv.org/abs/hep-ph/9307241} {arXiv:hep-ph/9307241} \BibitemShut
  {NoStop}%
\bibitem [{\citenamefont {Dashen}\ and\ \citenamefont
  {Manohar}(1993{\natexlab{b}})}]{Dashen:1993ac}%
  \BibitemOpen
  \bibfield  {author} {\bibinfo {author} {\bibfnamefont {R.~F.}\ \bibnamefont
  {Dashen}}\ and\ \bibinfo {author} {\bibfnamefont {A.~V.}\ \bibnamefont
  {Manohar}},\ }\bibfield  {title} {\enquote {\bibinfo {title} {{$1/N_c$
  corrections to the baryon axial currents in QCD}},}\ }\href {\doibase
  10.1016/0370-2693(93)91637-3} {\bibfield  {journal} {\bibinfo  {journal}
  {Phys. Lett. B}\ }\textbf {\bibinfo {volume} {315}},\ \bibinfo {pages}
  {438--440} (\bibinfo {year} {1993}{\natexlab{b}})},\ \Eprint
  {http://arxiv.org/abs/hep-ph/9307242} {arXiv:hep-ph/9307242 [hep-ph]}
  \BibitemShut {NoStop}%
\bibitem [{\citenamefont {Calle~Cordon}\ and\ \citenamefont
  {Goity}(2013)}]{CalleCordon:2012xz}%
  \BibitemOpen
  \bibfield  {author} {\bibinfo {author} {\bibfnamefont {A.}~\bibnamefont
  {Calle~Cordon}}\ and\ \bibinfo {author} {\bibfnamefont {J.~L.}\ \bibnamefont
  {Goity}},\ }\bibfield  {title} {\enquote {\bibinfo {title} {{Baryon Masses
  and Axial Couplings in the Combined 1/$N_c$ and Chiral Expansions}},}\ }\href
  {\doibase 10.1103/PhysRevD.87.016019} {\bibfield  {journal} {\bibinfo
  {journal} {Phys. Rev. D}\ }\textbf {\bibinfo {volume} {87}},\ \bibinfo
  {pages} {016019} (\bibinfo {year} {2013})},\ \Eprint
  {http://arxiv.org/abs/1210.2364} {arXiv:1210.2364 [nucl-th]} \BibitemShut
  {NoStop}%
\bibitem [{\citenamefont {Hemmert}\ \emph {et~al.}(2003)\citenamefont
  {Hemmert}, \citenamefont {Procura},\ and\ \citenamefont
  {Weise}}]{Hemmert:2003cb}%
  \BibitemOpen
  \bibfield  {author} {\bibinfo {author} {\bibfnamefont {T.~R.}\ \bibnamefont
  {Hemmert}}, \bibinfo {author} {\bibfnamefont {M.}~\bibnamefont {Procura}}, \
  and\ \bibinfo {author} {\bibfnamefont {W.}~\bibnamefont {Weise}},\ }\bibfield
   {title} {\enquote {\bibinfo {title} {{Quark mass dependence of the nucleon
  axial vector coupling constant}},}\ }\href {\doibase
  10.1103/PhysRevD.68.075009} {\bibfield  {journal} {\bibinfo  {journal} {Phys.
  Rev. D}\ }\textbf {\bibinfo {volume} {68}},\ \bibinfo {pages} {075009}
  (\bibinfo {year} {2003})},\ \Eprint {http://arxiv.org/abs/hep-lat/0303002}
  {arXiv:hep-lat/0303002 [hep-lat]} \BibitemShut {NoStop}%
\bibitem [{\citenamefont {Kambor}\ and\ \citenamefont {Moj{\v z}i{\v
  s}}(1999)}]{Kambor:1998pi}%
  \BibitemOpen
  \bibfield  {author} {\bibinfo {author} {\bibfnamefont {J.}~\bibnamefont
  {Kambor}}\ and\ \bibinfo {author} {\bibfnamefont {M.}~\bibnamefont {Moj{\v
  z}i{\v s}}},\ }\bibfield  {title} {\enquote {\bibinfo {title} {{Field
  redefinitions and wave function renormalization to ${\mathcal O}(p^4)$ in
  heavy baryon chiral perturbation theory}},}\ }\href {\doibase
  10.1088/1126-6708/1999/04/031} {\bibfield  {journal} {\bibinfo  {journal}
  {JHEP}\ }\textbf {\bibinfo {volume} {04}},\ \bibinfo {pages} {031} (\bibinfo
  {year} {1999})},\ \Eprint {http://arxiv.org/abs/hep-ph/9901235}
  {arXiv:hep-ph/9901235 [hep-ph]} \BibitemShut {NoStop}%
\bibitem [{\citenamefont {Becher}\ and\ \citenamefont
  {Leutwyler}(2001)}]{Becher:2001hv}%
  \BibitemOpen
  \bibfield  {author} {\bibinfo {author} {\bibfnamefont {T.}~\bibnamefont
  {Becher}}\ and\ \bibinfo {author} {\bibfnamefont {H.}~\bibnamefont
  {Leutwyler}},\ }\bibfield  {title} {\enquote {\bibinfo {title} {{Low energy
  analysis of $\pi N \to\pi N$}},}\ }\href {\doibase
  10.1088/1126-6708/2001/06/017} {\bibfield  {journal} {\bibinfo  {journal}
  {JHEP}\ }\textbf {\bibinfo {volume} {06}},\ \bibinfo {pages} {017} (\bibinfo
  {year} {2001})},\ \Eprint {http://arxiv.org/abs/hep-ph/0103263}
  {arXiv:hep-ph/0103263} \BibitemShut {NoStop}%
\bibitem [{\citenamefont {Gasser}\ \emph {et~al.}(2002)\citenamefont {Gasser},
  \citenamefont {Ivanov}, \citenamefont {Lipartia}, \citenamefont {Moj{\v
  z}i{\v s}},\ and\ \citenamefont {Rusetsky}}]{Gasser:2002am}%
  \BibitemOpen
  \bibfield  {author} {\bibinfo {author} {\bibfnamefont {J.}~\bibnamefont
  {Gasser}}, \bibinfo {author} {\bibfnamefont {M.~A.}\ \bibnamefont {Ivanov}},
  \bibinfo {author} {\bibfnamefont {E.}~\bibnamefont {Lipartia}}, \bibinfo
  {author} {\bibfnamefont {M.}~\bibnamefont {Moj{\v z}i{\v s}}}, \ and\
  \bibinfo {author} {\bibfnamefont {A.}~\bibnamefont {Rusetsky}},\ }\bibfield
  {title} {\enquote {\bibinfo {title} {{Ground state energy of pionic hydrogen
  to one loop}},}\ }\href {\doibase 10.1007/s10052-002-1013-z} {\bibfield
  {journal} {\bibinfo  {journal} {Eur. Phys. J. C}\ }\textbf {\bibinfo {volume}
  {26}},\ \bibinfo {pages} {13--34} (\bibinfo {year} {2002})},\ \Eprint
  {http://arxiv.org/abs/hep-ph/0206068} {arXiv:hep-ph/0206068} \BibitemShut
  {NoStop}%
\bibitem [{\citenamefont {Dashen}\ \emph {et~al.}(1994)\citenamefont {Dashen},
  \citenamefont {Jenkins},\ and\ \citenamefont {Manohar}}]{Dashen:1993jt}%
  \BibitemOpen
  \bibfield  {author} {\bibinfo {author} {\bibfnamefont {R.~F.}\ \bibnamefont
  {Dashen}}, \bibinfo {author} {\bibfnamefont {E.~E.}\ \bibnamefont {Jenkins}},
  \ and\ \bibinfo {author} {\bibfnamefont {A.~V.}\ \bibnamefont {Manohar}},\
  }\bibfield  {title} {\enquote {\bibinfo {title} {{The $1/N_c$ expansion for
  baryons}},}\ }\href {\doibase 10.1103/PhysRevD.51.2489} {\bibfield  {journal}
  {\bibinfo  {journal} {Phys. Rev. D}\ }\textbf {\bibinfo {volume} {49}},\
  \bibinfo {pages} {4713} (\bibinfo {year} {1994})},\ \bibinfo {note}
  {[Erratum: Phys. Rev. D {\bf 51}, 2489 (1995)]},\ \Eprint
  {http://arxiv.org/abs/hep-ph/9310379} {arXiv:hep-ph/9310379} \BibitemShut
  {NoStop}%
\bibitem [{\citenamefont {Dashen}\ \emph {et~al.}(1995)\citenamefont {Dashen},
  \citenamefont {Jenkins},\ and\ \citenamefont {Manohar}}]{Dashen:1994qi}%
  \BibitemOpen
  \bibfield  {author} {\bibinfo {author} {\bibfnamefont {R.~F.}\ \bibnamefont
  {Dashen}}, \bibinfo {author} {\bibfnamefont {E.~E.}\ \bibnamefont {Jenkins}},
  \ and\ \bibinfo {author} {\bibfnamefont {A.~V.}\ \bibnamefont {Manohar}},\
  }\bibfield  {title} {\enquote {\bibinfo {title} {{Spin flavor structure of
  large-$N_c$ baryons}},}\ }\href {\doibase 10.1103/PhysRevD.51.3697}
  {\bibfield  {journal} {\bibinfo  {journal} {Phys. Rev. D}\ }\textbf {\bibinfo
  {volume} {51}},\ \bibinfo {pages} {3697--3727} (\bibinfo {year} {1995})},\
  \Eprint {http://arxiv.org/abs/hep-ph/9411234} {arXiv:hep-ph/9411234}
  \BibitemShut {NoStop}%
\bibitem [{\citenamefont {Karl}\ and\ \citenamefont
  {Paton}(1984)}]{Karl:1984cz}%
  \BibitemOpen
  \bibfield  {author} {\bibinfo {author} {\bibfnamefont {G.}~\bibnamefont
  {Karl}}\ and\ \bibinfo {author} {\bibfnamefont {J.~E.}\ \bibnamefont
  {Paton}},\ }\bibfield  {title} {\enquote {\bibinfo {title} {{Naive Quark
  Model for an Arbitrary Number of Colors}},}\ }\href {\doibase
  10.1103/PhysRevD.30.238} {\bibfield  {journal} {\bibinfo  {journal} {Phys.
  Rev. D}\ }\textbf {\bibinfo {volume} {30}},\ \bibinfo {pages} {238} (\bibinfo
  {year} {1984})}\BibitemShut {NoStop}%
\bibitem [{\citenamefont {Jenkins}(1998)}]{Jenkins:1998wy}%
  \BibitemOpen
  \bibfield  {author} {\bibinfo {author} {\bibfnamefont {E.~E.}\ \bibnamefont
  {Jenkins}},\ }\bibfield  {title} {\enquote {\bibinfo {title} {{Large-$N_c$
  baryons}},}\ }\href {\doibase 10.1146/annurev.nucl.48.1.81} {\bibfield
  {journal} {\bibinfo  {journal} {Ann. Rev. Nucl. Part. Sci.}\ }\textbf
  {\bibinfo {volume} {48}},\ \bibinfo {pages} {81--119} (\bibinfo {year}
  {1998})},\ \Eprint {http://arxiv.org/abs/hep-ph/9803349}
  {arXiv:hep-ph/9803349} \BibitemShut {NoStop}%
\bibitem [{\citenamefont {Flores-Mendieta}\ and\ \citenamefont
  {Hofmann}(2006)}]{Flores-Mendieta:2006ojy}%
  \BibitemOpen
  \bibfield  {author} {\bibinfo {author} {\bibfnamefont {R.}~\bibnamefont
  {Flores-Mendieta}}\ and\ \bibinfo {author} {\bibfnamefont {C.~P.}\
  \bibnamefont {Hofmann}},\ }\bibfield  {title} {\enquote {\bibinfo {title}
  {{Renormalization of the baryon axial vector current in large-$N_c$ chiral
  perturbation theory}},}\ }\href {\doibase 10.1103/PhysRevD.74.094001}
  {\bibfield  {journal} {\bibinfo  {journal} {Phys. Rev. D}\ }\textbf {\bibinfo
  {volume} {74}},\ \bibinfo {pages} {094001} (\bibinfo {year} {2006})},\
  \Eprint {http://arxiv.org/abs/hep-ph/0609120} {arXiv:hep-ph/0609120}
  \BibitemShut {NoStop}%
\bibitem [{\citenamefont {Flores-Mendieta}\ \emph {et~al.}(2012)\citenamefont
  {Flores-Mendieta}, \citenamefont {Hernandez-Ruiz},\ and\ \citenamefont
  {Hofmann}}]{Flores-Mendieta:2012fxp}%
  \BibitemOpen
  \bibfield  {author} {\bibinfo {author} {\bibfnamefont {R.}~\bibnamefont
  {Flores-Mendieta}}, \bibinfo {author} {\bibfnamefont {M.~A.}\ \bibnamefont
  {Hernandez-Ruiz}}, \ and\ \bibinfo {author} {\bibfnamefont {C.~P.}\
  \bibnamefont {Hofmann}},\ }\bibfield  {title} {\enquote {\bibinfo {title}
  {{Renormalization of the baryon axial vector current in large-$N_c$ chiral
  perturbation theory: Effects of the decuplet-octet mass difference and flavor
  symmetry breaking}},}\ }\href {\doibase 10.1103/PhysRevD.86.094041}
  {\bibfield  {journal} {\bibinfo  {journal} {Phys. Rev. D}\ }\textbf {\bibinfo
  {volume} {86}},\ \bibinfo {pages} {094041} (\bibinfo {year} {2012})},\
  \Eprint {http://arxiv.org/abs/1210.8445} {arXiv:1210.8445 [hep-ph]}
  \BibitemShut {NoStop}%
\bibitem [{\citenamefont {Flores-Mendieta}\ \emph {et~al.}(2021)\citenamefont
  {Flores-Mendieta}, \citenamefont {Garcia},\ and\ \citenamefont
  {Hernandez}}]{Flores-Mendieta:2021wzh}%
  \BibitemOpen
  \bibfield  {author} {\bibinfo {author} {\bibfnamefont {R.}~\bibnamefont
  {Flores-Mendieta}}, \bibinfo {author} {\bibfnamefont {C.~I.}\ \bibnamefont
  {Garcia}}, \ and\ \bibinfo {author} {\bibfnamefont {J.}~\bibnamefont
  {Hernandez}},\ }\bibfield  {title} {\enquote {\bibinfo {title} {{Baryon axial
  vector current in large-$N_c$ chiral perturbation theory: Complete analysis
  for $N_c=3$}},}\ }\href {\doibase 10.1103/PhysRevD.103.094032} {\bibfield
  {journal} {\bibinfo  {journal} {Phys. Rev. D}\ }\textbf {\bibinfo {volume}
  {103}},\ \bibinfo {pages} {094032} (\bibinfo {year} {2021})},\ \Eprint
  {http://arxiv.org/abs/2102.06100} {arXiv:2102.06100 [hep-ph]} \BibitemShut
  {NoStop}%
\bibitem [{\citenamefont {Flores-Mendieta}\ and\ \citenamefont
  {Sanchez-Almanza}(2024)}]{Flores-Mendieta:2024kvj}%
  \BibitemOpen
  \bibfield  {author} {\bibinfo {author} {\bibfnamefont {R.}~\bibnamefont
  {Flores-Mendieta}}\ and\ \bibinfo {author} {\bibfnamefont {G.}~\bibnamefont
  {Sanchez-Almanza}},\ }\bibfield  {title} {\enquote {\bibinfo {title}
  {{Universality of the baryon axial vector current operator in large-$N_c$
  chiral perturbation theory}},}\ }\href@noop {} {\  (\bibinfo {year}
  {2024})},\ \Eprint {http://arxiv.org/abs/2411.19838} {arXiv:2411.19838
  [hep-ph]} \BibitemShut {NoStop}%
\bibitem [{\citenamefont {Bulava}\ \emph {et~al.}(2023)\citenamefont {Bulava},
  \citenamefont {Hanlon}, \citenamefont {H\"orz}, \citenamefont {Morningstar},
  \citenamefont {Nicholson}, \citenamefont {Romero-L\'opez}, \citenamefont
  {Skinner}, \citenamefont {Vranas},\ and\ \citenamefont
  {Walker-Loud}}]{Bulava:2022vpq}%
  \BibitemOpen
  \bibfield  {author} {\bibinfo {author} {\bibfnamefont {J.}~\bibnamefont
  {Bulava}}, \bibinfo {author} {\bibfnamefont {A.~D.}\ \bibnamefont {Hanlon}},
  \bibinfo {author} {\bibfnamefont {B.}~\bibnamefont {H\"orz}}, \bibinfo
  {author} {\bibfnamefont {C.}~\bibnamefont {Morningstar}}, \bibinfo {author}
  {\bibfnamefont {A.}~\bibnamefont {Nicholson}}, \bibinfo {author}
  {\bibfnamefont {F.}~\bibnamefont {Romero-L\'opez}}, \bibinfo {author}
  {\bibfnamefont {S.}~\bibnamefont {Skinner}}, \bibinfo {author} {\bibfnamefont
  {P.}~\bibnamefont {Vranas}}, \ and\ \bibinfo {author} {\bibfnamefont
  {A.}~\bibnamefont {Walker-Loud}},\ }\bibfield  {title} {\enquote {\bibinfo
  {title} {{Elastic nucleon-pion scattering at m\ensuremath{\pi}=200 MeV from
  lattice QCD}},}\ }\href {\doibase 10.1016/j.nuclphysb.2023.116105} {\bibfield
   {journal} {\bibinfo  {journal} {Nucl. Phys. B}\ }\textbf {\bibinfo {volume}
  {987}},\ \bibinfo {pages} {116105} (\bibinfo {year} {2023})},\ \Eprint
  {http://arxiv.org/abs/2208.03867} {arXiv:2208.03867 [hep-lat]} \BibitemShut
  {NoStop}%
\bibitem [{\citenamefont {Barca}\ \emph {et~al.}(2023)\citenamefont {Barca},
  \citenamefont {Bali},\ and\ \citenamefont {Collins}}]{Barca:2022uhi}%
  \BibitemOpen
  \bibfield  {author} {\bibinfo {author} {\bibfnamefont {L.}~\bibnamefont
  {Barca}}, \bibinfo {author} {\bibfnamefont {G.}~\bibnamefont {Bali}}, \ and\
  \bibinfo {author} {\bibfnamefont {S.}~\bibnamefont {Collins}},\ }\bibfield
  {title} {\enquote {\bibinfo {title} {{Toward N to N\ensuremath{\pi} matrix
  elements from lattice QCD}},}\ }\href {\doibase 10.1103/PhysRevD.107.L051505}
  {\bibfield  {journal} {\bibinfo  {journal} {Phys. Rev. D}\ }\textbf {\bibinfo
  {volume} {107}},\ \bibinfo {pages} {L051505} (\bibinfo {year} {2023})},\
  \Eprint {http://arxiv.org/abs/2211.12278} {arXiv:2211.12278 [hep-lat]}
  \BibitemShut {NoStop}%
\bibitem [{\citenamefont {Bernard}\ and\ \citenamefont
  {Mei{\ss}ner}(2006)}]{Bernard:2006te}%
  \BibitemOpen
  \bibfield  {author} {\bibinfo {author} {\bibfnamefont {V.}~\bibnamefont
  {Bernard}}\ and\ \bibinfo {author} {\bibfnamefont {U.-G.}\ \bibnamefont
  {Mei{\ss}ner}},\ }\bibfield  {title} {\enquote {\bibinfo {title} {{The
  Nucleon axial-vector coupling beyond one loop}},}\ }\href {\doibase
  10.1016/j.physletb.2006.06.018} {\bibfield  {journal} {\bibinfo  {journal}
  {Phys. Lett. B}\ }\textbf {\bibinfo {volume} {639}},\ \bibinfo {pages}
  {278--282} (\bibinfo {year} {2006})},\ \Eprint
  {http://arxiv.org/abs/hep-lat/0605010} {arXiv:hep-lat/0605010 [hep-lat]}
  \BibitemShut {NoStop}%
\bibitem [{\citenamefont {Miller}\ \emph {et~al.}(2021)\citenamefont {Miller}
  \emph {et~al.}}]{Miller:2020evg}%
  \BibitemOpen
  \bibfield  {author} {\bibinfo {author} {\bibfnamefont {N.}~\bibnamefont
  {Miller}} \emph {et~al.},\ }\bibfield  {title} {\enquote {\bibinfo {title}
  {{Scale setting the M\"obius domain wall fermion on gradient-flowed HISQ
  action using the omega baryon mass and the gradient-flow scales $t_0$ and
  $w_0$}},}\ }\href {\doibase 10.1103/PhysRevD.103.054511} {\bibfield
  {journal} {\bibinfo  {journal} {Phys. Rev. D}\ }\textbf {\bibinfo {volume}
  {103}},\ \bibinfo {pages} {054511} (\bibinfo {year} {2021})},\ \Eprint
  {http://arxiv.org/abs/2011.12166} {arXiv:2011.12166 [hep-lat]} \BibitemShut
  {NoStop}%
\bibitem [{\citenamefont {Renner}\ \emph {et~al.}(2005)\citenamefont {Renner},
  \citenamefont {Schroers}, \citenamefont {Edwards}, \citenamefont {Fleming},
  \citenamefont {Hagler}, \citenamefont {Negele}, \citenamefont {Orginos},
  \citenamefont {Pochinski},\ and\ \citenamefont {Richards}}]{Renner:2004ck}%
  \BibitemOpen
  \bibfield  {author} {\bibinfo {author} {\bibfnamefont {D.~B.}\ \bibnamefont
  {Renner}}, \bibinfo {author} {\bibfnamefont {W.}~\bibnamefont {Schroers}},
  \bibinfo {author} {\bibfnamefont {R.}~\bibnamefont {Edwards}}, \bibinfo
  {author} {\bibfnamefont {G.~T.}\ \bibnamefont {Fleming}}, \bibinfo {author}
  {\bibfnamefont {P.}~\bibnamefont {Hagler}}, \bibinfo {author} {\bibfnamefont
  {J.~W.}\ \bibnamefont {Negele}}, \bibinfo {author} {\bibfnamefont
  {K.}~\bibnamefont {Orginos}}, \bibinfo {author} {\bibfnamefont {A.~V.}\
  \bibnamefont {Pochinski}}, \ and\ \bibinfo {author} {\bibfnamefont
  {D.}~\bibnamefont {Richards}} (\bibinfo {collaboration} {LHP}),\ }\bibfield
  {title} {\enquote {\bibinfo {title} {{Hadronic physics with domain-wall
  valence and improved staggered sea quarks}},}\ }\href {\doibase
  10.1016/j.nuclphysbps.2004.11.357} {\bibfield  {journal} {\bibinfo  {journal}
  {Nucl. Phys. B Proc. Suppl.}\ }\textbf {\bibinfo {volume} {140}},\ \bibinfo
  {pages} {255--260} (\bibinfo {year} {2005})},\ \Eprint
  {http://arxiv.org/abs/hep-lat/0409130} {arXiv:hep-lat/0409130} \BibitemShut
  {NoStop}%
\bibitem [{\citenamefont {B{\"a}r}\ \emph {et~al.}(2003)\citenamefont
  {B{\"a}r}, \citenamefont {Rupak},\ and\ \citenamefont
  {Shoresh}}]{Bar:2002nr}%
  \BibitemOpen
  \bibfield  {author} {\bibinfo {author} {\bibfnamefont {O.}~\bibnamefont
  {B{\"a}r}}, \bibinfo {author} {\bibfnamefont {G.}~\bibnamefont {Rupak}}, \
  and\ \bibinfo {author} {\bibfnamefont {N.}~\bibnamefont {Shoresh}},\
  }\bibfield  {title} {\enquote {\bibinfo {title} {{Simulations with different
  lattice Dirac operators for valence and sea quarks}},}\ }\href {\doibase
  10.1103/PhysRevD.67.114505} {\bibfield  {journal} {\bibinfo  {journal} {Phys.
  Rev. D}\ }\textbf {\bibinfo {volume} {67}},\ \bibinfo {pages} {114505}
  (\bibinfo {year} {2003})},\ \Eprint {http://arxiv.org/abs/hep-lat/0210050}
  {arXiv:hep-lat/0210050 [hep-lat]} \BibitemShut {NoStop}%
\bibitem [{\citenamefont {Follana}\ \emph {et~al.}(2007)\citenamefont
  {Follana}, \citenamefont {Mason}, \citenamefont {Davies}, \citenamefont
  {Hornbostel}, \citenamefont {Lepage}, \citenamefont {Shigemitsu},
  \citenamefont {Trottier},\ and\ \citenamefont {Wong}}]{Follana:2006rc}%
  \BibitemOpen
  \bibfield  {author} {\bibinfo {author} {\bibfnamefont {E.}~\bibnamefont
  {Follana}}, \bibinfo {author} {\bibfnamefont {Q.}~\bibnamefont {Mason}},
  \bibinfo {author} {\bibfnamefont {C.}~\bibnamefont {Davies}}, \bibinfo
  {author} {\bibfnamefont {K.}~\bibnamefont {Hornbostel}}, \bibinfo {author}
  {\bibfnamefont {G.~P.}\ \bibnamefont {Lepage}}, \bibinfo {author}
  {\bibfnamefont {J.}~\bibnamefont {Shigemitsu}}, \bibinfo {author}
  {\bibfnamefont {H.}~\bibnamefont {Trottier}}, \ and\ \bibinfo {author}
  {\bibfnamefont {K.}~\bibnamefont {Wong}} (\bibinfo {collaboration} {HPQCD,
  UKQCD}),\ }\bibfield  {title} {\enquote {\bibinfo {title} {{Highly improved
  staggered quarks on the lattice, with applications to charm physics}},}\
  }\href {\doibase 10.1103/PhysRevD.75.054502} {\bibfield  {journal} {\bibinfo
  {journal} {Phys. Rev. D}\ }\textbf {\bibinfo {volume} {75}},\ \bibinfo
  {pages} {054502} (\bibinfo {year} {2007})},\ \Eprint
  {http://arxiv.org/abs/hep-lat/0610092} {arXiv:hep-lat/0610092 [hep-lat]}
  \BibitemShut {NoStop}%
\bibitem [{\citenamefont {Brower}\ \emph {et~al.}(2005)\citenamefont {Brower},
  \citenamefont {Neff},\ and\ \citenamefont {Orginos}}]{Brower:2004xi}%
  \BibitemOpen
  \bibfield  {author} {\bibinfo {author} {\bibfnamefont {R.~C.}\ \bibnamefont
  {Brower}}, \bibinfo {author} {\bibfnamefont {H.}~\bibnamefont {Neff}}, \ and\
  \bibinfo {author} {\bibfnamefont {K.}~\bibnamefont {Orginos}},\ }\bibfield
  {title} {\enquote {\bibinfo {title} {{Mobius fermions: Improved domain wall
  chiral fermions}},}\ }\href {\doibase 10.1016/j.nuclphysbps.2004.11.180}
  {\bibfield  {journal} {\bibinfo  {journal} {Nucl. Phys. B Proc. Suppl.}\
  }\textbf {\bibinfo {volume} {140}},\ \bibinfo {pages} {686--688} (\bibinfo
  {year} {2005})},\ \Eprint {http://arxiv.org/abs/hep-lat/0409118}
  {arXiv:hep-lat/0409118} \BibitemShut {NoStop}%
\bibitem [{\citenamefont {Brower}\ \emph {et~al.}(2006)\citenamefont {Brower},
  \citenamefont {Neff},\ and\ \citenamefont {Orginos}}]{Brower:2005qw}%
  \BibitemOpen
  \bibfield  {author} {\bibinfo {author} {\bibfnamefont {R.~C.}\ \bibnamefont
  {Brower}}, \bibinfo {author} {\bibfnamefont {H.}~\bibnamefont {Neff}}, \ and\
  \bibinfo {author} {\bibfnamefont {K.}~\bibnamefont {Orginos}},\ }\bibfield
  {title} {\enquote {\bibinfo {title} {{Mobius fermions}},}\ }\href {\doibase
  10.1016/j.nuclphysbps.2006.01.047} {\bibfield  {journal} {\bibinfo  {journal}
  {Nucl. Phys. B Proc. Suppl.}\ }\textbf {\bibinfo {volume} {153}},\ \bibinfo
  {pages} {191--198} (\bibinfo {year} {2006})},\ \Eprint
  {http://arxiv.org/abs/hep-lat/0511031} {arXiv:hep-lat/0511031} \BibitemShut
  {NoStop}%
\bibitem [{\citenamefont {{L\"uscher}}(2010)}]{Luscher:2010iy}%
  \BibitemOpen
  \bibfield  {author} {\bibinfo {author} {\bibfnamefont {M.}~\bibnamefont
  {{L\"uscher}}},\ }\bibfield  {title} {\enquote {\bibinfo {title} {{Properties
  and uses of the Wilson flow in lattice QCD}},}\ }\href {\doibase
  10.1007/JHEP08(2010)071} {\bibfield  {journal} {\bibinfo  {journal} {JHEP}\
  }\textbf {\bibinfo {volume} {1008}},\ \bibinfo {pages} {071} (\bibinfo {year}
  {2010})},\ \Eprint {http://arxiv.org/abs/1006.4518} {arXiv:1006.4518
  [hep-lat]} \BibitemShut {NoStop}%
\bibitem [{\citenamefont {Lohmayer}\ and\ \citenamefont
  {Neuberger}(2011)}]{Lohmayer:2011si}%
  \BibitemOpen
  \bibfield  {author} {\bibinfo {author} {\bibfnamefont {R.}~\bibnamefont
  {Lohmayer}}\ and\ \bibinfo {author} {\bibfnamefont {H.}~\bibnamefont
  {Neuberger}},\ }\bibfield  {title} {\enquote {\bibinfo {title} {{Continuous
  smearing of Wilson Loops}},}\ }\href {\doibase 10.22323/1.139.0249}
  {\bibfield  {journal} {\bibinfo  {journal} {PoS}\ }\textbf {\bibinfo {volume}
  {LATTICE2011}},\ \bibinfo {pages} {249} (\bibinfo {year} {2011})},\ \Eprint
  {http://arxiv.org/abs/1110.3522} {arXiv:1110.3522 [hep-lat]} \BibitemShut
  {NoStop}%
\bibitem [{\citenamefont {Berkowitz}\ \emph
  {et~al.}(2017{\natexlab{a}})\citenamefont {Berkowitz}, \citenamefont
  {Bouchard}, \citenamefont {Chang}, \citenamefont {Clark}, \citenamefont
  {Joo}, \citenamefont {Kurth}, \citenamefont {Monahan}, \citenamefont
  {Nicholson}, \citenamefont {Orginos}, \citenamefont {Rinaldi}, \citenamefont
  {Vranas},\ and\ \citenamefont {Walker-Loud}}]{Berkowitz:2017opd}%
  \BibitemOpen
  \bibfield  {author} {\bibinfo {author} {\bibfnamefont {E.}~\bibnamefont
  {Berkowitz}}, \bibinfo {author} {\bibfnamefont {C.}~\bibnamefont {Bouchard}},
  \bibinfo {author} {\bibfnamefont {C.}~\bibnamefont {Chang}}, \bibinfo
  {author} {\bibfnamefont {M.}~\bibnamefont {Clark}}, \bibinfo {author}
  {\bibfnamefont {B.}~\bibnamefont {Joo}}, \bibinfo {author} {\bibfnamefont
  {T.}~\bibnamefont {Kurth}}, \bibinfo {author} {\bibfnamefont
  {C.}~\bibnamefont {Monahan}}, \bibinfo {author} {\bibfnamefont
  {A.}~\bibnamefont {Nicholson}}, \bibinfo {author} {\bibfnamefont
  {K.}~\bibnamefont {Orginos}}, \bibinfo {author} {\bibfnamefont
  {E.}~\bibnamefont {Rinaldi}}, \bibinfo {author} {\bibfnamefont
  {P.}~\bibnamefont {Vranas}}, \ and\ \bibinfo {author} {\bibfnamefont
  {A.}~\bibnamefont {Walker-Loud}},\ }\bibfield  {title} {\enquote {\bibinfo
  {title} {{M\"obius Domain-Wall fermions on gradient-flowed dynamical HISQ
  ensembles}},}\ }\href {\doibase 10.1103/PhysRevD.96.054513} {\bibfield
  {journal} {\bibinfo  {journal} {Phys. Rev. D}\ }\textbf {\bibinfo {volume}
  {96}},\ \bibinfo {pages} {054513} (\bibinfo {year} {2017}{\natexlab{a}})},\
  \Eprint {http://arxiv.org/abs/1701.07559} {arXiv:1701.07559 [hep-lat]}
  \BibitemShut {NoStop}%
\bibitem [{\citenamefont {Bazavov}\ \emph {et~al.}(2010)\citenamefont {Bazavov}
  \emph {et~al.}}]{MILC:2010pul}%
  \BibitemOpen
  \bibfield  {author} {\bibinfo {author} {\bibfnamefont {A.}~\bibnamefont
  {Bazavov}} \emph {et~al.} (\bibinfo {collaboration} {MILC}),\ }\bibfield
  {title} {\enquote {\bibinfo {title} {{Scaling studies of QCD with the
  dynamical HISQ action}},}\ }\href {\doibase 10.1103/PhysRevD.82.074501}
  {\bibfield  {journal} {\bibinfo  {journal} {Phys. Rev. D}\ }\textbf {\bibinfo
  {volume} {82}},\ \bibinfo {pages} {074501} (\bibinfo {year} {2010})},\
  \Eprint {http://arxiv.org/abs/1004.0342} {arXiv:1004.0342 [hep-lat]}
  \BibitemShut {NoStop}%
\bibitem [{\citenamefont {Bazavov}\ \emph {et~al.}(2013)\citenamefont {Bazavov}
  \emph {et~al.}}]{Bazavov:2012xda}%
  \BibitemOpen
  \bibfield  {author} {\bibinfo {author} {\bibfnamefont {A.}~\bibnamefont
  {Bazavov}} \emph {et~al.} (\bibinfo {collaboration} {MILC}),\ }\bibfield
  {title} {\enquote {\bibinfo {title} {{Lattice QCD ensembles with four flavors
  of highly improved staggered quarks}},}\ }\href {\doibase
  10.1103/PhysRevD.87.054505} {\bibfield  {journal} {\bibinfo  {journal} {Phys.
  Rev. D}\ }\textbf {\bibinfo {volume} {87}},\ \bibinfo {pages} {054505}
  (\bibinfo {year} {2013})},\ \Eprint {http://arxiv.org/abs/1212.4768}
  {arXiv:1212.4768 [hep-lat]} \BibitemShut {NoStop}%
\bibitem [{\citenamefont {Miller}\ \emph {et~al.}(2020)\citenamefont {Miller}
  \emph {et~al.}}]{Miller:2020xhy}%
  \BibitemOpen
  \bibfield  {author} {\bibinfo {author} {\bibfnamefont {N.}~\bibnamefont
  {Miller}} \emph {et~al.},\ }\bibfield  {title} {\enquote {\bibinfo {title}
  {{$F_K / F_\pi$ from M\"obius Domain-Wall fermions solved on gradient-flowed
  HISQ ensembles}},}\ }\href {\doibase 10.1103/PhysRevD.102.034507} {\bibfield
  {journal} {\bibinfo  {journal} {Phys. Rev. D}\ }\textbf {\bibinfo {volume}
  {102}},\ \bibinfo {pages} {034507} (\bibinfo {year} {2020})},\ \Eprint
  {http://arxiv.org/abs/2005.04795} {arXiv:2005.04795 [hep-lat]} \BibitemShut
  {NoStop}%
\bibitem [{\citenamefont {Clark}\ \emph {et~al.}(2010)\citenamefont {Clark},
  \citenamefont {Babich}, \citenamefont {Barros}, \citenamefont {Brower},\ and\
  \citenamefont {Rebbi}}]{Clark:2009wm}%
  \BibitemOpen
  \bibfield  {author} {\bibinfo {author} {\bibfnamefont {M.~A.}\ \bibnamefont
  {Clark}}, \bibinfo {author} {\bibfnamefont {R.}~\bibnamefont {Babich}},
  \bibinfo {author} {\bibfnamefont {K.}~\bibnamefont {Barros}}, \bibinfo
  {author} {\bibfnamefont {R.~C.}\ \bibnamefont {Brower}}, \ and\ \bibinfo
  {author} {\bibfnamefont {C.}~\bibnamefont {Rebbi}} (\bibinfo {collaboration}
  {QUDA}),\ }\bibfield  {title} {\enquote {\bibinfo {title} {{Solving Lattice
  QCD systems of equations using mixed precision solvers on GPUs}},}\ }\href
  {\doibase 10.1016/j.cpc.2010.05.002} {\bibfield  {journal} {\bibinfo
  {journal} {Comput. Phys. Commun.}\ }\textbf {\bibinfo {volume} {181}},\
  \bibinfo {pages} {1517--1528} (\bibinfo {year} {2010})},\ \Eprint
  {http://arxiv.org/abs/0911.3191} {arXiv:0911.3191 [hep-lat]} \BibitemShut
  {NoStop}%
\bibitem [{\citenamefont {Babich}\ \emph {et~al.}(2011)\citenamefont {Babich},
  \citenamefont {Clark}, \citenamefont {Joo}, \citenamefont {Shi},
  \citenamefont {Brower} \emph {et~al.}}]{Babich:2011np}%
  \BibitemOpen
  \bibfield  {author} {\bibinfo {author} {\bibfnamefont {R.}~\bibnamefont
  {Babich}}, \bibinfo {author} {\bibfnamefont {M.}~\bibnamefont {Clark}},
  \bibinfo {author} {\bibfnamefont {B.}~\bibnamefont {Joo}}, \bibinfo {author}
  {\bibfnamefont {G.}~\bibnamefont {Shi}}, \bibinfo {author} {\bibfnamefont
  {R.}~\bibnamefont {Brower}},  \emph {et~al.},\ }\bibfield  {title} {\enquote
  {\bibinfo {title} {{Scaling Lattice QCD beyond 100 GPUs}},}\ }\href@noop {}
  {\  (\bibinfo {year} {2011})},\ \Eprint {http://arxiv.org/abs/1109.2935}
  {arXiv:1109.2935 [hep-lat]} \BibitemShut {NoStop}%
\bibitem [{\citenamefont {Bors\'anyi}\ \emph {et~al.}(2012)\citenamefont
  {Bors\'anyi}, \citenamefont {D\"urr}, \citenamefont {Fodor}, \citenamefont
  {Hoelbling}, \citenamefont {Katz}, \citenamefont {Krieg}, \citenamefont
  {Kurth}, \citenamefont {Lellouch}, \citenamefont {Lippert},\ and\
  \citenamefont {McNeile}}]{BMW:2012hcm}%
  \BibitemOpen
  \bibfield  {author} {\bibinfo {author} {\bibfnamefont {S.}~\bibnamefont
  {Bors\'anyi}}, \bibinfo {author} {\bibfnamefont {S.}~\bibnamefont {D\"urr}},
  \bibinfo {author} {\bibfnamefont {Z.}~\bibnamefont {Fodor}}, \bibinfo
  {author} {\bibfnamefont {C.}~\bibnamefont {Hoelbling}}, \bibinfo {author}
  {\bibfnamefont {S.~D.}\ \bibnamefont {Katz}}, \bibinfo {author}
  {\bibfnamefont {S.}~\bibnamefont {Krieg}}, \bibinfo {author} {\bibfnamefont
  {T.}~\bibnamefont {Kurth}}, \bibinfo {author} {\bibfnamefont
  {L.}~\bibnamefont {Lellouch}}, \bibinfo {author} {\bibfnamefont
  {T.}~\bibnamefont {Lippert}}, \ and\ \bibinfo {author} {\bibfnamefont
  {C.}~\bibnamefont {McNeile}} (\bibinfo {collaboration} {BMW}),\ }\bibfield
  {title} {\enquote {\bibinfo {title} {{High-precision scale setting in lattice
  QCD}},}\ }\href {\doibase 10.1007/JHEP09(2012)010} {\bibfield  {journal}
  {\bibinfo  {journal} {JHEP}\ }\textbf {\bibinfo {volume} {09}},\ \bibinfo
  {pages} {010} (\bibinfo {year} {2012})},\ \Eprint
  {http://arxiv.org/abs/1203.4469} {arXiv:1203.4469 [hep-lat]} \BibitemShut
  {NoStop}%
\bibitem [{\citenamefont {Bhattacharya}\ \emph {et~al.}(2015)\citenamefont
  {Bhattacharya}, \citenamefont {Cirigliano}, \citenamefont {Cohen},
  \citenamefont {Gupta}, \citenamefont {Joseph}, \citenamefont {Lin},\ and\
  \citenamefont {Yoon}}]{Bhattacharya:2015wna}%
  \BibitemOpen
  \bibfield  {author} {\bibinfo {author} {\bibfnamefont {T.}~\bibnamefont
  {Bhattacharya}}, \bibinfo {author} {\bibfnamefont {V.}~\bibnamefont
  {Cirigliano}}, \bibinfo {author} {\bibfnamefont {S.}~\bibnamefont {Cohen}},
  \bibinfo {author} {\bibfnamefont {R.}~\bibnamefont {Gupta}}, \bibinfo
  {author} {\bibfnamefont {A.}~\bibnamefont {Joseph}}, \bibinfo {author}
  {\bibfnamefont {H.-W.}\ \bibnamefont {Lin}}, \ and\ \bibinfo {author}
  {\bibfnamefont {B.}~\bibnamefont {Yoon}} (\bibinfo {collaboration} {PNDME}),\
  }\bibfield  {title} {\enquote {\bibinfo {title} {{Iso-vector and Iso-scalar
  Tensor Charges of the Nucleon from Lattice QCD}},}\ }\href {\doibase
  10.1103/PhysRevD.92.094511} {\bibfield  {journal} {\bibinfo  {journal} {Phys.
  Rev. D}\ }\textbf {\bibinfo {volume} {92}},\ \bibinfo {pages} {094511}
  (\bibinfo {year} {2015})},\ \Eprint {http://arxiv.org/abs/1506.06411}
  {arXiv:1506.06411 [hep-lat]} \BibitemShut {NoStop}%
\bibitem [{\citenamefont {Bouchard}\ \emph {et~al.}(2017)\citenamefont
  {Bouchard}, \citenamefont {Chang}, \citenamefont {Kurth}, \citenamefont
  {Orginos},\ and\ \citenamefont {Walker-Loud}}]{Bouchard:2016heu}%
  \BibitemOpen
  \bibfield  {author} {\bibinfo {author} {\bibfnamefont {C.}~\bibnamefont
  {Bouchard}}, \bibinfo {author} {\bibfnamefont {C.~C.}\ \bibnamefont {Chang}},
  \bibinfo {author} {\bibfnamefont {T.}~\bibnamefont {Kurth}}, \bibinfo
  {author} {\bibfnamefont {K.}~\bibnamefont {Orginos}}, \ and\ \bibinfo
  {author} {\bibfnamefont {A.}~\bibnamefont {Walker-Loud}},\ }\bibfield
  {title} {\enquote {\bibinfo {title} {{On the Feynman-Hellmann Theorem in
  Quantum Field Theory and the Calculation of Matrix Elements}},}\ }\href
  {\doibase 10.1103/PhysRevD.96.014504} {\bibfield  {journal} {\bibinfo
  {journal} {Phys. Rev. D}\ }\textbf {\bibinfo {volume} {96}},\ \bibinfo
  {pages} {014504} (\bibinfo {year} {2017})},\ \Eprint
  {http://arxiv.org/abs/1612.06963} {arXiv:1612.06963 [hep-lat]} \BibitemShut
  {NoStop}%
\bibitem [{\citenamefont {Symanzik}(1983{\natexlab{a}})}]{Symanzik:1983gh}%
  \BibitemOpen
  \bibfield  {author} {\bibinfo {author} {\bibfnamefont {K.}~\bibnamefont
  {Symanzik}},\ }\bibfield  {title} {\enquote {\bibinfo {title} {{Continuum
  Limit and Improved Action in Lattice Theories. 2. $O(N)$ Nonlinear Sigma
  Model in Perturbation Theory}},}\ }\href {\doibase
  10.1016/0550-3213(83)90469-8} {\bibfield  {journal} {\bibinfo  {journal}
  {Nucl. Phys. B}\ }\textbf {\bibinfo {volume} {226}},\ \bibinfo {pages}
  {205--227} (\bibinfo {year} {1983}{\natexlab{a}})}\BibitemShut {NoStop}%
\bibitem [{\citenamefont {Symanzik}(1983{\natexlab{b}})}]{Symanzik:1983dc}%
  \BibitemOpen
  \bibfield  {author} {\bibinfo {author} {\bibfnamefont {K.}~\bibnamefont
  {Symanzik}},\ }\bibfield  {title} {\enquote {\bibinfo {title} {{Continuum
  Limit and Improved Action in Lattice Theories. 1. Principles and $\phi^4$
  Theory}},}\ }\href {\doibase 10.1016/0550-3213(83)90468-6} {\bibfield
  {journal} {\bibinfo  {journal} {Nucl. Phys. B}\ }\textbf {\bibinfo {volume}
  {226}},\ \bibinfo {pages} {187--204} (\bibinfo {year}
  {1983}{\natexlab{b}})}\BibitemShut {NoStop}%
\bibitem [{\citenamefont {B{\"a}r}\ \emph {et~al.}(2005)\citenamefont
  {B{\"a}r}, \citenamefont {Bernard}, \citenamefont {Rupak},\ and\
  \citenamefont {Shoresh}}]{Bar:2005tu}%
  \BibitemOpen
  \bibfield  {author} {\bibinfo {author} {\bibfnamefont {O.}~\bibnamefont
  {B{\"a}r}}, \bibinfo {author} {\bibfnamefont {C.}~\bibnamefont {Bernard}},
  \bibinfo {author} {\bibfnamefont {G.}~\bibnamefont {Rupak}}, \ and\ \bibinfo
  {author} {\bibfnamefont {N.}~\bibnamefont {Shoresh}},\ }\bibfield  {title}
  {\enquote {\bibinfo {title} {{Chiral perturbation theory for staggered sea
  quarks and Ginsparg-Wilson valence quarks}},}\ }\href {\doibase
  10.1103/PhysRevD.72.054502} {\bibfield  {journal} {\bibinfo  {journal} {Phys.
  Rev. D}\ }\textbf {\bibinfo {volume} {72}},\ \bibinfo {pages} {054502}
  (\bibinfo {year} {2005})},\ \Eprint {http://arxiv.org/abs/hep-lat/0503009}
  {arXiv:hep-lat/0503009 [hep-lat]} \BibitemShut {NoStop}%
\bibitem [{\citenamefont {Aoki}\ \emph {et~al.}(2024)\citenamefont {Aoki} \emph
  {et~al.}}]{FlavourLatticeAveragingGroupFLAG:2024oxs}%
  \BibitemOpen
  \bibfield  {author} {\bibinfo {author} {\bibfnamefont {Y.}~\bibnamefont
  {Aoki}} \emph {et~al.} (\bibinfo {collaboration} {Flavour Lattice Averaging
  Group (FLAG)}),\ }\bibfield  {title} {\enquote {\bibinfo {title} {{FLAG
  Review 2024}},}\ }\href@noop {} {\  (\bibinfo {year} {2024})},\ \Eprint
  {http://arxiv.org/abs/2411.04268} {arXiv:2411.04268 [hep-lat]} \BibitemShut
  {NoStop}%
\bibitem [{\citenamefont {Berkowitz}\ \emph
  {et~al.}(2017{\natexlab{b}})\citenamefont {Berkowitz}, \citenamefont
  {Bouchard}, \citenamefont {Brantley}, \citenamefont {Chang}, \citenamefont
  {Clark}, \citenamefont {Garron}, \citenamefont {Jo\'{o}}, \citenamefont
  {Kurth}, \citenamefont {Monahan}, \citenamefont {Monge-Camacho},
  \citenamefont {Nicholson}, \citenamefont {Orginos}, \citenamefont {Rinaldi},
  \citenamefont {Vranas},\ and\ \citenamefont
  {Walker-Loud}}]{Berkowitz:2017gql}%
  \BibitemOpen
  \bibfield  {author} {\bibinfo {author} {\bibfnamefont {E.}~\bibnamefont
  {Berkowitz}}, \bibinfo {author} {\bibfnamefont {C.}~\bibnamefont {Bouchard}},
  \bibinfo {author} {\bibfnamefont {D.~B.}\ \bibnamefont {Brantley}}, \bibinfo
  {author} {\bibfnamefont {C.}~\bibnamefont {Chang}}, \bibinfo {author}
  {\bibfnamefont {M.}~\bibnamefont {Clark}}, \bibinfo {author} {\bibfnamefont
  {N.}~\bibnamefont {Garron}}, \bibinfo {author} {\bibfnamefont
  {B.}~\bibnamefont {Jo\'{o}}}, \bibinfo {author} {\bibfnamefont
  {T.}~\bibnamefont {Kurth}}, \bibinfo {author} {\bibfnamefont
  {C.}~\bibnamefont {Monahan}}, \bibinfo {author} {\bibfnamefont
  {H.}~\bibnamefont {Monge-Camacho}}, \bibinfo {author} {\bibfnamefont
  {A.}~\bibnamefont {Nicholson}}, \bibinfo {author} {\bibfnamefont
  {K.}~\bibnamefont {Orginos}}, \bibinfo {author} {\bibfnamefont
  {E.}~\bibnamefont {Rinaldi}}, \bibinfo {author} {\bibfnamefont
  {P.}~\bibnamefont {Vranas}}, \ and\ \bibinfo {author} {\bibfnamefont
  {A.}~\bibnamefont {Walker-Loud}},\ }\bibfield  {title} {\enquote {\bibinfo
  {title} {{An accurate calculation of the nucleon axial charge with lattice
  QCD}},}\ }\href@noop {} {\  (\bibinfo {year} {2017}{\natexlab{b}})},\ \Eprint
  {http://arxiv.org/abs/1704.01114} {arXiv:1704.01114 [hep-lat]} \BibitemShut
  {NoStop}%
\bibitem [{\citenamefont {Yamanaka}\ \emph {et~al.}(2018)\citenamefont
  {Yamanaka}, \citenamefont {Hashimoto}, \citenamefont {Kaneko},\ and\
  \citenamefont {Ohki}}]{Yamanaka:2018uud}%
  \BibitemOpen
  \bibfield  {author} {\bibinfo {author} {\bibfnamefont {N.}~\bibnamefont
  {Yamanaka}}, \bibinfo {author} {\bibfnamefont {S.}~\bibnamefont {Hashimoto}},
  \bibinfo {author} {\bibfnamefont {T.}~\bibnamefont {Kaneko}}, \ and\ \bibinfo
  {author} {\bibfnamefont {H.}~\bibnamefont {Ohki}} (\bibinfo {collaboration}
  {JLQCD}),\ }\bibfield  {title} {\enquote {\bibinfo {title} {{Nucleon charges
  with dynamical overlap fermions}},}\ }\href {\doibase
  10.1103/PhysRevD.98.054516} {\bibfield  {journal} {\bibinfo  {journal} {Phys.
  Rev. D}\ }\textbf {\bibinfo {volume} {98}},\ \bibinfo {pages} {054516}
  (\bibinfo {year} {2018})},\ \Eprint {http://arxiv.org/abs/1805.10507}
  {arXiv:1805.10507 [hep-lat]} \BibitemShut {NoStop}%
\bibitem [{\citenamefont {Liang}\ \emph {et~al.}(2018)\citenamefont {Liang},
  \citenamefont {Yang}, \citenamefont {Draper}, \citenamefont {Gong},\ and\
  \citenamefont {Liu}}]{Liang:2018pis}%
  \BibitemOpen
  \bibfield  {author} {\bibinfo {author} {\bibfnamefont {J.}~\bibnamefont
  {Liang}}, \bibinfo {author} {\bibfnamefont {Y.-B.}\ \bibnamefont {Yang}},
  \bibinfo {author} {\bibfnamefont {T.}~\bibnamefont {Draper}}, \bibinfo
  {author} {\bibfnamefont {M.}~\bibnamefont {Gong}}, \ and\ \bibinfo {author}
  {\bibfnamefont {K.-F.}\ \bibnamefont {Liu}},\ }\bibfield  {title} {\enquote
  {\bibinfo {title} {{Quark spins and Anomalous Ward Identity}},}\ }\href
  {\doibase 10.1103/PhysRevD.98.074505} {\bibfield  {journal} {\bibinfo
  {journal} {Phys. Rev. D}\ }\textbf {\bibinfo {volume} {98}},\ \bibinfo
  {pages} {074505} (\bibinfo {year} {2018})},\ \Eprint
  {http://arxiv.org/abs/1806.08366} {arXiv:1806.08366 [hep-ph]} \BibitemShut
  {NoStop}%
\bibitem [{\citenamefont {Ishikawa}\ \emph {et~al.}(2018)\citenamefont
  {Ishikawa}, \citenamefont {Kuramashi}, \citenamefont {Sasaki}, \citenamefont
  {Tsukamoto}, \citenamefont {Ukawa},\ and\ \citenamefont
  {Yamazaki}}]{Ishikawa:2018rew}%
  \BibitemOpen
  \bibfield  {author} {\bibinfo {author} {\bibfnamefont {K.-I.}\ \bibnamefont
  {Ishikawa}}, \bibinfo {author} {\bibfnamefont {Y.}~\bibnamefont {Kuramashi}},
  \bibinfo {author} {\bibfnamefont {S.}~\bibnamefont {Sasaki}}, \bibinfo
  {author} {\bibfnamefont {N.}~\bibnamefont {Tsukamoto}}, \bibinfo {author}
  {\bibfnamefont {A.}~\bibnamefont {Ukawa}}, \ and\ \bibinfo {author}
  {\bibfnamefont {T.}~\bibnamefont {Yamazaki}} (\bibinfo {collaboration}
  {PACS}),\ }\bibfield  {title} {\enquote {\bibinfo {title} {{Nucleon form
  factors on a large volume lattice near the physical point in 2+1 flavor
  QCD}},}\ }\href {\doibase 10.1103/PhysRevD.98.074510} {\bibfield  {journal}
  {\bibinfo  {journal} {Phys. Rev. D}\ }\textbf {\bibinfo {volume} {98}},\
  \bibinfo {pages} {074510} (\bibinfo {year} {2018})},\ \Eprint
  {http://arxiv.org/abs/1807.03974} {arXiv:1807.03974 [hep-lat]} \BibitemShut
  {NoStop}%
\bibitem [{\citenamefont {Shintani}\ \emph {et~al.}(2019)\citenamefont
  {Shintani}, \citenamefont {Ishikawa}, \citenamefont {Kuramashi},
  \citenamefont {Sasaki},\ and\ \citenamefont {Yamazaki}}]{Shintani:2018ozy}%
  \BibitemOpen
  \bibfield  {author} {\bibinfo {author} {\bibfnamefont {E.}~\bibnamefont
  {Shintani}}, \bibinfo {author} {\bibfnamefont {K.-I.}\ \bibnamefont
  {Ishikawa}}, \bibinfo {author} {\bibfnamefont {Y.}~\bibnamefont {Kuramashi}},
  \bibinfo {author} {\bibfnamefont {S.}~\bibnamefont {Sasaki}}, \ and\ \bibinfo
  {author} {\bibfnamefont {T.}~\bibnamefont {Yamazaki}},\ }\bibfield  {title}
  {\enquote {\bibinfo {title} {{Nucleon form factors and root-mean-square radii
  on a (10.8 fm)$^4$ lattice at the physical point}},}\ }\href {\doibase
  10.1103/PhysRevD.99.014510} {\bibfield  {journal} {\bibinfo  {journal} {Phys.
  Rev. D}\ }\textbf {\bibinfo {volume} {99}},\ \bibinfo {pages} {014510}
  (\bibinfo {year} {2019})},\ \bibinfo {note} {[Erratum: Phys. Rev. D {\bf
  102}, 019902 (2020)]},\ \Eprint {http://arxiv.org/abs/1811.07292}
  {arXiv:1811.07292 [hep-lat]} \BibitemShut {NoStop}%
\bibitem [{\citenamefont {Harris}\ \emph {et~al.}(2019)\citenamefont {Harris},
  \citenamefont {von Hippel}, \citenamefont {Junnarkar}, \citenamefont {Meyer},
  \citenamefont {Ottnad}, \citenamefont {Wilhelm}, \citenamefont {Wittig},\
  and\ \citenamefont {Wrang}}]{Harris:2019bih}%
  \BibitemOpen
  \bibfield  {author} {\bibinfo {author} {\bibfnamefont {T.}~\bibnamefont
  {Harris}}, \bibinfo {author} {\bibfnamefont {G.}~\bibnamefont {von Hippel}},
  \bibinfo {author} {\bibfnamefont {P.}~\bibnamefont {Junnarkar}}, \bibinfo
  {author} {\bibfnamefont {H.~B.}\ \bibnamefont {Meyer}}, \bibinfo {author}
  {\bibfnamefont {K.}~\bibnamefont {Ottnad}}, \bibinfo {author} {\bibfnamefont
  {J.}~\bibnamefont {Wilhelm}}, \bibinfo {author} {\bibfnamefont
  {H.}~\bibnamefont {Wittig}}, \ and\ \bibinfo {author} {\bibfnamefont
  {L.}~\bibnamefont {Wrang}},\ }\bibfield  {title} {\enquote {\bibinfo {title}
  {{Nucleon isovector charges and twist-2 matrix elements with $N_f=2+1$
  dynamical Wilson quarks}},}\ }\href {\doibase 10.1103/PhysRevD.100.034513}
  {\bibfield  {journal} {\bibinfo  {journal} {Phys. Rev. D}\ }\textbf {\bibinfo
  {volume} {100}},\ \bibinfo {pages} {034513} (\bibinfo {year} {2019})},\
  \Eprint {http://arxiv.org/abs/1905.01291} {arXiv:1905.01291 [hep-lat]}
  \BibitemShut {NoStop}%
\bibitem [{\citenamefont {Hasan}\ \emph {et~al.}(2019)\citenamefont {Hasan},
  \citenamefont {Green}, \citenamefont {Meinel}, \citenamefont {Engelhardt},
  \citenamefont {Krieg}, \citenamefont {Negele}, \citenamefont {Pochinsky},\
  and\ \citenamefont {Syritsyn}}]{Hasan:2019noy}%
  \BibitemOpen
  \bibfield  {author} {\bibinfo {author} {\bibfnamefont {N.}~\bibnamefont
  {Hasan}}, \bibinfo {author} {\bibfnamefont {J.}~\bibnamefont {Green}},
  \bibinfo {author} {\bibfnamefont {S.}~\bibnamefont {Meinel}}, \bibinfo
  {author} {\bibfnamefont {M.}~\bibnamefont {Engelhardt}}, \bibinfo {author}
  {\bibfnamefont {S.}~\bibnamefont {Krieg}}, \bibinfo {author} {\bibfnamefont
  {J.}~\bibnamefont {Negele}}, \bibinfo {author} {\bibfnamefont
  {A.}~\bibnamefont {Pochinsky}}, \ and\ \bibinfo {author} {\bibfnamefont
  {S.}~\bibnamefont {Syritsyn}},\ }\bibfield  {title} {\enquote {\bibinfo
  {title} {{Nucleon axial, scalar, and tensor charges using lattice QCD at the
  physical pion mass}},}\ }\href {\doibase 10.1103/PhysRevD.99.114505}
  {\bibfield  {journal} {\bibinfo  {journal} {Phys. Rev. D}\ }\textbf {\bibinfo
  {volume} {99}},\ \bibinfo {pages} {114505} (\bibinfo {year} {2019})},\
  \Eprint {http://arxiv.org/abs/1903.06487} {arXiv:1903.06487 [hep-lat]}
  \BibitemShut {NoStop}%
\bibitem [{\citenamefont {Bali}\ \emph {et~al.}(2020)\citenamefont {Bali},
  \citenamefont {Barca}, \citenamefont {Collins}, \citenamefont {Gruber},
  \citenamefont {L\"offler}, \citenamefont {Sch\"afer}, \citenamefont
  {S\"oldner}, \citenamefont {Wein}, \citenamefont {Weish\"aupl},\ and\
  \citenamefont {Wurm}}]{RQCD:2019jai}%
  \BibitemOpen
  \bibfield  {author} {\bibinfo {author} {\bibfnamefont {G.~S.}\ \bibnamefont
  {Bali}}, \bibinfo {author} {\bibfnamefont {L.}~\bibnamefont {Barca}},
  \bibinfo {author} {\bibfnamefont {S.}~\bibnamefont {Collins}}, \bibinfo
  {author} {\bibfnamefont {M.}~\bibnamefont {Gruber}}, \bibinfo {author}
  {\bibfnamefont {M.}~\bibnamefont {L\"offler}}, \bibinfo {author}
  {\bibfnamefont {A.}~\bibnamefont {Sch\"afer}}, \bibinfo {author}
  {\bibfnamefont {W.}~\bibnamefont {S\"oldner}}, \bibinfo {author}
  {\bibfnamefont {P.}~\bibnamefont {Wein}}, \bibinfo {author} {\bibfnamefont
  {S.}~\bibnamefont {Weish\"aupl}}, \ and\ \bibinfo {author} {\bibfnamefont
  {T.}~\bibnamefont {Wurm}} (\bibinfo {collaboration} {RQCD}),\ }\bibfield
  {title} {\enquote {\bibinfo {title} {{Nucleon axial structure from lattice
  QCD}},}\ }\href {\doibase 10.1007/JHEP05(2020)126} {\bibfield  {journal}
  {\bibinfo  {journal} {JHEP}\ }\textbf {\bibinfo {volume} {05}},\ \bibinfo
  {pages} {126} (\bibinfo {year} {2020})},\ \Eprint
  {http://arxiv.org/abs/1911.13150} {arXiv:1911.13150 [hep-lat]} \BibitemShut
  {NoStop}%
\bibitem [{\citenamefont {Park}\ \emph {et~al.}(2022)\citenamefont {Park},
  \citenamefont {Gupta}, \citenamefont {Yoon}, \citenamefont {Mondal},
  \citenamefont {Bhattacharya}, \citenamefont {Jang}, \citenamefont {Jo\'o},\
  and\ \citenamefont {Winter}}]{Park:2021ypf}%
  \BibitemOpen
  \bibfield  {author} {\bibinfo {author} {\bibfnamefont {S.}~\bibnamefont
  {Park}}, \bibinfo {author} {\bibfnamefont {R.}~\bibnamefont {Gupta}},
  \bibinfo {author} {\bibfnamefont {B.}~\bibnamefont {Yoon}}, \bibinfo {author}
  {\bibfnamefont {S.}~\bibnamefont {Mondal}}, \bibinfo {author} {\bibfnamefont
  {T.}~\bibnamefont {Bhattacharya}}, \bibinfo {author} {\bibfnamefont {Y.-C.}\
  \bibnamefont {Jang}}, \bibinfo {author} {\bibfnamefont {B.}~\bibnamefont
  {Jo\'o}}, \ and\ \bibinfo {author} {\bibfnamefont {F.}~\bibnamefont {Winter}}
  (\bibinfo {collaboration} {Nucleon Matrix Elements (NME)}),\ }\bibfield
  {title} {\enquote {\bibinfo {title} {{Precision nucleon charges and form
  factors using (2+1)-flavor lattice QCD}},}\ }\href {\doibase
  10.1103/PhysRevD.105.054505} {\bibfield  {journal} {\bibinfo  {journal}
  {Phys. Rev. D}\ }\textbf {\bibinfo {volume} {105}},\ \bibinfo {pages}
  {054505} (\bibinfo {year} {2022})},\ \Eprint
  {http://arxiv.org/abs/2103.05599} {arXiv:2103.05599 [hep-lat]} \BibitemShut
  {NoStop}%
\bibitem [{\citenamefont {Djukanovic}\ \emph {et~al.}(2022)\citenamefont
  {Djukanovic}, \citenamefont {von Hippel}, \citenamefont {Koponen},
  \citenamefont {Meyer}, \citenamefont {Ottnad}, \citenamefont {Schulz},\ and\
  \citenamefont {Wittig}}]{Djukanovic:2022wru}%
  \BibitemOpen
  \bibfield  {author} {\bibinfo {author} {\bibfnamefont {D.}~\bibnamefont
  {Djukanovic}}, \bibinfo {author} {\bibfnamefont {G.}~\bibnamefont {von
  Hippel}}, \bibinfo {author} {\bibfnamefont {J.}~\bibnamefont {Koponen}},
  \bibinfo {author} {\bibfnamefont {H.~B.}\ \bibnamefont {Meyer}}, \bibinfo
  {author} {\bibfnamefont {K.}~\bibnamefont {Ottnad}}, \bibinfo {author}
  {\bibfnamefont {T.}~\bibnamefont {Schulz}}, \ and\ \bibinfo {author}
  {\bibfnamefont {H.}~\bibnamefont {Wittig}},\ }\bibfield  {title} {\enquote
  {\bibinfo {title} {{Isovector axial form factor of the nucleon from lattice
  QCD}},}\ }\href {\doibase 10.1103/PhysRevD.106.074503} {\bibfield  {journal}
  {\bibinfo  {journal} {Phys. Rev. D}\ }\textbf {\bibinfo {volume} {106}},\
  \bibinfo {pages} {074503} (\bibinfo {year} {2022})},\ \Eprint
  {http://arxiv.org/abs/2207.03440} {arXiv:2207.03440 [hep-lat]} \BibitemShut
  {NoStop}%
\bibitem [{\citenamefont {Tsuji}\ \emph {et~al.}(2022)\citenamefont {Tsuji},
  \citenamefont {Tsukamoto}, \citenamefont {Aoki}, \citenamefont {Ishikawa},
  \citenamefont {Kuramashi}, \citenamefont {Sasaki}, \citenamefont {Shintani},\
  and\ \citenamefont {Yamazaki}}]{Tsuji:2022ric}%
  \BibitemOpen
  \bibfield  {author} {\bibinfo {author} {\bibfnamefont {R.}~\bibnamefont
  {Tsuji}}, \bibinfo {author} {\bibfnamefont {N.}~\bibnamefont {Tsukamoto}},
  \bibinfo {author} {\bibfnamefont {Y.}~\bibnamefont {Aoki}}, \bibinfo {author}
  {\bibfnamefont {K.-I.}\ \bibnamefont {Ishikawa}}, \bibinfo {author}
  {\bibfnamefont {Y.}~\bibnamefont {Kuramashi}}, \bibinfo {author}
  {\bibfnamefont {S.}~\bibnamefont {Sasaki}}, \bibinfo {author} {\bibfnamefont
  {E.}~\bibnamefont {Shintani}}, \ and\ \bibinfo {author} {\bibfnamefont
  {T.}~\bibnamefont {Yamazaki}} (\bibinfo {collaboration} {PACS}),\ }\bibfield
  {title} {\enquote {\bibinfo {title} {{Nucleon isovector couplings in
  $N_f=2+1$ lattice QCD at the physical point}},}\ }\href {\doibase
  10.1103/PhysRevD.106.094505} {\bibfield  {journal} {\bibinfo  {journal}
  {Phys. Rev. D}\ }\textbf {\bibinfo {volume} {106}},\ \bibinfo {pages}
  {094505} (\bibinfo {year} {2022})},\ \Eprint
  {http://arxiv.org/abs/2207.11914} {arXiv:2207.11914 [hep-lat]} \BibitemShut
  {NoStop}%
\bibitem [{\citenamefont {Alexandrou}\ \emph
  {et~al.}(2024{\natexlab{b}})\citenamefont {Alexandrou}, \citenamefont
  {Bacchio}, \citenamefont {Constantinou}, \citenamefont {Finkenrath},
  \citenamefont {Frezzotti}, \citenamefont {Kostrzewa}, \citenamefont
  {Koutsou}, \citenamefont {Spanoudes},\ and\ \citenamefont
  {Urbach}}]{Alexandrou:2023qbg}%
  \BibitemOpen
  \bibfield  {author} {\bibinfo {author} {\bibfnamefont {C.}~\bibnamefont
  {Alexandrou}}, \bibinfo {author} {\bibfnamefont {S.}~\bibnamefont {Bacchio}},
  \bibinfo {author} {\bibfnamefont {M.}~\bibnamefont {Constantinou}}, \bibinfo
  {author} {\bibfnamefont {J.}~\bibnamefont {Finkenrath}}, \bibinfo {author}
  {\bibfnamefont {R.}~\bibnamefont {Frezzotti}}, \bibinfo {author}
  {\bibfnamefont {B.}~\bibnamefont {Kostrzewa}}, \bibinfo {author}
  {\bibfnamefont {G.}~\bibnamefont {Koutsou}}, \bibinfo {author} {\bibfnamefont
  {G.}~\bibnamefont {Spanoudes}}, \ and\ \bibinfo {author} {\bibfnamefont
  {C.}~\bibnamefont {Urbach}} (\bibinfo {collaboration} {Extended Twisted
  Mass}),\ }\bibfield  {title} {\enquote {\bibinfo {title} {{Nucleon axial and
  pseudoscalar form factors using twisted-mass fermion ensembles at the
  physical point}},}\ }\href {\doibase 10.1103/PhysRevD.109.034503} {\bibfield
  {journal} {\bibinfo  {journal} {Phys. Rev. D}\ }\textbf {\bibinfo {volume}
  {109}},\ \bibinfo {pages} {034503} (\bibinfo {year} {2024}{\natexlab{b}})},\
  \Eprint {http://arxiv.org/abs/2309.05774} {arXiv:2309.05774 [hep-lat]}
  \BibitemShut {NoStop}%
\bibitem [{\citenamefont {Smail}\ \emph {et~al.}(2023)\citenamefont {Smail}
  \emph {et~al.}}]{QCDSFUKQCDCSSM:2023qlx}%
  \BibitemOpen
  \bibfield  {author} {\bibinfo {author} {\bibfnamefont {R.~E.}\ \bibnamefont
  {Smail}} \emph {et~al.} (\bibinfo {collaboration} {QCDSF/UKQCD/CSSM}),\
  }\bibfield  {title} {\enquote {\bibinfo {title} {{Constraining beyond the
  standard model nucleon isovector charges}},}\ }\href {\doibase
  10.1103/PhysRevD.108.094511} {\bibfield  {journal} {\bibinfo  {journal}
  {Phys. Rev. D}\ }\textbf {\bibinfo {volume} {108}},\ \bibinfo {pages}
  {094511} (\bibinfo {year} {2023})},\ \Eprint
  {http://arxiv.org/abs/2304.02866} {arXiv:2304.02866 [hep-lat]} \BibitemShut
  {NoStop}%
\bibitem [{\citenamefont {Tsuji}\ \emph {et~al.}(2024)\citenamefont {Tsuji},
  \citenamefont {Aoki}, \citenamefont {Ishikawa}, \citenamefont {Kuramashi},
  \citenamefont {Sasaki}, \citenamefont {Sato}, \citenamefont {Shintani},
  \citenamefont {Watanabe},\ and\ \citenamefont {Yamazaki}}]{Tsuji:2023llh}%
  \BibitemOpen
  \bibfield  {author} {\bibinfo {author} {\bibfnamefont {R.}~\bibnamefont
  {Tsuji}}, \bibinfo {author} {\bibfnamefont {Y.}~\bibnamefont {Aoki}},
  \bibinfo {author} {\bibfnamefont {K.-I.}\ \bibnamefont {Ishikawa}}, \bibinfo
  {author} {\bibfnamefont {Y.}~\bibnamefont {Kuramashi}}, \bibinfo {author}
  {\bibfnamefont {S.}~\bibnamefont {Sasaki}}, \bibinfo {author} {\bibfnamefont
  {K.}~\bibnamefont {Sato}}, \bibinfo {author} {\bibfnamefont {E.}~\bibnamefont
  {Shintani}}, \bibinfo {author} {\bibfnamefont {H.}~\bibnamefont {Watanabe}},
  \ and\ \bibinfo {author} {\bibfnamefont {T.}~\bibnamefont {Yamazaki}}
  (\bibinfo {collaboration} {PACS}),\ }\bibfield  {title} {\enquote {\bibinfo
  {title} {{Nucleon form factors in $N_f=2+1$ lattice QCD at the physical
  point: Finite lattice spacing effect on the root-mean-square radii}},}\
  }\href {\doibase 10.1103/PhysRevD.109.094505} {\bibfield  {journal} {\bibinfo
   {journal} {Phys. Rev. D}\ }\textbf {\bibinfo {volume} {109}},\ \bibinfo
  {pages} {094505} (\bibinfo {year} {2024})},\ \Eprint
  {http://arxiv.org/abs/2311.10345} {arXiv:2311.10345 [hep-lat]} \BibitemShut
  {NoStop}%
\bibitem [{\citenamefont {Jang}\ \emph {et~al.}(2024)\citenamefont {Jang},
  \citenamefont {Gupta}, \citenamefont {Bhattacharya}, \citenamefont {Yoon},\
  and\ \citenamefont {Lin}}]{Jang:2023zts}%
  \BibitemOpen
  \bibfield  {author} {\bibinfo {author} {\bibfnamefont {Y.-C.}\ \bibnamefont
  {Jang}}, \bibinfo {author} {\bibfnamefont {R.}~\bibnamefont {Gupta}},
  \bibinfo {author} {\bibfnamefont {T.}~\bibnamefont {Bhattacharya}}, \bibinfo
  {author} {\bibfnamefont {B.}~\bibnamefont {Yoon}}, \ and\ \bibinfo {author}
  {\bibfnamefont {H.-W.}\ \bibnamefont {Lin}} (\bibinfo {collaboration}
  {Precision Neutron Decay Matrix Elements (PNDME)}),\ }\bibfield  {title}
  {\enquote {\bibinfo {title} {{Nucleon isovector axial form factors}},}\
  }\href {\doibase 10.1103/PhysRevD.109.014503} {\bibfield  {journal} {\bibinfo
   {journal} {Phys. Rev. D}\ }\textbf {\bibinfo {volume} {109}},\ \bibinfo
  {pages} {014503} (\bibinfo {year} {2024})},\ \Eprint
  {http://arxiv.org/abs/2305.11330} {arXiv:2305.11330 [hep-lat]} \BibitemShut
  {NoStop}%
\bibitem [{\citenamefont {Epelbaum}\ \emph {et~al.}(2020)\citenamefont
  {Epelbaum}, \citenamefont {Krebs},\ and\ \citenamefont
  {Reinert}}]{Epelbaum:2019kcf}%
  \BibitemOpen
  \bibfield  {author} {\bibinfo {author} {\bibfnamefont {E.}~\bibnamefont
  {Epelbaum}}, \bibinfo {author} {\bibfnamefont {H.}~\bibnamefont {Krebs}}, \
  and\ \bibinfo {author} {\bibfnamefont {P.}~\bibnamefont {Reinert}},\
  }\bibfield  {title} {\enquote {\bibinfo {title} {{High-precision nuclear
  forces from chiral EFT: State-of-the-art, challenges and outlook}},}\ }\href
  {\doibase 10.3389/fphy.2020.00098} {\bibfield  {journal} {\bibinfo  {journal}
  {Front. in Phys.}\ }\textbf {\bibinfo {volume} {8}},\ \bibinfo {pages} {98}
  (\bibinfo {year} {2020})},\ \Eprint {http://arxiv.org/abs/1911.11875}
  {arXiv:1911.11875 [nucl-th]} \BibitemShut {NoStop}%
\bibitem [{\citenamefont {Hoferichter}\ \emph {et~al.}(2015)\citenamefont
  {Hoferichter}, \citenamefont {Ruiz~de Elvira}, \citenamefont {Kubis},\ and\
  \citenamefont {Mei\ss{}ner}}]{Hoferichter:2015tha}%
  \BibitemOpen
  \bibfield  {author} {\bibinfo {author} {\bibfnamefont {M.}~\bibnamefont
  {Hoferichter}}, \bibinfo {author} {\bibfnamefont {J.}~\bibnamefont {Ruiz~de
  Elvira}}, \bibinfo {author} {\bibfnamefont {B.}~\bibnamefont {Kubis}}, \ and\
  \bibinfo {author} {\bibfnamefont {U.-G.}\ \bibnamefont {Mei\ss{}ner}},\
  }\bibfield  {title} {\enquote {\bibinfo {title} {{Matching pion--nucleon
  Roy--Steiner equations to chiral perturbation theory}},}\ }\href {\doibase
  10.1103/PhysRevLett.115.192301} {\bibfield  {journal} {\bibinfo  {journal}
  {Phys. Rev. Lett.}\ }\textbf {\bibinfo {volume} {115}},\ \bibinfo {pages}
  {192301} (\bibinfo {year} {2015})},\ \Eprint
  {http://arxiv.org/abs/1507.07552} {arXiv:1507.07552 [nucl-th]} \BibitemShut
  {NoStop}%
\bibitem [{\citenamefont {Hoferichter}\ \emph {et~al.}(2016)\citenamefont
  {Hoferichter}, \citenamefont {Ruiz~de Elvira}, \citenamefont {Kubis},\ and\
  \citenamefont {Mei\ss{}ner}}]{Hoferichter:2015hva}%
  \BibitemOpen
  \bibfield  {author} {\bibinfo {author} {\bibfnamefont {M.}~\bibnamefont
  {Hoferichter}}, \bibinfo {author} {\bibfnamefont {J.}~\bibnamefont {Ruiz~de
  Elvira}}, \bibinfo {author} {\bibfnamefont {B.}~\bibnamefont {Kubis}}, \ and\
  \bibinfo {author} {\bibfnamefont {U.-G.}\ \bibnamefont {Mei\ss{}ner}},\
  }\bibfield  {title} {\enquote {\bibinfo {title}
  {{Roy\textendash{}Steiner-equation analysis of pion\textendash{}nucleon
  scattering}},}\ }\href {\doibase 10.1016/j.physrep.2016.02.002} {\bibfield
  {journal} {\bibinfo  {journal} {Phys. Rept.}\ }\textbf {\bibinfo {volume}
  {625}},\ \bibinfo {pages} {1--88} (\bibinfo {year} {2016})},\ \Eprint
  {http://arxiv.org/abs/1510.06039} {arXiv:1510.06039 [hep-ph]} \BibitemShut
  {NoStop}%
\bibitem [{\citenamefont {Hebeler}(2021)}]{Hebeler:2020ocj}%
  \BibitemOpen
  \bibfield  {author} {\bibinfo {author} {\bibfnamefont {K.}~\bibnamefont
  {Hebeler}},\ }\bibfield  {title} {\enquote {\bibinfo {title} {{Three-nucleon
  forces: Implementation and applications to atomic nuclei and dense
  matter}},}\ }\href {\doibase 10.1016/j.physrep.2020.08.009} {\bibfield
  {journal} {\bibinfo  {journal} {Phys. Rept.}\ }\textbf {\bibinfo {volume}
  {890}},\ \bibinfo {pages} {1--116} (\bibinfo {year} {2021})},\ \Eprint
  {http://arxiv.org/abs/2002.09548} {arXiv:2002.09548 [nucl-th]} \BibitemShut
  {NoStop}%
\bibitem [{\citenamefont {Krebs}(2020)}]{Krebs:2020pii}%
  \BibitemOpen
  \bibfield  {author} {\bibinfo {author} {\bibfnamefont {H.}~\bibnamefont
  {Krebs}},\ }\bibfield  {title} {\enquote {\bibinfo {title} {{Nuclear Currents
  in Chiral Effective Field Theory}},}\ }\href {\doibase
  10.1140/epja/s10050-020-00230-9} {\bibfield  {journal} {\bibinfo  {journal}
  {Eur. Phys. J. A}\ }\textbf {\bibinfo {volume} {56}},\ \bibinfo {pages} {234}
  (\bibinfo {year} {2020})},\ \Eprint {http://arxiv.org/abs/2008.00974}
  {arXiv:2008.00974 [nucl-th]} \BibitemShut {NoStop}%
\bibitem [{\citenamefont {Tews}\ \emph {et~al.}(2022)\citenamefont {Tews} \emph
  {et~al.}}]{Tews:2022yfb}%
  \BibitemOpen
  \bibfield  {author} {\bibinfo {author} {\bibfnamefont {I.}~\bibnamefont
  {Tews}} \emph {et~al.},\ }\bibfield  {title} {\enquote {\bibinfo {title}
  {{Nuclear Forces for Precision Nuclear Physics: A Collection of
  Perspectives}},}\ }\href {\doibase 10.1007/s00601-022-01749-x} {\bibfield
  {journal} {\bibinfo  {journal} {Few Body Syst.}\ }\textbf {\bibinfo {volume}
  {63}},\ \bibinfo {pages} {67} (\bibinfo {year} {2022})},\ \Eprint
  {http://arxiv.org/abs/2202.01105} {arXiv:2202.01105 [nucl-th]} \BibitemShut
  {NoStop}%
\bibitem [{\citenamefont {Hoferichter}\ \emph {et~al.}(2020)\citenamefont
  {Hoferichter}, \citenamefont {Men\'endez},\ and\ \citenamefont
  {Schwenk}}]{Hoferichter:2020osn}%
  \BibitemOpen
  \bibfield  {author} {\bibinfo {author} {\bibfnamefont {M.}~\bibnamefont
  {Hoferichter}}, \bibinfo {author} {\bibfnamefont {J.}~\bibnamefont
  {Men\'endez}}, \ and\ \bibinfo {author} {\bibfnamefont {A.}~\bibnamefont
  {Schwenk}},\ }\bibfield  {title} {\enquote {\bibinfo {title} {{Coherent
  elastic neutrino--nucleus scattering: EFT analysis and nuclear responses}},}\
  }\href {\doibase 10.1103/PhysRevD.102.074018} {\bibfield  {journal} {\bibinfo
   {journal} {Phys. Rev. D}\ }\textbf {\bibinfo {volume} {102}},\ \bibinfo
  {pages} {074018} (\bibinfo {year} {2020})},\ \Eprint
  {http://arxiv.org/abs/2007.08529} {arXiv:2007.08529 [hep-ph]} \BibitemShut
  {NoStop}%
\bibitem [{\citenamefont {Manohar}\ and\ \citenamefont
  {Georgi}(1984)}]{Manohar:1983md}%
  \BibitemOpen
  \bibfield  {author} {\bibinfo {author} {\bibfnamefont {A.}~\bibnamefont
  {Manohar}}\ and\ \bibinfo {author} {\bibfnamefont {H.}~\bibnamefont
  {Georgi}},\ }\bibfield  {title} {\enquote {\bibinfo {title} {{Chiral Quarks
  and the Nonrelativistic Quark Model}},}\ }\href {\doibase
  10.1016/0550-3213(84)90231-1} {\bibfield  {journal} {\bibinfo  {journal}
  {Nucl. Phys. B}\ }\textbf {\bibinfo {volume} {234}},\ \bibinfo {pages}
  {189--212} (\bibinfo {year} {1984})}\BibitemShut {NoStop}%
\bibitem [{\citenamefont {Hoferichter}\ \emph {et~al.}()\citenamefont
  {Hoferichter}, \citenamefont {Mereghetti}, \citenamefont {Ruiz~de Elvira},
  \citenamefont {Siemens},\ and\ \citenamefont {Walker-Loud}}]{gA_inprep}%
  \BibitemOpen
  \bibfield  {author} {\bibinfo {author} {\bibfnamefont {M.}~\bibnamefont
  {Hoferichter}}, \bibinfo {author} {\bibfnamefont {E.}~\bibnamefont
  {Mereghetti}}, \bibinfo {author} {\bibfnamefont {J.}~\bibnamefont {Ruiz~de
  Elvira}}, \bibinfo {author} {\bibfnamefont {D.}~\bibnamefont {Siemens}}, \
  and\ \bibinfo {author} {\bibfnamefont {A.}~\bibnamefont {Walker-Loud}},\
  }\href@noop {} {\ }\bibinfo {note} {In preparation}\BibitemShut {NoStop}%
\bibitem [{lal()}]{lalibe}%
  \BibitemOpen
  \href@noop {} {\enquote {\bibinfo {title} {{LALIBE: CalLat C}ollaboration
  code for lattice qcd calculations},}\ }\bibinfo {note}
  {\url{https://github.com/callat-qcd/lalibe}}\BibitemShut {NoStop}%
\bibitem [{\citenamefont {Edwards}\ and\ \citenamefont
  {Joo}(2005)}]{Edwards:2004sx}%
  \BibitemOpen
  \bibfield  {author} {\bibinfo {author} {\bibfnamefont {R.~G.}\ \bibnamefont
  {Edwards}}\ and\ \bibinfo {author} {\bibfnamefont {B.}~\bibnamefont {Joo}}
  (\bibinfo {collaboration} {SciDAC, LHPC, UKQCD}),\ }\bibfield  {title}
  {\enquote {\bibinfo {title} {{The Chroma software system for lattice QCD}},}\
  }\href {\doibase 10.1016/j.nuclphysbps.2004.11.254} {\bibfield  {journal}
  {\bibinfo  {journal} {Nucl. Phys. B Proc. Suppl.}\ }\textbf {\bibinfo
  {volume} {140}},\ \bibinfo {pages} {832} (\bibinfo {year} {2005})},\ \Eprint
  {http://arxiv.org/abs/hep-lat/0409003} {arXiv:hep-lat/0409003} \BibitemShut
  {NoStop}%
\bibitem [{\citenamefont {{The HDF Group}}(NNNN)}]{hdf5}%
  \BibitemOpen
  \bibfield  {author} {\bibinfo {author} {\bibnamefont {{The HDF Group}}},\
  }\href@noop {} {\enquote {\bibinfo {title} {{Hierarchical Data Format,
  version 5}},}\ } (\bibinfo {year} {1997-NNNN}),\ \bibinfo {note}
  {\url{http://www.hdfgroup.org/HDF5/}}\BibitemShut {NoStop}%
\bibitem [{\citenamefont {Kurth}\ \emph {et~al.}(2015)\citenamefont {Kurth},
  \citenamefont {Pochinsky}, \citenamefont {Sarje}, \citenamefont {Syritsyn},\
  and\ \citenamefont {Walker-Loud}}]{Kurth:2015mqa}%
  \BibitemOpen
  \bibfield  {author} {\bibinfo {author} {\bibfnamefont {T.}~\bibnamefont
  {Kurth}}, \bibinfo {author} {\bibfnamefont {A.}~\bibnamefont {Pochinsky}},
  \bibinfo {author} {\bibfnamefont {A.}~\bibnamefont {Sarje}}, \bibinfo
  {author} {\bibfnamefont {S.}~\bibnamefont {Syritsyn}}, \ and\ \bibinfo
  {author} {\bibfnamefont {A.}~\bibnamefont {Walker-Loud}},\ }\bibfield
  {title} {\enquote {\bibinfo {title} {{High-Performance I/O: HDF5 for Lattice
  QCD}},}\ }\href {\doibase 10.22323/1.214.0045} {\bibfield  {journal}
  {\bibinfo  {journal} {PoS}\ }\textbf {\bibinfo {volume} {LATTICE2014}},\
  \bibinfo {pages} {045} (\bibinfo {year} {2015})},\ \Eprint
  {http://arxiv.org/abs/1501.06992} {arXiv:1501.06992 [hep-lat]} \BibitemShut
  {NoStop}%
\bibitem [{\citenamefont {Berkowitz}(2017)}]{Berkowitz:2017vcp}%
  \BibitemOpen
  \bibfield  {author} {\bibinfo {author} {\bibfnamefont {E.}~\bibnamefont
  {Berkowitz}},\ }\bibfield  {title} {\enquote {\bibinfo {title} {{METAQ:
  Bundle Supercomputing Tasks}},}\ }\href@noop {} {\  (\bibinfo {year}
  {2017})},\ \bibinfo {note} {\url{https://github.com/evanberkowitz/metaq}},\
  \Eprint {http://arxiv.org/abs/1702.06122} {arXiv:1702.06122
  [physics.comp-ph]} \BibitemShut {NoStop}%
\bibitem [{\citenamefont {Berkowitz}\ \emph {et~al.}(2018)\citenamefont
  {Berkowitz}, \citenamefont {Jansen}, \citenamefont {McElvain},\ and\
  \citenamefont {Walker-Loud}}]{Berkowitz:2017xna}%
  \BibitemOpen
  \bibfield  {author} {\bibinfo {author} {\bibfnamefont {E.}~\bibnamefont
  {Berkowitz}}, \bibinfo {author} {\bibfnamefont {G.~R.}\ \bibnamefont
  {Jansen}}, \bibinfo {author} {\bibfnamefont {K.}~\bibnamefont {McElvain}}, \
  and\ \bibinfo {author} {\bibfnamefont {A.}~\bibnamefont {Walker-Loud}},\
  }\bibfield  {title} {\enquote {\bibinfo {title} {{Job Management and Task
  Bundling}},}\ }\href {\doibase 10.1051/epjconf/201817509007} {\bibfield
  {journal} {\bibinfo  {journal} {EPJ Web Conf.}\ }\textbf {\bibinfo {volume}
  {175}},\ \bibinfo {pages} {09007} (\bibinfo {year} {2018})},\ \Eprint
  {http://arxiv.org/abs/1710.01986} {arXiv:1710.01986 [hep-lat]} \BibitemShut
  {NoStop}%
\bibitem [{mil()}]{milc:code}%
  \BibitemOpen
  \href@noop {} {\enquote {\bibinfo {title} {{MILC C}ollaboration code for
  lattice qcd calculations},}\ }\bibinfo {note}
  {\url{https://github.com/milc-qcd/milc_qcd}}\BibitemShut {NoStop}%
\bibitem [{\citenamefont {Lepage}(2020{\natexlab{a}})}]{gvar:11.2}%
  \BibitemOpen
  \bibfield  {author} {\bibinfo {author} {\bibfnamefont {P.}~\bibnamefont
  {Lepage}},\ }\href@noop {} {\enquote {\bibinfo {title} {gplepage/gvar},}\ }
  (\bibinfo {year} {2020}{\natexlab{a}}),\ \bibinfo {note}
  {\url{https://github.com/gplepage/gvar}}\BibitemShut {NoStop}%
\bibitem [{\citenamefont {Lepage}(2020{\natexlab{b}})}]{lsqfit:11.5.1}%
  \BibitemOpen
  \bibfield  {author} {\bibinfo {author} {\bibfnamefont {P.}~\bibnamefont
  {Lepage}},\ }\href@noop {} {\enquote {\bibinfo {title} {gplepage/lsqfit},}\ }
  (\bibinfo {year} {2020}{\natexlab{b}}),\ \bibinfo {note}
  {\url{https://github.com/gplepage/lsqfit}}\BibitemShut {NoStop}%
\bibitem [{\citenamefont {Gupta}\ \emph {et~al.}(2021)\citenamefont {Gupta},
  \citenamefont {Park}, \citenamefont {Hoferichter}, \citenamefont
  {Mereghetti}, \citenamefont {Yoon},\ and\ \citenamefont
  {Bhattacharya}}]{Gupta:2021ahb}%
  \BibitemOpen
  \bibfield  {author} {\bibinfo {author} {\bibfnamefont {R.}~\bibnamefont
  {Gupta}}, \bibinfo {author} {\bibfnamefont {S.}~\bibnamefont {Park}},
  \bibinfo {author} {\bibfnamefont {M.}~\bibnamefont {Hoferichter}}, \bibinfo
  {author} {\bibfnamefont {E.}~\bibnamefont {Mereghetti}}, \bibinfo {author}
  {\bibfnamefont {B.}~\bibnamefont {Yoon}}, \ and\ \bibinfo {author}
  {\bibfnamefont {T.}~\bibnamefont {Bhattacharya}},\ }\bibfield  {title}
  {\enquote {\bibinfo {title} {{Pion\textendash{}Nucleon Sigma Term from
  Lattice QCD}},}\ }\href {\doibase 10.1103/PhysRevLett.127.242002} {\bibfield
  {journal} {\bibinfo  {journal} {Phys. Rev. Lett.}\ }\textbf {\bibinfo
  {volume} {127}},\ \bibinfo {pages} {242002} (\bibinfo {year} {2021})},\
  \Eprint {http://arxiv.org/abs/2105.12095} {arXiv:2105.12095 [hep-lat]}
  \BibitemShut {NoStop}%
\bibitem [{\citenamefont {Gupta}\ \emph {et~al.}(2024)\citenamefont {Gupta},
  \citenamefont {Bhattacharya}, \citenamefont {Hoferichter}, \citenamefont
  {Mereghetti}, \citenamefont {Park},\ and\ \citenamefont
  {Yoon}}]{Gupta:2022aba}%
  \BibitemOpen
  \bibfield  {author} {\bibinfo {author} {\bibfnamefont {R.}~\bibnamefont
  {Gupta}}, \bibinfo {author} {\bibfnamefont {T.}~\bibnamefont {Bhattacharya}},
  \bibinfo {author} {\bibfnamefont {M.}~\bibnamefont {Hoferichter}}, \bibinfo
  {author} {\bibfnamefont {E.}~\bibnamefont {Mereghetti}}, \bibinfo {author}
  {\bibfnamefont {S.}~\bibnamefont {Park}}, \ and\ \bibinfo {author}
  {\bibfnamefont {B.}~\bibnamefont {Yoon}},\ }\bibfield  {title} {\enquote
  {\bibinfo {title} {{The pion--nucleon sigma term from Lattice QCD}},}\ }\href
  {\doibase 10.22323/1.413.0060} {\bibfield  {journal} {\bibinfo  {journal}
  {PoS}\ }\textbf {\bibinfo {volume} {CD2021}},\ \bibinfo {pages} {060}
  (\bibinfo {year} {2024})},\ \Eprint {http://arxiv.org/abs/2203.13862}
  {arXiv:2203.13862 [hep-lat]} \BibitemShut {NoStop}%
\bibitem [{\citenamefont {Siemens}(2020)}]{Siemens:2020vop}%
  \BibitemOpen
  \bibfield  {author} {\bibinfo {author} {\bibfnamefont {D.}~\bibnamefont
  {Siemens}},\ }\bibfield  {title} {\enquote {\bibinfo {title} {{Elastic
  pion--nucleon scattering in chiral perturbation theory: Explicit
  $\Delta$(1232) degrees of freedom}},}\ }\href@noop {} {\  (\bibinfo {year}
  {2020})},\ \Eprint {http://arxiv.org/abs/2001.03906} {arXiv:2001.03906
  [nucl-th]} \BibitemShut {NoStop}%
\end{thebibliography}%
